\documentstyle{article}
\def\brr{\begin{array}}
\def\beq{\begin{equation}}
\def\ben{\begin{enumerate}}
\def\een{\end{enumerate}}
\def\err{\end{array}}
\def\eeq{\end{equation}}
\def\bea{\begin{eqnarray}}
\def\eea{\end{eqnarray}}
\def\bs{\bigskip}
\def\tr{\mbox{Tr}\, }
\def\ni{\noindent}
\def\wt{\widetilde}
\def\wh{\widehat}
\def\ul{\underline}

\def\nn{\nonumber}
\def\ms{\medskip}

\def\cL{{\cal L}}
\def\be{\begin{equation}}
\def\ee{\end{equation}}
\def\bea{\begin{eqnarray}}
\def\eea{\end{eqnarray}}
\def\tr{{\rm tr}\, }
\def\nn{\nonumber \\}
\def\gd{g^\dagger}
\def\e{{\rm e}}
\def\cF{{\cal F}}
\def\det{{\rm det\,}}
\def\Tr{{\rm Tr\,}}
\def\tr{{\rm tr\,}}
\def\e{{\rm e}}
\def\etal{{\it et al.}}
\def\erp2{{\rm e}^{2\rho}}
\def\erm2{{\rm e}^{-2\rho}}
\def\er4{{\rm e}^{4\rho}}

\begin{document}

\begin{titlepage}
\begin{center}

\ 

\vfill

{\Huge {\bf Quantum dilatonic gravity in $d=$2, 4 and 5 dimensions}}

\vfill

{\sc Shin'ichi NOJIRI}\footnote{\scriptsize 
e-mail: nojiri@cc.nda.ac.jp, snojiri@yukawa.kyoto-u.ac.jp} and
{\sc Sergei D. ODINTSOV$^{\spadesuit}$}\footnote{\scriptsize 
e-mail: odintsov@ifug5.ugto.mx, odintsov@mail.tomsknet.ru}

\vfill

{\sl Department of Mathematics and Physics \\
National Defence Academy, 
Hashirimizu Yokosuka 239, JAPAN}

\ 

{\sl $\spadesuit$ 
Instituto de Fisica de la Universidad de Guanajuato, \\
Lomas del Bosque 103, Apdo. Postal E-143, 37150 Leon, Gto.,
 MEXICO \\
and \\
Tomsk State Pedagogical University, 634041 Tomsk, RUSSIA }

\vfill

{\bf Abstract}

\end{center}

We review (mainly) quantum effects in the theories where 
gravity sector is described by metric and dilaton. 
The one-loop effective action for dilatonic gravity in 
two and four dimensions is evaluated.
Renormalization group equations are constructed. 
The conformal anomaly and induced effective action 
for 2d and 4d dilaton coupled theories are 
found. It is applied to study of quantum aspects of 
black hole thermodynamics, like calculation of Hawking radiation 
and quantum corrections to black hole parameters and investigation
of quantum instability for such objects with multiple horizons.
The use of above effective action in the construction of 
non-singular cosmological models in Einstein or Brans-Dicke 
(super)gravity and investigation of induced wormholes 
in supersymmetric Yang-Mills theory are given. 

5d dilatonic gravity (bosonic sector of compactified IIB 
supergravity) is discussed in connection with bulk/boundary 
(or AdS/CFT) correspondence. 
Running gauge coupling and quark-antiquark potential 
for boundary  gauge theory at zero or non-zero 
temperature are calculated from $d=5$ dilatonic Anti-de 
Sitter-like background solution which 
represents Anti-de Sitter black hole for periodic time.

\end{titlepage}

\newpage
\pagenumbering{roman}
\tableofcontents
\newpage

\setcounter{page}{1}
\pagenumbering{arabic}

\section{Introduction: Why dilatonic gravity?}

Dilatonic gravity represents very natural generalization of general 
relativity where gravitational sector is described by metric tensor 
and also by the scalar partner (dilaton). In principle, any covariant 
theory of gravity interacting in some way with scalar which is 
assumed to describe not matter may be called the dilatonic gravity. 
The simplest form of dilatonic gravity is given by general 
relativity plus scalar field Lagrangian where 
scalar sector may contain kinetic term and some potential. It is 
then clear that for constant scalar such a theory is reduced to 
general relativity (with cosmological constant in case of presence 
of scalar potential). One may think about more complicated 
generalizations where curvature non-minimally 
couples with scalar. 

Historically, one of the first theories of this sort has been 
suggested by Brans and Dicke. It is still considered as an 
alternative to general relativity. Its action in so-called 
Jordan frame is given as 
\be
S_{BD}=\frac{1}{16\pi}\int d^4x \sqrt{-g} \left[ 
\phi R -\frac{\omega}{\phi}
(\nabla_\mu \phi)(\nabla^\mu \phi) \right]+S_M\,,
\label{bdj0}
\ee
where $\phi$ is scalar (Brans-Dicke dilaton), $R$ is curvature, 
$\omega$ is Brans-Dicke parameter and $S_M$ is usual matter action. 
As one can see in Jordan frame the gravitational field is described 
by the metric tensor and by dilaton.

The important role in dilatonic gravity is played by conformal
transformations.
Generally speaking, starting from the spacetime $M$ with metric 
$g_{\mu\nu}$ one can make (non-singular) coordinates dependent 
rescaling of metric
$g_{\mu\nu}\rightarrow \tilde g_{\mu\nu} =\exp\left(
\sigma(x)\right)g_{\mu\nu}$ where $\sigma(x)$ is some arbitrary 
function of coordinates. Such rescaling is called Weyl or 
conformal transformation. It leaves the light cones 
unchanged, in other words, the spacetime $M$ with metric $g_{\mu\nu}$ 
and the one with metric $\tilde g_{\mu\nu}$ have the same causal 
structure.

This conformal transformation is extremely important as it permits 
to map one classical theory to another theory which is sometimes 
easier to study. 
The explicit example of such mapping (dilaton is also transformed) 
is given in section 5.2 (see also section 2.1)
 where Brans-Dicke-matter theory in 
Jordan frame is transformed to the one in Einstein frame. As a 
result, in gravity sector the curvature appears as in Einstein 
theory (minimal coupling with dilaton) while matter is getting 
dilaton coupled. There may be given arguments that Einstein frame 
is the physical one. 

Conformal transformations are especially important in conformal 
quantum theory. It means that on classical level such 
theories do not change under conformal transformations. The 
manifestation of this symmetry on the conservation laws level 
is the vanishing of energy-momentum tensor trace, i.e. 
${T_\mu}^\mu=0$. However, on quantum level one needs 
to apply some regularization and (or) renormalization for calculation 
of formally divergent, infinite quantities. Preserving covariance in 
such regularized calculations, one breaks the conformal invariance 
on quantum level. As a result, trace of quantum energy-momentum 
tensor is not zero. There appears so-called conformal or Weyl anomaly 
(for a review, see \cite{Dff1}). The practical importance of 
conformal anomaly is related with the possibility to integrate 
over it and to construct the anomaly induced, finite effective 
action. (In other words, instead of taking into account quantum 
effects one is forced to add to classical gravitational action 
some (non-local) finite gravitational action which 
is classical one and which role is to describe the effects 
related with quantum theory). This property will be extensively 
used in the present review for conformally invariant, dilaton 
coupled theories where anomaly induced effective action on the 
arbitrary dilaton-gravitational background is found.
(Note that usually one can calculate the effective action or on some 
specific background or approximately, see book \cite{BOS1} for 
a review).  

Another big class of theories where classical dilatonic gravity 
naturally appears is related with the process of compactification 
(or reduction). As an example, let us consider six-dimensional 
Einstein gravity (with cosmological constant). Imagine that one 
is looking to the background of the sort: $M_4\times S_2$ where 
$M_4$ is an arbitrary four-dimensional manifold 
and $S_2$ is two-dimensional sphere with the radius $r$ which 
depends explicitly from the coordinates of the space $M_4$. 
The metric of the background $M_4\times S_2$ may be substituted 
to the action of above six-dimensional Kaluza-Klein gravity. 
After integration over coordinates of two-dimensional sphere one 
is left with effective 4d dilatonic gravity of special form 
where Kaluza-Klein dilaton (scalar) originates from the radius of 
$S_2$ and 6d cosmological constant (if presents) gives rise to 
dilatonic potential.
Similarly, after spherical (or hyperbolic, or torus) reduction of 
4d Einstein gravity one gets two-dimensional dilatonic gravity. 
That is why in many cases 
usual four-dimensional backgrounds of Einstein gravity may be also 
understood as 2d dilatonic backgrounds. In this sense, standard 
Schwarzschild black hole (no dilaton) is the same as 2d dilatonic 
black hole of special form. Then, the calculations performed for 
2d dilatonic gravity may very often give the answers for 4d 
Einstein gravity and vice-versa. 

 From another side this property gives additional motivation 
to study the dilatonic gravity. Indeed, even the hypothetical 
proof of non-existence of dilaton in Nature would be given 
one may still work with 2d dilatonic gravity 
having in mind that this is just another way to describe Einstein
gravity after 
reduction. Similarly, if Universe was multi-dimensional at some stage 
one can argue the presence of Kaluza-Klein dilaton at early Universe. 
Of course, such dilaton should quickly decay.

Nevertheless, the most natural way to introduce the dilaton 
comes presumably from supergravity and (super)strings. First of all,
in many cases supersymmetry dictates the presence of dilaton in 
supermultiplet as in 2d dilatonic supergravity of section 3.6 
 or IIB supergravity
in section 6.2. 
For strings the situation is even more dramatical: all versions of 
string theory predict the presence of scalar partner to usual tensor 
graviton. It is assumed that such scalar should acquire a mass 
(thanks to some, not yet well-understood dynamical mechanism). 
Then, macroscopic consequences of Einstein and string dilaton 
gravity are more-less equivalent. 
Otherwise, if dilaton is massless the theory may become 
non-consistent due to possible violation of (weak) equivalence 
principle and cosmological variations of gauge coupling 
constants, etc.(However, some mechanisms to
make theory consistent even for massless dilaton may be suggested).
 We will not discuss 
this circle of questions here as we are mainly dealing with quantum 
dilatonic gravity  but not with macroscopic 
manifestations. Moreover, even for classical dilatonic gravity 
there remain still some unresolved questions. For example, the 
dilaton may serve sometimes as panacea to
explain new observational effects like Pioner effect\cite{Mbelek}.

The purpose of this work is to give the review of results and methods 
in (mainly quantum) dilatonic gravity in various dimensions. 
We attempt to present the details of typical calculations in 
the given examples in a self-consistent way so that in most cases 
there is no necessity to consult 
original works in order to understand the main points.

In the next chapter we review the background field method evaluation 
of the divergent part of the one-loop effective action in 2d and 4d
dilatonic gravities where dilaton is dimensionless scalar. 
Consideration is done of general (non-renormalizable) model of 2d dilatonic
gravity 
(with matter) which includes string-inspired models and the ones 
reduced from Einstein gravity. 
The obtained one-loop effective action is applied to show that 
2d dilatonic gravities which are 
classically equivalent are not equivalent on quantum level, generally
speaking. 
The equivalence is restored only on-shell. That alerts on the danger 
of making quantum calculations in convenient parameterization and then 
using results of these calculations in another (physical) 
parameterization of fields.
As another application of the effective action we give the 
renormalization group study of dilatonic gravity near two 
dimensions, using $\epsilon$-expansion technique. It is shown 
that theory under consideration may have stable 
fixed points for all coupling functions. 

In the third chapter the calculation of conformal anomaly for Weyl 
invariant, dilaton coupled 2d and 4d scalars and spinors as well 
as for 4d dilaton coupled vector is presented. Such $n$-dimensional 
dilaton coupled matter appears from usual $n+2$-dimensional matter 
(no dilaton) after reduction. Integrating over such conformal 
anomaly one is able to construct finite, anomaly induced action 
which is not limited to some specific background. This is very 
powerful method to get the effective action on the arbitrary 
dilaton-gravitational background. Unfortunately, it cannot be 
applied to dilatonic gravity itself
as such theory is not, as a rule, Weyl invariant. Of course, 
working in 
large $N$ approximation one can neglect the proper quantum gravity
contribution to the effective action. As one example the large $N$, 
anomaly induced effective action is presented for 2d dilatonic 
supergravity with dilaton coupled Wess-Zumino model.

 Fourth chapter is devoted to study of quantum aspects of (dilatonic) 
black holes using above
induced effective action. The overview of two simple models of
2d dilatonic black holes where quantum effects of minimal matter 
induce Hawking radiation is given. In these models the dilaton 
is classical one as it 
does not appear in quantum effective action. Then, the calculation of
quantum corrections to thermodynamics of 4d Schwarzschild and
Schwarzschild-(anti) de Sitter black hole is presented where 
effective action of third chapter (in
large $N$ and $s$-wave approximation) is extensively used. 
In particular, quantum corrections to black hole horizon radii, 
temperature, mass and entropy are found. The important property 
of such results is that such quantum corrections are also 
the quantum corrections to 2d dilatonic black holes.
This double role of results is discussed. 
At the final section of fourth chapter we discuss the effect of
anti-evaporation (i.e. increase of horizon radius) 
of multiply horizon black holes due to quantum effects of matter.
As an example, again Schwarzschild-de Sitter and 2d charged dilatonic 
black holes are taken and anomaly induced effective action is used.
Above effect is opposite to the well-known Hawking radiation. 

Fifth chapter is devoted to presentation of some elements of 
(dilatonic) quantum cosmology. Again with large $N$ anomaly 
induced effective action due to quantum conformally invariant matter  
we construct non-singular 4d Kantowski-Sachs Universe. 
Such Universe may be also presented as 2d dilatonic cosmology. 
For 2d dilatonic supergravity (or some its modifications) with 
Wess-Zumino matter the role of quantum effects of matter to 
early Universe evolution is carefully discussed.
For Brans-Dicke theory with dilaton coupled matter (Einstein frame) 
the quantum matter effects lead to the possibility of 
inflationary Universe with quickly decaying dilaton. 
 In the last section 
we show the possibility to induce the 
primordial wormholes 
which are hard to realize in terms of classical gravity  
by quantum effects of Grand Unified Theories in the early Universe.

In the last, sixth chapter we study mainly classical aspects 
of odd-dimensional 
(mainly five-dimensional) dilatonic gravity descending 
from IIB supergravity (bosonic sector). While we look for classical 
Anti-de Sitter-like solutions 
of such theory and it may look distant from quantum dilatonic gravity,
it is not so. The reason is well-accepted now bulk/boundary 
correspondence.
With it from classical solutions of IIB supergravity we get 
the description of dual gauge theory which lives on the boundary.
In the first section of this chapter from one approximate solution 
of 3d and 5d axion-dilatonic gravity we get correct conformal 
anomaly studied from quantum field theory point of view in 
chapter 3 for 2d dilaton coupled theories and for maximally 
supersymmetric Yang-Mills theory 
conformally coupled with conformal supergravity. In the next section 
we present the 5d dilatonic solution interpolating between 
Anti-de Sitter space and flat space with singular dilaton. 
This solution may be interpreted as the one describing running 
gauge coupling in boundary gauge theory. In addition, the Wilson 
loop is obtained and the possibility of confinement is 
mentioned. The generalization of this solution (approximate 
Anti-de Sitter black hole) to include non-zero
temperature is done. Then, from calculation of entropy 
and free energy  of such 5d black hole one can get the entropy 
and free energy of boundary gauge theory at strong 
coupling limit. The role of higher derivative terms in such 
correspondence 
is discussed and free energy of  super Yang-Mills  theory
(with two supersymmetries) is found.

Some resume and open questions are listed in the Discussion. 
In this review, we mainly use the following conventions of 
curvatures : 
\bea
\label{curv0}
&& R=g^{\mu\nu}R_{\mu\nu} \ ,\quad 
R_{\mu\nu}= -\Gamma^\lambda_{\mu\lambda,\kappa}
+ \Gamma^\lambda_{\mu\kappa,\lambda}
- \Gamma^\eta_{\mu\lambda}\Gamma^\lambda_{\kappa\eta}
+ \Gamma^\eta_{\mu\kappa}\Gamma^\lambda_{\lambda\eta} \nn
&& \Gamma^\eta_{\mu\lambda}={1 \over 2}g^{\eta\nu}\left(
g_{\mu\nu,\lambda} + g_{\lambda\nu,\mu} - g_{\mu\lambda,\nu} 
\right)
\eea
and we choose the Einstein action in the following form :
\be
\label{E0}
S_{\rm Einstein}=-{1 \over \kappa^2}\int d^dx \sqrt{-g}R
\ee
However, in some cases we use also Euclidean notations
\be
\label{EE0}
S_{\rm Euclid}={1 \over \kappa^2}\int d^dx \sqrt{g}R\ .
\ee
The reason is that we follow to results of different works, often done 
with different notations.

\section{One-loop effective action in dilatonic gravity}

In the present Chapter we discuss the perturbative approach to
dilatonic gravity. The one-loop effective action is calculated and 
renormalization structure is discussed. Some immediate consequences 
of the effective action like renormalization group equations near two 
and in four dimensions are presented. The comparison of 
off-shell and on-shell ( S-matrix) effective actions for 
classically equivalent models is made.

\subsection{Perturbative renormalization of two-dimensional 
dilatonic gravity}

In this section, we discuss the covariant effective action
corresponding to a very general, multiplicatively 
renormalizable \cite{OS1,OS2,OS3,BO} (in generalized sense)  model 
of 2d dilatonic gravity with matter. Its action has the following
form
\be
\label{IIIi}
S = - \int d^2x \sqrt{g} \left[ \frac{1}{2} Z( \Phi) g^{\mu\nu} 
\partial_\mu \Phi \partial_\nu \Phi + C (\Phi ) R 
 -\frac{1}{2} f (\Phi) g^{\mu\nu} \partial_\mu \chi_i 
\partial_\nu \chi_i + V(\Phi, \chi) \right].
\ee
It includes a dilaton field $\Phi$ and real scalar $\chi_i$.

This action describes and generalizes many well-known dilaton
models. For instance, the celebrated bosonic string
effective action corresponds to
\begin{equation}
\label{IIIii}
Z(\Phi )= 8 \e^{-2\Phi}, \ C(\Phi) =  \e^{-2\Phi}, \ V(\Phi)  =4
\lambda^2  \e^{-2\Phi}, \  f(\Phi)
=1.
\end{equation}
On the other hand, in the absence of matter our action for
\begin{equation}
\label{IIIiii}
Z=1, \ \ \ \ C(\Phi) =  \Phi, \ \ \ \ V(\Phi)  = \Lambda \Phi,
\end{equation}
coincides with the Jackiw-Teitelboim action \cite{Jcw1,Ttm1}. 

In principle, at the classical level the theory defined by the
action (\ref{IIIi}) can be transformed into an equivalent theory 
whose corresponding action is  more simple. Indeed, this can be
done by choosing the field $\varphi_1$ as defined through the
equation
\begin{equation}
\label{IIIiv}
Z^{1/2} (\Phi) \partial_\mu \Phi = \partial_\mu \varphi_1,
\end{equation}
and by expressing $\Phi$ as $\Phi = \Phi (\varphi_1) $. Next,
let us introduce a new field, $\varphi_2$, via
\begin{equation}
\label{IIIv}
c \varphi_2 = C ( \Phi (\varphi_1) ),
\end{equation}
and write then  $\Phi = \Phi (\varphi_2) $. After having done
this, by making the transformation \cite{RT1}
\begin{equation}
\label{IIIvi}
g_{\mu\nu} \longrightarrow \e^{2 \rho (\varphi_2)}
\bar{g}_{\mu\nu},
\end{equation}
with a properly chosen $ \rho (\varphi_2)$ \cite{RT1}, one can see
that the theory (\ref{IIIi}) with the transformed metric 
(\ref{IIIvi}) is classically equivalent to more simple, 
particular case
\begin{equation}
\label{IIIvii}
Z=1, \ \ C= c\varphi_2, \ \ V= \e^{2 \rho (\varphi_2)} V(
\varphi_2, \chi ), \ \ f (\Phi ) = f(\varphi_2).
\end{equation}
However, the model (\ref{IIIvii}) (which, generally speaking, may be
considered as a representative of the general class (\ref{IIIi})) is
still complicated enough. Moreover, it still includes arbitrary
functions of the dilaton (now $\varphi_2$), as (\ref{IIIi}) does. 
Finally, the classical equivalence may be lost at the quantum level. 
For all these reasons, we choose to consider the quantum effective 
action corresponding to  more general theory (\ref{IIIi}).

Let us briefly remind the definition and the properties of 
the effective action. Let $\phi$ is set of fields, 
$S(\phi)$ is classical action. Imagine for the moment 
that classical fields $\phi$ are not gauge fields. 
Then, the effective action $\Gamma(\phi)$ is defined via the 
following integro-differential equation:
\be
\label{In1}
\exp\left({i \over \hbar}\Gamma(\phi)\right)
= \int {\rm D}\varphi \left[{i \over\hbar}
\left( S[\varphi + \phi] - \int d^4x \varphi(x)
{\delta\Gamma(\phi) \over \delta\phi(x)}\right)\right]\ .
\ee 
Here $\hbar$ is Planck constant, $\varphi$ is set of quantum 
fields corresponding to $\phi$. Actually, to derive 
the above equation, one uses the background field method where 
the quantum fields transformation $\varphi\rightarrow \varphi 
+ \phi$ is supposed. Here, $\varphi$ are called quantum fields 
and $\phi$ are called the background fields. In case when gauge 
fields are present, the correspoonding gauge-fixing term should 
be included into the classical action. It also induces the 
so-called ghosts action. We will not give more details of 
presentation of effective action for gauge fields as it may be 
found in the most of modern textbooks on quantum field theory. 
It is enough to consider the effective action for non-gauge 
theories to demonstrate the qualitative properties. 

The equation (\ref{In1}) may be further represented as follows 
(after the functional Taylor series expansion of 
$S[\phi + \varphi]$ is done):
\bea
\label{In2}
\exp\left({i \over \hbar}\Gamma(\phi)\right) 
&=&\int {\rm D}\varphi \exp\left\{{i \over \hbar}
\left(S(\phi) + {1 \over 2}S_2(\phi)\varphi^2 \right.\right.\nn
&& \left.\left. + \sum_{n=3}^\infty {\hbar^{{n \over 2}-1} \over n!} 
S_n(\phi)\varphi^n - \hbar^{-{1 \over 2}}\varphi\left(
\Gamma_1(\phi) - S_1(\phi)\right)\right)\right\}\ .
\eea
Here $S_n={\delta^n S(\phi) \over \delta\phi(x_1)
\cdots \delta\phi(x_n)}$, $\varphi\Gamma(\phi)=\int d^4x 
\varphi(x) {\delta \Gamma(\phi) \over \delta \phi(x)}$. 
Eq.(\ref{In2}) gives the loop expansion of the effective action. 
Limiting to only one-loop approximation (second term in right 
hand side of Eq.(\ref{In2})), introducing the notation
\be
\label{In3}
\Gamma(\phi)=S(\phi) + \bar\Gamma(\phi)=S(\phi) 
+ \sum_{n=1}^\infty \hbar^n\bar\Gamma^{(n)}(\phi)
\ee
and making functional integration over quadratic on $\varphi$ 
terms one gets;
\be
\label{In4}
\Gamma^{(1)}(\phi)=S(\phi) + \bar\Gamma^{(1)}(\phi)=S(\phi) 
+ {i \over 2}{\rm Tr}\log S_2(\phi)\ .
\ee
Eq.(\ref{In4}) gives the one-loop effective action for 
non-gauge theories. For gauge theories, the key idea is 
very similar, but due to difference in technique more 
${\rm Tr}\log$ terms (ghosts contribution) appear. As we 
will be dealing with only one-loop approximation, we often do 
not write explicitly the index ``1'' or hat under $\Gamma(\phi)$. 

It is usually happens in QFT that loop corrections to the 
effective action are divergent (they contain infinities). To avoid these 
divergences the standard prescription is to use some regularization. 
For example, in one-loop approximation we can present
\be
\label{In5}
\bar\Gamma^{(1)}(\phi) = \Gamma^{(1)}_{fin}(\phi) 
+ \Gamma^{(1)}_{div}(\phi)\ .
\ee
Depending on regularization under consideration, now
the separation to finite and divergent pieces in Eq.(\ref{In5}) 
is ambiguous. The simplest regularization could be just omitting 
of the divergent part in Eq.(\ref{In5}). 

There is one extremely 
important case of the theories where the structure of divergences of 
effective action (in all orders of loop expansion) repeats the 
structure of classical action. In this case (for details, one can 
consult any modern textbook on QFT) $\Gamma_{div}(\phi)$ which may 
be calculated via Schwinger-De Witt expansion is in some sense 
universal and represents the main interest. First of all, 
the explicit
calculation of $\Gamma_{div}$ (which can be done normally only at 
one-loop, at best at two- or three-loops) indicates explicitly 
if theory is renormalizable or not. This is especially important for 
non-linear theories like (dilatonic) gravity. Second, if theory is 
renormalizable (even at one-loop) then $\Gamma_{div}$ defines 
beta-functions. Using these beta-functions (for details, see 
\cite{BOS1})  one can construct the finite part of effective 
action (actually, even in better approximation as so-called RG 
improvement may be applied). The example of such finite effective 
action which satisfies to RG equations is presented in Section 
\ref{2s4}. Third, divergent part of effective action maybe used in the 
construction of conformal anomaly as we see in next Chapter.

Taking into account above remarks, we give the 
explicit calculation of one-loop effective action 
divergences in dilatonic gravity (with matter). We show 
that theory could be one-loop renormalizable but only for 
special values of dilatonic couplings and dilatonic 
potential. In this case, finite part of effective action may be 
restored also. Before going to direct calculation, let us briefly 
describe the generalization of above effective action for 
gauge theories (in background gauge) as this is the case of dilatonic 
gravity.
 
Having the gauge theory with action $S(\phi)$, closed algebra and 
linearly independent generators one introduces the generating 
functional of connected Green functions as follows:
\bea
\label{In6}
\exp\left[iW[J,\tilde\phi]\right]&=& \int {\rm D}\varphi 
{\rm D}\bar C {\rm D}C\exp\Bigl\{i\Bigl(S(\varphi) \nn
&& + {1 \over 2}G_{\alpha\beta}[\tilde\phi]t_i^\alpha[\tilde\phi]
t_j^\beta[\tilde\phi]\varphi^i\varphi^j \nn
&& + \bar C_\alpha t_i^\alpha [\tilde\phi] R_\beta^i[\tilde\phi]C^\beta 
 - {1 \over 2}{\rm Tr}\log G_{\alpha\beta}[\tilde\phi] 
+ \varphi^iJ_i\Bigr)\Bigr\}
\eea
where $\varphi^i$ is set of gauge fields, $\tilde\phi^i$ are 
correspondent background fields, $R^i_\alpha$ are generators of gauge 
group, linear gauge is chosen: $\chi^\alpha[\varphi]
=t^\alpha_i\varphi^i$, $C^\alpha$, $\bar C^\alpha$ are 
ghost and antighost. The Planck constant is chosen to be equal to unity.
The gauge dependence is 
defined through the choice of functions $G_{\alpha\beta}$, $t^\alpha_i$. 
One also defines the mean field $\bar\phi^i$: 
\be
\label{In7}
\bar\phi^i={\delta W[J,\tilde\phi] \over \delta J^i}\ .
\ee
The effective action is introduced via Legendre transformation:
\be
\label{In8}
\Gamma[\bar\phi, \tilde\phi]=W[J,\tilde\phi]- \bar\phi^iJ_i
\ee
where the source $J_i$ is expressed in terms of mean field. 

Changing the variable $\varphi\rightarrow \varphi + \bar\phi$ 
in Eq.(\ref{In6}) and putting $\bar\phi=\tilde\phi$ as 
special condition, one can get
\bea
\label{In9}
\exp\left(i\Gamma[\bar\phi]\right)&=& \int {\rm D}\varphi 
{\rm D}\bar C {\rm D}C\exp\Bigl\{i\Bigl(S(\bar\phi+\varphi) \nn
&& + {1 \over 2}G_{\alpha\beta}[\bar\phi]t_i^\alpha[\bar\phi]
t_j^\beta[\bar\phi]\left(\bar\phi^i+\varphi^i\right)
\left(\bar\phi^j+\varphi^j\right) \nn
&& + \bar C_\alpha t_i^\alpha [\bar\phi] 
R_\beta^i[\bar\phi+\varphi]C^\beta 
 - {1 \over 2}{\rm Tr}\log G_{\alpha\beta}[\bar\phi] 
+ \varphi^i\left.{\delta\Gamma[\bar\phi,\tilde\phi] \over 
\delta\varphi^i}\right|_{\bar\phi=\tilde\phi}
\Bigr)\Bigr\}
\eea
where $\Gamma[\bar\phi]=\Gamma[\bar\phi,
\tilde\phi]|_{\tilde\phi=\bar\phi}$. This effective action 
in background gauge represents the generalization of the one 
for non-gauge theories in Eq.(\ref{In1}). It is important 
that one can prove that
\be
\label{In10}
\bar\Gamma_{,i}\left[\bar\phi\right]R^i_\alpha[\bar\phi]=0\ .
\ee
In other words, such effective action is gauge invariant one. The 
above approach is called background field method. 

In the one-loop approximation it may be found in 
analogy with Eq.(\ref{In4}) as follows:
\bea
\label{In11}
\bar\Gamma^{(1)}[\bar\phi]&=& {i \over 2}\Tr\log \left\{
S_{,ij}[\bar\phi]+ G_{\alpha\beta}[\bar\phi]t_i^\alpha[\bar\phi]
t_j^\beta[\bar\phi]\right\} \nn
&& -i \Tr \log\left\{ t_i^\alpha [\bar\phi]
R_\beta^i[\bar\phi]\right\} 
 - {i \over 2}{\rm Tr}\ln G_{\alpha\beta}[\bar\phi] \ .
\eea
Second term in Eq.(\ref{In11}) gives the one-loop ghosts contribution 
while third term usually gives no contribution to divergent 
part of the effective action (except the case of higher 
derivative theories). As one sees below the Eq.(\ref{In11}) 
(no last term) is typical one-loop divergent EA structure 
in dilatonic gravity.

Let us now briefly describe the calculation of one-loop effective 
action in dilatonic gravity (\ref{IIIi}). 
The covariant gauge fixing condition is chosen as follows (see also 
next section,Eq.(42) and below for definition of background and quantum 
fields) \be
S_{g.f.}=-{1\over2}\int c_{\mu\nu}\chi^\mu\chi^\nu \ ,  \quad 
\chi^\mu =-\nabla_\nu\bar{h}^{\mu\nu}+{C'\over C}\nabla^\mu
\varphi \ , \quad c_{\mu\nu}=-C\sqrt{g}\,g_{\mu\nu} \ , 
\label{gauge_fix}
\ee
The total quadratic contribution to the action takes the form
$$
S^{(2)}_{tot}=-{1\over2}\int
d^2x\,\sqrt{g}\phi^i\widehat{H}_{ij}\phi^j \ ,
$$
where $\widehat{H}$ is the second order minimal operator,
 and $\phi^i$ is the set of all quantum fields including graviton, dilaton
and matter fields (if present). 
The one-loop effective action in terms of supertraces becomes
\begin{equation}
\Gamma={i\over2}\mbox{Tr}\,\log\widehat{H}
        -i\mbox{Tr}\,\log\widehat{\cal M} \ ,
\label{def_eff_action}
\end{equation}
where $\widehat{H}=-\widehat{K}\Delta+\widehat{L}^\lambda
\nabla_\lambda+\widehat{P}$, and the last term in
(\ref{def_eff_action}) is the ghost operator corresponding via 
(\ref{gauge_fix}) to diffeomorphisms. Here $\widehat K$ and 
$\widehat L$ and $\widehat P$ are some matrices which form 
follows from making second functional derivative of the action. 
 We use the same technique as in refs. \cite{OS1,OS2,OS3},
representing
\begin{equation}
\label{IIIxx}
\widehat{H}=- \widehat{K} \left( \widehat{1} \Delta +
2 \widehat{E}^\lambda \nabla_\lambda+ \widehat{\Pi} \right), \
\ \ \ \widehat{E}^\lambda =-\frac{1}{2} \widehat{K}^{-1}
 \widehat{L}^\lambda, \ \ \ \
 \widehat{\Pi} =-  \widehat{K}^{-1}  \widehat{P}.
\end{equation}
After that, the standard algorithm can be used, namely
\begin{equation}
\label{IIIxxxi}
\frac{i}{2} \mbox{Tr}\, \log  \widehat{H}=
\left. \frac{i}{2} \mbox{Tr}\, \log  \left( \widehat{1} \Delta
+ 2 \widehat{E}^\lambda \nabla_\lambda+ \widehat{\Pi}
\right)\right|_{div} = \frac{1}{2\epsilon} \int d^2 x \,
\sqrt{g} \,
\mbox{Tr}\, \left( \widehat{\Pi}+ \frac{R}{6} \widehat{1}
  - \widehat{E}^\lambda  \widehat{E}_\lambda - \nabla_\lambda
 \widehat{E}^\lambda \right),
\end{equation}
where $\epsilon = 2\pi (n-2)$.
To obtain the divergent part, $\Gamma_{2-div}$, we have to
evaluate the functional traces of the matrices above according to
(\ref{IIIxxxi}). After some tedious algebra \cite{OS3} we arrive at
\begin{eqnarray}
\label{IIIxxiii}
\Gamma_{2-div} =\frac{i}{2} \mbox{Tr}\, \log  \widehat{H}
=-{1\over 2\epsilon}\int d^2x\,\sqrt{g}\,\Biggl\{{4 \over 3}R
+{2\over C}V+{2\over C'}V'+\left({2C'\over C}
  -{Z\over C'} \right)(\Delta\Phi)  && \nn
  -\left({3{C'}^2\over2C^2}+{C''Z\over{C'}^2}\right)
(\nabla^\lambda \Phi)(\nabla_\lambda\Phi)\Biggr\}. &&
\end{eqnarray}

To complete the calculation one has to consider the ghost
operator
$$
\widehat{\cal
M}^\mu_\nu\equiv{\delta_{\vphantom{p}_R}\chi^\mu\over\delta
                              \phi^j}\nabla^j_\nu \ ,
$$
where $\delta_{\vphantom{p}_R}$ stands for the right
functional
derivative and the generators of the diffeomorphisms are
\begin{equation}
\label{IIIxxv}
\nabla^\varphi_\nu=-(\nabla_\nu \Phi) \ , \qquad
\nabla^{\bar{h}_{\rho\sigma}}_\nu=g_{\rho\sigma}\nabla_\nu
 -g_{\nu
\rho}\nabla_\sigma-g_{\nu\sigma}\nabla_\rho
\ .
\end{equation}
Explicitly,
\begin{equation}
\label{IIIxxvi}
\widehat{\cal M}^\mu_\nu=g^\mu_\nu\Delta-{C'\over
C}(\nabla_\nu\Phi)\nabla^\mu-{C'\over
C}(\nabla^\mu\nabla_\nu\Phi)+ R^\mu_\nu\ ,
\end{equation}
which leads to
\begin{equation}
\label{IIIxxvii}
\Gamma_{gh-div}=-{1\over 2\epsilon}\int
d^2x\,\sqrt{g}\,\Biggl\{{8\over3}R
                -{C'\over C}(\Delta\Phi)+\left({C''\over
C}-{3{C'}^2\over
2C^2}\right)(\nabla^\lambda\Phi)(\nabla_\lambda\Phi)\Biggr\} \ .
\end{equation}
The final answer is \cite{OS1,OS3}
\begin{eqnarray}
\label{IIIxxviii}
\Gamma_{div}=-{1\over 2\epsilon}\int
d^2x\,\sqrt{g}\,\Biggl\{4R
+{2\over C}V+{2\over C'}V'+\left({C'\over C}-{Z\over C'}\right)
             (\Delta\Phi)      \qquad\qquad  \cr\cr
+\left({C''\over C}-3{{C'}^2\over C^2}-{C''Z\over{C'}^2}\right)
(\nabla^\lambda \Phi)(\nabla_\lambda\Phi)\Biggr\} \ .
\end{eqnarray}
\smallskip
Notice that all surface terms have been kept in Eq. (\ref{IIIxxviii}).
Similar calculation of one-loop effective action in 2d dilatonic gravity 
(without matter) has been discussed also in refs.\cite{KM1,
MM1,Ich1,RT1,ENO2}.
                                           
Having performed the above calculation, now it is not 
difficult to take into account the scalar fields $\chi_i$, 
and to repeat it for the
effective action (\ref{IIIi}). In fact, the introduction of 
the scalars leads to minor changes. In the background field 
notation $\chi_i\rightarrow \chi_i + \sigma_i$ the quantum fields 
are $\phi^i= \{\varphi; h; \bar{h}_{\mu\nu}; \sigma_i\}$. Scalars 
do not spoil the minimality of the second functional derivative 
operator, so that the gauge condition and ghost action may be left untouched. 
The 't Hooft-Veltman procedure \cite{HV1} is assumed to be 
used when it is necessary. Thus, the complete one-loop 
divergences for the theory (\ref{IIIi}) become
\begin{eqnarray}
\label{IIIxxxiii}
&& \Gamma_{div}=-{1\over2\epsilon}\int d^2x\,\sqrt{g}\,\Biggl\{
 {48+2m\over12}R+{2\over C} V+ {2\over C'} V' -
{V_{,ii}\over f} \\
&& + \left( {C''\over C}-{3{C'}^2\over C^2}- {C''Z\over {C'}^2} -
{m{f'}^2\over4f^2}+ {mf''\over2f} \right)(\nabla^\lambda\Phi)
(\nabla_\lambda\Phi) + \left( {C'\over C}- {Z\over C'}
+{mf'\over2f}\right)\Delta\Phi\Biggr\}\ .\nonumber 
\end{eqnarray}
This is the main result of the present section ---the one-loop
effective action for 2d dilaton
gravity with scalar matter. Furthermore, one can
easily generalize this expression to the case when Maxwell fields
are added, namely, when one considers action (1) plus the Maxwell
action:
\begin{equation}
\label{IIIxxxiv}
S= -\int d^2x\,\sqrt{g}\,\left[ \cdots + \frac{1}{4} f_1 (\Phi)
F^2_{\mu\nu} \right]. \nonumber
\end{equation}
With the background versus quantum field separation
$A_\mu\rightarrow A_\mu +Q_\mu$, in the Lorentz gauge,
\begin{equation}
\label{IIIxxxv}
S_{Lorentz}= -\int d^2x\,\sqrt{g}\, f_1 (\Phi) (\nabla_\mu Q_\mu)^2,
\end{equation}
the extra contributions are known to split into the $F^2$-terms and
total divergences (see  ref. \cite{EO1}).
Thus, the total divergent contribution to the one-loop effective
action of the theory (\ref{IIIi}) plus the Maxwell terms 
(\ref{IIIxxxiv}) is given by
\begin{eqnarray}
\label{IIIxxxvii}
\Gamma_{div}=-{1\over2\epsilon}\int d^2x\,\sqrt{g}\,\Biggl\{
 {-2m+60\over12}R+{2\over C} V+ {2\over C'} V' -
{V_{,ii}\over f}  + \left(\frac{{f_1}'}{2C'}- \frac{f_1}{2C}
\right) F^2_{\mu\nu} && \nn 
+ \left( {C''\over C}-{3{C'}^2\over C^2}- {C''Z\over {C'}^2} -
{m{f'}^2\over4f^2}+ {mf''\over2f}+\frac{{f_1}''}{f_1}-
\frac{{f_1'}^2}{f_1^2} \right)(\nabla^\lambda\Phi
)(\nabla_\lambda\Phi) && \nn 
+ \left( {C'\over C}-{Z\over C'}+{mf'\over2f}
+\frac{{f_1}'}{f_1}\right)\Delta\Phi\Biggr\}\ .&&
\end{eqnarray}
This gives the divergences of the covariant effective action 
(general review is given in \cite{BOS1}). 
Gauge dependence of Eq.(\ref{IIIxxxvii}) has been studied in 
ref.\cite{ENO3}. It is interesting to note that such one-loop 
EA may be also calculated in gauge-fixing condition independent 
formulation \cite{ENO3}. 

In the similar way, one can calculate the one-loop effective
action for the theory (\ref{IIIi}) in the case when the 
gravitational field is a classical one, but the dilaton and the 
scalar fields are quantized. Such a calculation is much simpler 
since there are no gauge-fixing terms and corresponding ghosts.
The effective action may be found as following
\begin{eqnarray}
\label{IIIxxxxiv}
&& \Gamma_{div}=-{1\over2\epsilon}\int d^2x\,\sqrt{g}\,\Biggl\{
\left( {C'' \over Z}  -{2m+2\over12}\right)R+{V''\over Z} -
{V_{,ii}\over f} + \left( {{f'}^2\over2fZ}- {f''\over2Z}
\right)(\nabla^\lambda\chi
)(\nabla_\lambda\chi) \nn 
&& + \left( {mf''\over2f}- {m{f'}^2\over4f^2}
 -{{Z'}^2\over4Z^2}\right)(\nabla^\lambda\Phi)(\nabla_\lambda\Phi)
 + \left( {mf'\over2f}- {Z'\over2Z}\right)
\Delta \Phi\Biggr\}\ .
\end{eqnarray}

Expression (\ref{IIIxxxxiv}) gives the one-loop effective action 
of the system composed of  quantized dilaton, scalars in an
external classical gravitational field. 

Let us make here remarks on the issue of renormalization, 
to one-loop order. Without loss of generality, we will restrict 
ourselves to the case of pure 2d dilaton gravity. By adding to
the classical action  
the corresponding
counterterms ($\Gamma_{div} $  with opposite sign),
one obtains the one-loop renormalized effective action.

Choosing the renormalization of the metric tensor in the
following
form
\begin{equation}
\label{IIIxxxxxxvi}
g_{\mu\nu} = \exp\left\{ \frac{1}{\epsilon} \left[
\frac{1}{C(\Phi
)} + \frac{Z(\Phi)}{2{C'}^2 (\Phi)} \right] \right\} \,
g^R_{\mu\nu}
\end{equation}
(this choice absorbs all the divergences of the dilaton kinetic
term), one can obtain the renormalized effective action
as (for simplicity, we drop the superscript `R' off
$g_{\mu\nu}$ and $R$):
\begin{equation}
\label{IIIxxxxxxvii}
S_R = - \int d^2x\,\sqrt{g}\,\left\{ \frac{1}{2} Z
g^{\mu\nu}
\partial_\mu \Phi \partial_\nu \Phi + CR +V +
\frac{1}{\epsilon}
\left( - \frac{V'}{C'} + \frac{ZV}{2{C'}^2} \right) \right\}.
\end{equation}

Now, the condition of multiplicative renormalizability of the
theory in the usual sense has the following form
\begin{equation}
\label{IIIxxxxxxviii}
 - \frac{V'}{C'} + \frac{ZV}{2{C'}^2} = a_1V,
\end{equation}
where $a_1$ is an arbitrary constant. This condition restricts 
the form of the functions under discussion.
Some sets of solutions of Eqs. (\ref{IIIxxxxxxviii}) can be obtained
explicitly:
\begin{equation}
\label{IIIxxxxxxix}
Z=1, \ \ \ \ C(\Phi) =C_1 \Phi, \ \ \ \ V=0.
\end{equation}
There are, of course, more complicated solutions \cite{OS1,OS2,OS3}. 
One can also look to fixed points solutions of generalized RG 
equations, finding some specific models of dilatonic gravity as 
fixed points of such RG equations \cite{ENO3}.

\subsection{Quantum equivalence in the models of 2d 
dilatonic gravity}

Let us limit to specific version of 2d dilatonic gravity 
with action:
\begin{equation}
\label{2iii}
S_1=-\int d^{2}x \,\sqrt{g}\left[ {1\over 2} g^{\mu  \nu}\partial_\mu
\Phi  \partial_\nu \Phi + CR \Phi +V(\Phi )\right],
\label{so1}
\end{equation}
where $C$ is a positive constant and $V(\Phi )$ an arbitrary
function. The one-loop renormalization of the theory (\ref{2iii}) 
has been discussed above. We will study  quantum equivalence in such 
model following to ref. \cite{ENO4}.

Let us here briefly summarize the  results 
concerning the one-loop effective action. We use the background
field method
\begin{equation}
\label{2iv}
g_{\mu\nu} \longrightarrow \bar{g}_{\mu\nu} =g_{\mu\nu}
+h_{\mu\nu}, \ \ \ \ \
\Phi \longrightarrow \bar{\Phi} = \Phi + \varphi,
\end{equation}
where $h_{\mu\nu}$ and $\varphi$ are the quantum fields. The
simplest minimal covariant gauge is given by
\begin{equation}
S_{GF}=-{1\over 2}\int d^2x\,c_{\mu\nu}\, \chi^\mu\chi^\nu,
\label{2v}
\end{equation}
where
\beq
\label{2vi}
c_{\mu\nu} = - C \Phi \sqrt{g}\, g_{\mu\nu}, \ \ \ \ \chi^\mu = -
\nabla^\nu \bar{h}^\mu_{\ \nu} + \frac{1}{\Phi} \nabla^\mu \varphi,
\eeq
and  $\bar{h}_{\mu\nu}=h_{\mu\nu}-\frac{1}{2} g_{\mu\nu} h$. The
divergences of the one-loop effective action (including all surface
terms) have been calculated in Eq.(\ref{IIIxxviii}) as following:
\beq
\label{2vii}
\Gamma_{div}=-{1\over 2\epsilon}\int d^2 x\,\sqrt g\,
\Biggl[4R+{2\over C\Phi}V+{2\over
C}V'+\biggl({1\over \Phi}-{1\over C} \biggr) (\Delta \Phi )- {3\over
\Phi^2}(\nabla^\lambda\Phi) (\nabla_\lambda\Phi)\Biggr].
\eeq

Let us now make in the theory (\ref{2iii}) the field transformation
\beq
\label{2viii}
\Psi^2=\frac{C}{\gamma} \Phi, \ \ \ \  g_{\mu  \nu} \longrightarrow
e^{-2\rho} \,  \wt{g}_{\mu \nu},
\eeq
where \[ \gamma >0, \ \ \ \rho = \frac{\gamma \Psi^2}{4C^2} -
\frac{1}{8\gamma} \ln \Psi. \] Then, action (\ref{2iii}) becomes
\begin{equation}
\label{2ix}
S_2=-\int d^{2}x \,\sqrt{\wt{g}}\left[ {1\over 2} \wt{g}^{\mu
\nu}\partial_\mu
\Psi  \partial_\nu \Psi + \gamma \wt{R} \Psi^2 + U (\Psi)  \right],
\label{so2}
\end{equation}
where we have defined $ U (\Psi)\equiv \e^{-2\rho} V(\Phi (\Psi) )$
and dropped off a total derivative term. The actions $S_1$ and $S_2$
(eqs. (\ref{2iii}) and (\ref{2ix}), respectively) 
are parameterized through different triplets of functions $\{ Z, C,
V\}$. They are classically equivalent and lead to the same
classical physics.

We shall now investigate the one-loop effective action for the
theory (\ref{2ix}). The calculation will be done in 
the same gauge (\ref{2v})-(\ref{2vi}),
that is also to be transformed in accordance with (\ref{2viii}). 
So, the natural prescription is:

a). For the background fields we shall make the transformation
(\ref{2viii}), where $\Psi$ and $\wt{g}_{\mu\nu}$ will 
be now the background fields of the theory (\ref{2ix}).

b). The quantum fields will be transformed according to the first
order Taylor expansion of eq. (\ref{2viii}) (local change of variables)
\beq
\label{2x}
\varphi \longrightarrow \frac{2\gamma}{C}\, \Psi \, \eta, \ \ \
h_{\mu\nu}  \longrightarrow \e^{-2\rho (\Psi )} \left[
\wt{h}_{\mu\nu}+ \left( \frac{1}{4\gamma\Psi}- \frac{\gamma}{C^2}
\Psi \right)\wt{g}_{\mu\nu} \eta \right].
\eeq
We should recall now that in the background field method for
the theory (\ref{2ix}),
\beq
\label{2xi}
\wt{g}_{\mu\nu} \longrightarrow \wt{g}_{\mu\nu} +\wt{h}_{\mu\nu},
\ \ \ \ \Psi  \longrightarrow \Psi + \eta,
\eeq
where $\wt{h}_{\mu\nu}$ and $\eta$ are the quantum fields.
Taking into account all the remarks above, we get the following
covariant (minimal) gauge for the theory ({\ref{2ix}):
\begin{equation}
\label{2xii}
S_{GF}=-{1\over 2}\int d^2x\,c_{\mu\nu}\, \chi^\mu\chi^\nu,
\end{equation}
where
\bea
\label{2xiii}
&& c_{\mu\nu} = -\gamma \Psi^2 \sqrt{g}\, g_{\mu\nu}, \\
&& \chi^\mu = -\nabla^\nu \bar{h}_{\mu\nu} +  \left(
\frac{\gamma}{C^2} \Psi - \frac{1}{4\gamma\Psi} \right)
(\nabla^\nu \Psi) \bar{h}^\mu_{\ \nu} + \frac{2}{\Psi}
\nabla^\mu \eta + \frac{2}{\Psi^2}(\nabla^\mu \Psi) \eta. 
\eea
In order to simplify notation,  in which follows we can suppress
tildes over $g_{\mu\nu}$ and $h_{\mu\nu}$.
The total one-loop divergence, $\Gamma_{div} =\Gamma_{2-
div}+\Gamma_{ghost-div}$, is 
\bea
\label{2xxiv}
\Gamma_{div}&=&-{1\over 2\epsilon}\int d^2 x\,\sqrt g\,
\left[4R+{2\over \gamma\Psi^2}U+{1\over
\gamma\Psi}U'+\biggl({4\gamma-3\over 2\gamma\Psi}+{4\gamma\Psi\over
C^2} \biggr)(\Delta \Psi) \right.\nn
&+& \left. \left( {4\gamma\over C^2}+  {1-20\gamma\over
2\gamma\Psi^2}\right) (\nabla^\lambda\Psi)
 (\nabla_\lambda\Psi)\right].
\eea
We must observe that in this final formula all surface terms 
have been kept.

Let us now discuss the on-shell limit of $\Gamma_{div}$, 
eq. (\ref{2xxiv}).
Keeping all the surface terms and using the classical field
equations resulting from the action (\ref{2ix}) we obtain the total derivative
\beq
\label{2xxxi}
\Gamma_{div}^{on-shell}=-{1\over 2\epsilon}\int d^2 x\,\sqrt g\,
\Biggl[ 2R+ \Delta \left(\frac{12\gamma -1}{2\gamma} \ln \Psi
+{2\gamma\over C^2} \Psi^2 \right) \Biggr].
\eeq

One may consider another form of  gauge
fixing action. Then, one-loop effective action in new gauge differs from
$\Gamma_{div}$, eq. (\ref{2xxiv}), only  
in surface terms. The
off-shell one-loop renormalization is the same in different gauges.
However, even on shell  they still {\it differ} in some total derivative
terms. 
Summing up,
we see that the on-shell effective action in different gauges is given
by surface divergences only (finiteness of the $S$ matrix), but
these terms depend yet on the choice of gauge condition.

We are now going to investigate the theory (\ref{2iii}) 
in the variables
(\ref{2viii}). Transforming   $\Gamma_{div}$, eq. (\ref{2vii}), 
to the new variables (\ref{2viii}), we get
\bea
\label{2xxxiv}
\Gamma_{div}&=&-{1\over 2\epsilon}\int d^2 x\,\sqrt g\,
\left[ 4R+ {2\over \gamma\Psi^2}U+e^{-2\rho (\Psi)}{2\over
\gamma} \, \frac{\partial V(\Psi)}{\partial \Psi}\right.\nn
&+&\left. \left( {4\gamma-2 \over 2\gamma \Psi}+ {2\gamma\Psi \over
C^2}\right) \Delta \Psi+ \left( {2\gamma \over C^2}+{2-
20\gamma\over 2\gamma\Psi^2}\right) (\nabla^\lambda\Psi)
(\nabla_\lambda\Psi)\right],
\eea
As we see, there is no perturbative quantum equivalence between the
two classically equivalent dilaton gravities (\ref{2iii}) and 
(\ref{2ix}). In fact, the one-loop effective action 
(\ref{2xxxiv}) which comes
from action (\ref{2iii}) in the gauge (\ref{2v}) 
does {\it not} coincide with
$\Gamma_{div}$, eq. (\ref{2xxiv}), 
which is obtained when starting from the
classically equivalent theory (\ref{2ix}) in the gauge 
(\ref{2xii}) ---that can be
made correspond with the gauge (\ref{2v}). 
However one can show that two classically equivalent
dilaton gravities  lead to the same class of multiplicatively
renormalizable potentials.

Let us now consider the on-shell effective action of the theory
(\ref{2iii}). Using classical field equations and keeping all the 
total derivative terms in (\ref{2vii}), we get
\beq
\label{2xxxvii}
\Gamma_{div}^{on-shell}=-{1\over 2\epsilon}\int d^2 x\,\sqrt g\,
\Biggl[ 2R+ \Delta \left(\frac{1}{C}  \Phi +3\ln \Phi
\right)\Biggr].
\eeq
Transforming the variables in (\ref{2xxxvii}) according 
to the change (\ref{2viii}), we obtain exactly eq. (\ref{2xxxi}).

 From the discussion above, we conclude that there is on-shell
perturbative quantum equivalence for the classically equivalent
dilaton gravities (\ref{2iii}) and (\ref{2ix}). 
This analysis is presented in detail in order to avoid the 
following confusion. It is very often to study some questions 
related with quantum cosmology or black holes the researchers 
come to ``convenient'' parameterization of the quantum model where 
quantum calculations are easier to do. After that, the result 
of such calculations is transformed back to original parameterization
(original quantum variables) 
and is used in subsequent analysis. As it follows from above results 
such approach is most likely wrong unless on-shell calculations are 
used.

\subsection{Dilatonic gravity with matter in 
$2+\epsilon$-dimensions}

It is quite an old idea
\cite{Wng1,GKT1,CD2,KH1,NTT1} to try to study Einstein gravity in
$2+\epsilon$ dimensions, which may improve its ultraviolet 
properties. 
The gravitational coupling constant in such a theory ---near
two dimensions--- shows an asymptotically free behavior 
\cite{Wng1,GKT1,CD2,KH1,NTT1},
which can actually be very exciting in attempts to solve the
problem
of non-renormalizability of 4-dimensional Einstein gravity
\cite{HV1}.
Unfortunately, it was shown that Einstein gravity in
$2+\epsilon$  dimensions is on its turn a non-renormalizable
theory \cite{JJ1} in the usual sense.

Furthermore, dynamical triangulations in more than two
dimensions (see, for example, \cite{AJ1,AMM1}) have clearly shown the
existence of a phase transition to a strong-coupling phase of
similar nature as 2-dimensional quantum gravity \cite{AJJK1}.
Because of this,  of considerable interest is still to try to
construct a consistent theory of QG  in  $2+\epsilon$  dimensions 
(on the next step, one can think on hypothetical continuation of 
small $\epsilon$ to $\epsilon =1$ or $\epsilon =2$).

It has been suggested \cite{KST1}, that a thing to do would be
to study dilatonic gravity near two dimensions (similarly, it
was the idea in ref. \cite{AKHN1} to consider Einstein gravity with
a conformal scalar field near two dimensions).
Notice, however, that unlike
Einstein gravity, dilatonic gravity possesses a smooth $\epsilon
\rightarrow 0$ limit. This difference manifests itself in the fact
that dilatonic gravity ---which after a proper field definition
can be presented as an Einsteinian theory with a scalar field--- is
{\it not} equivalent to Einstein gravity in {\it exactly}
two dimensions. The gravitational coupling constant \cite{KST1} 
in this theory also has an ultraviolet stable fixed point 
 for $\epsilon>0$
and $n<24$ ($n$ is the number of scalars or matter central charge).
Thus, the matter central charge in dilatonic gravity is bounded,
as it also happens in Einstein's theory.

In the present section, we give short overview of typical example:
dilatonic-Maxwell gravity with scalar
matter in $2+\epsilon$ dimensions. This theory, which includes a
dilatonic potential and a Maxwell term with an arbitrary
dilatonic-vector
coupling function, may be considered as a toy model for
unification of gravity with matter (scalar and vector fields). 
We follow to ref.\cite{EO3,EO4} where it was shown  that
in the ultraviolet stable fixed point the gravitational coupling
constant has the value $G^* = 3\epsilon /[2(30-n)]$ and, hence, 
owing to the contribution of the vectors, the matter central 
charge of our universe in such a model can be naturally 
increased from $0<n<24$ (for pure dilatonic gravity) to 
$0<n<30$ (for dilatonic-Maxwell gravity). 

We start from the theory of dilatonic gravity 
interacting with scalars and vectors via dilatonic couplings in
$2+\epsilon$ dimensions. 
The action is
\beq
S = \int d^d x\,  \sqrt{-g} \biggl[{1 \over 2} Z(\phi) g^{\mu\nu}
\partial_\mu \phi\partial_\nu \phi + {\mu^\epsilon \over 16\pi
G} R C(\phi )+ V(\phi)
 - {1 \over 2} f(\phi) g^{\mu\nu} \partial_\mu \chi_i \partial_\nu
\chi^i+ {1 \over 4} f_1(\phi) F_{\mu\nu}^2 \biggr], \label{1}
\eeq
where $g_{\mu\nu}$ is the ($2+\epsilon$)-dimensional metric, $R$
the corresponding curvature, $\chi_i$ are scalars ($i=1,2,
\ldots,
n$), $F_{\mu\nu} = \nabla_\mu A_\nu - \nabla_\nu A_\mu$, with
$A_\mu$ a vector, and where the smooth functions $Z(\phi)$,
$C(\phi)$, $V(\phi)$, $f(\phi)$ and $f_1(\phi)$ describe the
dilatonic interactions. Notice that $V(\phi)$ is a dimensional
function and it is therefore convenient to redefine $V
\rightarrow m^2V$, where $m$ is some parameter with dimension of
mass. 

One can easily prove that the above theory (\ref{1}) is
renormalizable in a generalized sense. The one-loop
counterterms corresponding to (\ref{1}) in two dimensions have
been calculated in refs. \cite{ENO2,EO1}, (see also 
previous section). Our purpose is to present 
the renormalization structure of (\ref{1})
in $2+\epsilon$ dimensions and in particular the
corresponding renormalization
group equations. The very remarkable point that gives sense to
this study is the fact that the theory (\ref{1}) (unlike Einstein
gravity \cite{Wng1,GKT1,CD2,KH1,NTT1}) has a smooth limit 
for $\epsilon \rightarrow
0$. This property allows for the possibility to study the behavior
of (\ref{1}) in $2+\epsilon$ dimensions by simply using the
counterterms  calculated already in 2 dimensions ---in close
analogy with quantum field theory in frames of the
$\epsilon$-expansion technique (for a review see \cite{WK1}
and \cite{Znn1,BGZ1}).

Action (\ref{1}) may be rewritten in another form. 
First, motivated by the form
of the string effective action, we choose the dilatonic
couplings in (\ref{1}) as the exponents of convenient dilatonic
functions. Using the rigid rescaling of scalars and vectors:
\beq
\chi_i \longrightarrow a_1 \chi_i, \ \ \ \ \ \
A_\mu \longrightarrow a_2 A_\mu, \label{2} \eeq
with $a_1$ and $a_2$ constants, one can always normalize the
dilatonic couplings in such a manner that
\beq
f(0) =1, \ \ \ \ \ \ f_1(0)=1. \label{3} \eeq
Second, as has been done in refs. \cite{RT1,OS3}, we can also use a
local Weyl transformation of the metric of the form
\beq
g_{\mu\nu} \longrightarrow \e^{-2\sigma (\phi)} \, g_{\mu\nu} ,
 \label{4} \eeq
in order to simplify the dilatonic sector of (\ref{1}). In
particular, one can use (\ref{4}) of a specific form such that
$Z(\phi)=0$. Finally, we can  perform a transformation of the
dilaton field in order to simplify the function $C(\phi)$, namely
to reduce it to the form $C(\phi)=e^{-2\phi}$. With all this taken
into account, the action (\ref{1}) can be written as follows
 \beq
S = \int d^d x\,  \sqrt{-g}\,  \biggl[ {\mu^\epsilon \over 16\pi
G}
R \e^{-2\phi}
 - {1 \over 2} g^{\mu\nu} \partial_\mu \chi_i \partial_\nu \chi^i
\e^{-2\Phi (\phi)}+ \mu^\epsilon m^2 \e^{-V(\phi )}+ {1 \over 4}
\e^{-f_2(\phi)} F_{\mu\nu}^2 \biggr], \label{5}
\eeq
being $\mu$ a mass parameter and where we have chosen the
dilatonic couplings so that $\Phi (0) =f_2(0)=V(0) =0$. The
first two terms in (\ref{5}) correspond to the dilatonic gravity
action in ref. \cite{KST1}, which was considered in $2+\epsilon$
dimensions. With our choice of (\ref{5}) the zero modes of the
dilatonic couplings are fixed with the help of reference
operators in the gravitational sector \cite{KST1} and with the
renormalization (by constants) of the scalar, vector, and mass
$m^2$ in the matter sector.

We can now present short resume of the divergences of the action
(\ref{5}). The study of the one-loop divergences of dilatonic
gravity in the covariant formalism, initiated in 
refs. \cite{OS1,OS2}, 
has been continued in refs. \cite{RT1,OS3,ENO2,
KM1}. We 
do not consider it necessary to repeat the details of such a
calculation for the case of the model (\ref{5}), 
see for more details, the previous section. 
The gauge-fixing conditions that are most convenient
to use are the following. The covariant gauge-fixing action is
\beq
S_{gf} = -  {\mu^\epsilon \over 32\pi G} \int d^d x\,
\sqrt{-g}\,
 g_{\mu\nu} \chi^\mu \chi^\nu  \e^{-2\phi}, \label{6}
\eeq
where
\beq
\chi^\mu =   \nabla_\nu \bar{h}^{\mu\nu} + 2 \nabla^\mu \varphi
\label{7}
\eeq
and $\bar{h}^{\mu\nu}$ is a traceless quantum gravitational field
and $\varphi$ a quantum scalar field, in the background field
method.
In the electromagnetic sector the gauge-fixing action is chosen
as
\beq
S_{L} = \int d^d x\,  \sqrt{-g}\,
 (\nabla_\mu Q^\mu)^2  \e^{-f_2(\phi)}, 
 \label{8b}
\eeq
where $Q_\mu$ is a quantum vector field. Notice that for the
background fields we  use the same notations as for the classical
fields in (\ref{5}).

The calculation of the one-loop effective action corresponding to
(\ref{5}) in the gauges (\ref{6}) and (\ref{8b}) can be performed
in close analogy with the one in refs. \cite{ENO2,EO1}. 
 This is reviewed in previous section.
Owing to the smooth behavior of (\ref{5}) for
$\epsilon \rightarrow 0$, the divergences of (\ref{5}) can be
also calculated in exactly two dimensions, which will also provide
the result for $2+\epsilon$ dimensions:
\bea
\Gamma_{div} &=& \frac{1}{4\pi \epsilon}  \int d^d x\,
\sqrt{-g}\,
\left\{ \frac{30-n}{6} R +m^2 \e^{2\phi - V(\phi)} 16\pi G [2 +
V'(\phi)] \right. \nn
&& -  g^{\mu\nu} \partial_\mu \phi \partial_\nu \phi 
[8-n\Phi'(\phi)^2] + \left.  
{4\pi G \over \mu^\epsilon} \e^{2\phi-f_2(\phi)}
F_{\mu\nu}^2 [f_2'(\phi) -2 ] \right\},
\label{9}
\eea
where $\epsilon =d-2$. As usually off-shell it is gauge dependent.

We turn to the study of the RG corresponding to our model in
$2+\epsilon$ dimensions. 
 The main idea of the whole approach is to use $G$ as
the coupling constant in perturbation theory and work at some
fixed power of $G$.
The counterterm  follows from (\ref{9}) 
\beq
\Gamma_{count} =- \mu^\epsilon  \int d^d x\,  \sqrt{-g}\,
\left[
RA_1(\phi) + g^{\mu\nu} \partial_\mu \phi \partial_\nu \phi
A_2(\phi) + m^2A_3(\phi) +  \frac{\mu^{-\epsilon}}{4}
F_{\mu\nu}^2 A_4(\phi) \right],
\label{11}
\eeq
where
\bea
&& A_1(\phi) = \frac{30-n}{24\pi\epsilon} \equiv A_1, \quad 
A_2(\phi) = \frac{1}{4\pi\epsilon} [n\Phi'(\phi)^2-8] , \nn
&& A_3(\phi) = \frac{4G}{\epsilon}[2 + V'(\phi)]\e^{2\phi -V(\phi)}
\equiv \e^{-V(\phi)} \wt{A}_3 (\phi), \nn
&& A_4(\phi) = \frac{4G}{\epsilon}[f_2'(\phi)-2]\e^{2\phi -
f_2(\phi)} \equiv \e^{-f_2(\phi)} \wt{A}_4 (\phi).
\label{12}
\eea

Renormalization is now performed in the standard way
\bea
S& =& S_{cl} + S_{count} =\int d^d x\,  \sqrt{-g}\,  \biggl[ {1
\over 16\pi G_0} R \e^{-2\phi_0}
 - {1 \over 2} g_0^{\mu\nu} \partial_\mu \chi_{0i} \partial_\nu
\chi_0^i \e^{-2\Phi_0 (\phi_0)} \nn
&&+  \mu^\epsilon m_0^2 \e^{-V_0(\phi_0 )}+ {1 \over 4} 
\e^{-f_{02}(\phi_0)} g_0^{\mu\alpha}g_0^{\nu\beta}F_{0\mu\nu}
F_{0\alpha\beta} \biggr]. \label{13}
\eea
The renormalization transformations can be defined as follows
\bea
&& \phi_0 = \phi + f(\phi), \ \ \ g_{0\mu\nu} =g_{\mu\nu}e^{-
2\Lambda (\phi)}, \ \ \ \Phi_0 (\phi_0)= \Phi(\phi) +
F(\phi), \ \ \ V_0 (\phi_0)= V(\phi) + F_V(\phi), \nn
&&  f_{02} (\phi_0)= f_{2}(\phi) + F_f(\phi), \ \ \ \chi_{0i} =
Z_\chi^{1/2} \chi, \ \ \ A_{0\mu} = Z_A^{1/2} A_\mu, \ \ \ m_0^2
=Z_m m^2. \label{14}
\eea
The functions $\Lambda$, $f$, $F$,... are chosen so that
$\Lambda(0)$, $f(0)$, $F(0)$,... can be set equal to zero.
Substituting the renormalization transformations (\ref{14}) into
the renormalized action (\ref{13}) (in close analogy with ref.
\cite{KST1}), we obtain
 \bea
S_0& =& \int d^d x\,  \sqrt{-g}\,  \biggl\{ {1 \over 16\pi G_0} R
e^{-2\phi-2f(\phi)-\epsilon \Lambda (\phi)} \nn &&
+ {\epsilon +1 \over 16\pi G_0} [4\Lambda'(\phi) + \epsilon
\Lambda'(\phi)^2+
4f'(\phi)\Lambda'(\phi)]e^{-2\phi-2f(\phi)-
\epsilon \Lambda (\phi)} g^{\mu\nu} \partial_\mu \phi
\partial_\nu \phi \nn
&&- {1 \over 2}Z_\chi g^{\mu\nu} \partial_\mu \chi_{i}
\partial_\nu
\chi^i \e^{-2\Phi(\phi)-2F(\phi)-\epsilon \Lambda (\phi)}+  \mu^\epsilon
m^2 Z_m \e^{-V(\phi )-F_V(\phi)-(2+\epsilon) \Lambda (\phi)} \nn
&&+ {Z_A \over 4} \e^{-f_{2}(\phi)-F_f (\phi)+(2-\epsilon)
\Lambda (\phi) } F_{\mu\nu}^2 \biggr\}. \label{15}
\eea
 From here, one can easily obtain
\bea
&& {1 \over 16\pi G_0}
\e^{-2\phi-2f(\phi)-\epsilon \Lambda (\phi)} =   \mu^\epsilon
\left( {1 \over 16\pi G}  \e^{-2\phi} -A_1 \right), \nn
&& {\epsilon +1 \over 16\pi G_0} [4\Lambda'(\phi) + \epsilon
\Lambda'(\phi)^2+ 4\epsilon f'(\phi)\Lambda'(\phi)]
e^{-2\phi-2f(\phi)-\epsilon \Lambda (\phi)} =  - \mu^\epsilon
A_2(\phi), \nn
&& Z_\chi \e^{-2\Phi(\phi)-2F (\phi)-\epsilon
\Lambda (\phi) }= \e^{-2\Phi(\phi)}, \quad 
Z_A \e^{-f_{2}(\phi)-F_f (\phi)+(2-\epsilon)
\Lambda (\phi) }= \e^{-f_{2}(\phi)} [1- \wt{A}_4 (\phi)], \nn
&& Z_m \e^{-V(\phi)-F_V (\phi)-(2+\epsilon)
\Lambda (\phi) }= \e^{-V(\phi)} [1- \wt{A}_3 (\phi)]. \label{16}
\eea
Working to leading order in $G$ (notice that the functions $f$,
$\Lambda$, $F$,..., $F_f$ are of first order in $G$), we may
follow ref. \cite{KST1} and obtain the renormalized parameters as
\bea
&& G_0^{-1}= \mu^\epsilon (G^{-1} -16 \pi A_1), \ \ \ Z_\chi =1,
\
\ \ \Lambda (\phi) = - \frac{4\pi G}{\epsilon +1} \int_0^\phi
d\phi' \, \e^{2\phi'}A_2(\phi'), \ \ \ \epsilon \Lambda (\phi) =-
2F(\phi), \nn
&& f(\phi) = 8\pi A_1 G(e^{2\phi} -1)+\frac{2\pi\epsilon
G}{\epsilon +1} \int_0^\phi d\phi' \, \e^{2\phi'}A_2(\phi'), \ \ \
Z_A =1-A_4 (0), \ \ \ Z_m =1-A_3 (0), \nn
&& F_f (\phi)= (2-\epsilon) \Lambda (\phi) + \wt{A}_4(\phi) -
\wt{A}_4(0), \ \ \ F_V (\phi)=-(2+\epsilon) \Lambda (\phi) +
\wt{A}_3(\phi) -\wt{A}_3(0). \label{17}
\eea
Hence, we see that the zero modes of the functions $f$, $f_2$ and
$V$ are indeed controlled by a constant renormalization of the
vector, the scalar and the mass.

We can now present the beta functions. We will
list below only the ones corresponding to the dilatonic couplings
and gravitational constant 
\bea
&& \beta_G = \mu {\partial G \over \partial \mu} = \epsilon G - 16
\pi \epsilon A_1 G^2, \quad 
\mu {\partial \Lambda (\phi_0) \over \partial \mu} =  -
\frac{4\pi
\epsilon G}{\epsilon +1} \int_0^\phi d\phi' \,
\e^{2\phi'}A_2(\phi'), \nn
&& \beta_\Phi (\phi_0) = \mu {\partial \Phi (\phi_0)  \over
\partial \mu} = 8\pi \epsilon A_1 G (\e^{2\phi_0}-1)\Phi'
(\phi_0) 
+ \frac{2\pi \epsilon^2 G}{\epsilon +1} [\Phi' (\phi_0) -1]
\int_0^{\phi_0} d\phi' \, \e^{2\phi'}A_2(\phi'), \nn
&& \beta_{f_2} (\phi_0) = \mu {\partial f_2 (\phi_0)  \over
\partial \mu} = 8\pi \epsilon A_1 G (\e^{2\phi_0}-1) f_2' (\phi_0)
\nn
&& \quad + \frac{2\pi \epsilon G}{\epsilon +1} 
[\epsilon f_2' (\phi_0)
+2(2-\epsilon)] \int_0^{\phi_0} d\phi' \, \e^{2\phi'}A_2(\phi')-
\epsilon [\wt{A}_4(\phi_0) -\wt{A}_4(0)] , \nn
&& \beta_V (\phi_0) = \mu {\partial V (\phi_0)  \over \partial
\mu} = 8\pi \epsilon A_1 G (e^{2\phi_0}-1) V' (\phi_0) \nn
&& \quad + \frac{2\pi \epsilon G}{\epsilon +1} [\epsilon V' (\phi_0)
 -2(2+\epsilon)] \int_0^{\phi_0} d\phi' \, \e^{2\phi'}A_2(\phi')-
\epsilon [\wt{A}_3(\phi_0) -\wt{A}_3(0)]. \label{18}
\eea
Similarly, one can obtain the $\gamma$-functions for the fields
and mass parameter, which are however of less importance,
due to the fact that they correspond to non-essential couplings.
Notice also that  the only conformal mode of the gravitational
field is renormalized in $2+\epsilon$ dimensions, which is quite
well known \cite{Wng1,GKT1,CD2,KH1,NTT1,KST1}.

The $\beta$-function for the gravitational constant has the form
(\ref{18})
\beq
\beta_G = \epsilon G - \frac{2(30-n)}{3} G^2. \label{19}
\eeq
Hence, for $n<30$ one obtains the infrared stable fixed point
$G=0$. There is also an ultraviolet stable fixed point as in 
\cite{Wng1,GKT1,CD2,KH1,NTT1}
\beq
G^* = \frac{3\epsilon}{2(30-n)}. \label{20}
\eeq
The theory is UV-stable (asymptotically free at exactly two dimensions).

One can start the search for fixed-point solutions corresponding to
the dilatonic couplings. There is following {\it Ansatz}:
\beq
\Phi (\phi) = \lambda \phi, \ \ \ \
f_2 (\phi) = \lambda_f \phi, \ \ \ \
V (\phi) = \lambda_V \phi.
\label{21}
\eeq
In that case the $\beta$-functions are simplified: 
\bea
\beta_\Phi &=& G(e^{2\phi} -1) \left[ \frac{30-n}{3} \lambda +
\frac{\epsilon}{4(1+\epsilon)} (\lambda -1) (n\lambda^2-8)
\right], \nn
\beta_{f_2} &=& G(e^{2\phi} -1) \left[ \frac{30-n}{3} \lambda_f +
\frac{n\lambda^2-8}{4(1+\epsilon)} (\epsilon\lambda_f+4 -
2\epsilon) -4\lambda_f +8 \right], \nn
\beta_V &=& G(e^{2\phi} -1) \left[ \frac{30-n}{3} \lambda_V +
\frac{n\lambda^2-8}{4(1+\epsilon)} (\epsilon\lambda_V-4 -
2\epsilon) -4\lambda_V -8 \right].
\label{22}
\eea

Equating these $\beta$-functions to zero we easily get 
the fixed-point solutions: 
\bea
&& \lambda^* = -\frac{6\epsilon}{30-n} + {\cal O} (\epsilon^2),
\label{23} \\
&& \lambda_f^* = -\frac{36\epsilon}{18-n} + {\cal O} 
(\epsilon^2), 
\ \ \ \ \lambda_V^* = \frac{12\epsilon}{18-n} + {\cal
O} (\epsilon^2). \label{24}
\eea
The fixed-point solutions (\ref{24}) are of the same nature as
solution (\ref{23}). Notice, however, that the denominator in
(\ref{24}) is different from the denominator for the case of a
purely dilatonic sector (\ref{23}). For the existence of the solutions
(\ref{24}) a new limitation appears: $n\neq 18$. Under a discontinuous
transition through the point $n=18$, the sign of the fixed points in
(\ref{24}) changes.

It is interesting to see in which way the model (\ref{5}) can be
rewritten at the fixed point. In particular, we will perform a
Weyl
rescaling (it is non-singular) in the manner suggested in ref.
\cite{KST1}
\beq
g_{\mu\nu} \longrightarrow g_{\mu\nu}  \, \exp \left( \frac{4
\lambda^*}{\epsilon} \phi \right). \label{25}
\eeq
Then, the classical action becomes
 \bea
S& =& \int d^d x\,  \sqrt{-g}\,  \biggl\{ {\mu^\epsilon \over
16\pi G^*}
e^{-2(1-\lambda^*)\phi} \left[ R - \frac{4(1+\epsilon)}{\epsilon}
\lambda^* (2- \lambda^*) \right] g^{\mu\nu} \partial_\mu \phi
\partial_\nu \phi  \nn
&&- {1 \over 2} g^{\mu\nu} \partial_\mu \chi_{i} \partial_\nu
\chi^i +  \mu^\epsilon
m^2  \e^{(2\lambda^* +4\lambda^*/\epsilon- \lambda^*_V)\phi}
+\frac{1}{4} F_{\mu\nu}^2 \e^{(2\lambda^* -4\lambda^*/\epsilon-
\lambda^*_f)\phi} \biggr\}. \label{26}
\eea
As one can see from expression (\ref{26}), without the
electromagnetic sector and setting  $m^2=0$, it describes the
CGHS action \cite{CGHS1} (at the limit $\epsilon \rightarrow 0$ it is finite
and scalars become non-interacting with the dilaton). As it stands,
action (\ref{26}) is of a  similar form as dilaton-Maxwell
gravity \cite{MNY1,LN1,EO1} with the Liouville potential. 
Notice that
such  kind of dilaton-Maxwell gravity (which can also be
considered as a charged string-inspired 
model \cite{MNY1,LN1,LK1}) admits charged
black hole solutions with multiple horizons, being in
this sense analogous to four or higher-dimensional
Einstein-Maxwell theories \cite{Gbn1,GM1}. 
Different forms of dilatonic
gravity can be easily obtained too, by transforming the metric 
and the
dilaton. 

One can study the 
stability of the fixed points (\ref{20}), (\ref{23}), (\ref{24}),
along the same lines as in ref. \cite{KST1}. 
Such analysis may be found in ref.\cite{EO3}. It shows that 
above fixed points may be stable along some RG trajectories. 
The gauge dependence analysis of above fixed points has been also 
done in ref.\cite{EO3,EO4}. It follows from this analysis that 
gravitational constant in UV fixed point is gauge independent. 

In this section we discussed dilatonic-Maxwell gravity with 
matter near two dimensions. The nice properties of this theory are: 
(i) its renormalizability in $2+\epsilon$ dimensions,  
(ii) The fact that there is 
a non-trivial fixed-point solution for the dilatonic couplings and 
that, at the fixed point, the theory may be represented in the 
standard form of the string-inspired models. (iii) The increase of 
the upper limit for the matter central charge ---due to the 
contribution of the vector field--- from 24 (pure dilatonic gravity) 
or 25 (Einstein gravity) to 30. 
More realistic models (like dilaton-Yang-Mills gravity) may be 
considered in a similar way \cite{EO3,EO4}, as well as supergravity 
models \cite{KST2}. It is also interesting that such models can be 
formulated as string theory.

Note that $2+\epsilon$-approach to quantum gravity did not 
finally succeeded to solve completely renormalizability issue 
of the quantum gravity in the usual sense. As a result the interest to this 
approach is minor now, it awaits until some new and fresh ideas will 
appear. 

\subsection{One-loop effective action in 4d higher derivative 
dilatonic gravity \label{2s4}}

\def\beq{\begin{equation}}
\def\eeq{\end{equation}}
\def\bce{\begin{center}}
\def\ece{\end{center}}
\def\bea{\begin{eqnarray}}
\def\eea{\end{eqnarray}}
\def\ben{\begin{enumerate}}
\def\een{\end{enumerate}}
\def\ul{\underline}
\def\ni{\noindent}
\def\nn{\nonumber}
\def\bs{\bigskip}
\def\ms{\medskip}
\def\wt{\widetilde}
\def\wh{\widehat}
\def\brr{\begin{array}}
\def\err{\end{array}}
\def\Tr{\mbox{Tr}\ }

In the present section we show how one can generalize 
the discussion of one-loop effective action for 4d dilatonic 
gravity. Supposing that 4d dilaton is still dimensionless 
we naturally come to higher derivative theory of sigma-model 
type which is again renormalizable in a generalized sense.
The most general form of such action looks like
$$
S= \int d^4x \sqrt{-g} \{ b_1 (\varphi ) ( \Box
\varphi )^2 +  b_2 (\varphi ) \left( \nabla_\mu \varphi
\right) \left( \nabla^\mu \varphi \right)
 \Box \varphi +  b_3 (\varphi ) [ ( \nabla_\mu \varphi
)( \nabla^\mu \varphi ) ]^2
 $$
$$
 +  b_4 (\varphi ) ( \nabla_\mu \varphi ) (
\nabla^\mu \varphi )+  b_5 (\varphi ) +  c_1 (\varphi ) R
( \nabla_\mu \varphi ) ( \nabla^\mu \varphi
) +  c_2 (\varphi ) R^{\mu\nu} ( \nabla_\mu \varphi
) ( \nabla_\nu \varphi )
 $$
\beq
+ c_3 (\varphi ) R \Box \varphi + a_1 (\varphi )
R^2_{\mu\nu\alpha\beta }  + a_2 (\varphi ) R^2_{\mu\nu} + a_3
(\varphi ) R^2 + a_4 (\varphi ) R \}
+ (\mbox{s.t.}),       \label{1x}
\eeq
where $s.t.$ means `surface terms'.
All generalized coupling constants are
dimensionless, except for $b_4$, $b_5$ and $a_4$, for which we
have:  $[b_4 (\varphi )]=2$,  $[b_5 (\varphi )]=4$,  $[a_4
(\varphi )]=2$. 

It is interesting to notice that, at the classical level and for some
particular choices of the generalized couplings, the
 action (\ref{1x}) may be viewed, in principle, as a superstring
theory effective action ---the only background fields being the
gravitational
field and the dilaton. 
The calculation of one-loop effective action may be done 
using similar algorithm as in 2d case (only some technical 
details are a bit more involved). Considering only case when 
dilaton is quantized and gravity is classical, one can get 
\cite{EJOS}
$$
\Gamma^{(1-loop)}_{div}= -{2\over\varepsilon}
\int d^4x\sqrt{-g}[A_1R^2_{\alpha\beta\gamma\tau}+A_2R^2_{\alpha\beta}+
A_3R^2+A_4Rm^2
$$
$$
+ C_1R(\nabla_\mu\varphi)^2+C_2R_{\mu\nu}
(\nabla^\mu\varphi)(\nabla^\nu\varphi) +C_3R(\Box\varphi)
$$
\beq
+ B_1(\Box\varphi)^2+B_2(\nabla_\mu\varphi)^2(\Box\varphi)+
B_3(\nabla_\mu\varphi)^4+B_4m^2(\nabla_\mu\varphi)^2+B_5m^4],
\label{z11}
\eeq
where functions $A_1$, $\cdots$, $B_5$ are explicitly given in 
ref. \cite{EJOS}. 

Since the theory is renormalizable, one can formulate the
renormalization group equations for the effective action 
and couplings and then explore its asymptotic behavior. 
The renormalization group equations for the effective action 
have the standard form, since the number (finite or infinite) 
of coupling constants is not essential. The general solution 
of this RG equation
has the form
\beq
\Gamma[e^{-2t}g_{\alpha\beta},a_i,b_j,c_k,\mu] =
\Gamma[g_{\alpha\beta},a_i(t),b_j(t),c_k(t),\mu ],
\label{z112}
\eeq
where $\mu$ is the renormalization parameter and the
effective couplings satisfy renormalization group equations
of the form
\be
{da_i(t)\over{dt}} = \beta_{a_i},\quad a_i=a_i(0), \quad 
{db_i(t)\over{dt}} = \beta_{b_i},\quad b_i=b_i(0), \quad 
{dc_i(t)\over{dt}} = \beta_{c_i},\quad c_i=c_i(0).
\label{z113}
\ee
Note that we do not take into account the dimensions of the functions
$a_4, b_4, b_5$. 
The derivation of the $\beta$-functions is pretty the same as in theories
with finite number of couplings, and we easily get
\beq
\beta_{a_i} = - (4\pi)^{-2} A_i,\;\;\;
\beta_{b_i} = - (4\pi)^{-2}B_i,\;\;\;
\beta_{c_i} = - (4\pi)^{-2}C_i.
\label{z115}
\eeq
The analysis of corresponding RG equations may be found 
in ref.\cite{EJOS}. 

As an example, one can consider the most general conformally-invariant
version of the theory (\ref{1x}):
\beq
S_c = \int d^4x \, \sqrt{-g} \, \left\{ f (\varphi ) \varphi \nabla^4
\varphi  +  q (\varphi ) C^2_{\mu\nu\alpha\beta}+ p(\varphi)\left[
 \left( \nabla_\mu \varphi
\right) \left( \nabla^\mu \varphi \right) \right]^2 \right\}.
\label{IIiv21}
\eeq
Here $ f(\varphi ), q( \varphi)$ and $ p( \varphi)$
are arbitrary functions, $\nabla^4 = \Box^2 +2R^{\mu\nu} \nabla_\mu
\nabla_\nu - \frac{2}{3} R \Box + \frac{1}{3} (\nabla^\mu R)
\nabla_\mu$ is a fourth-order conformally invariant operator, and we
should recall that due to the fact that $[\varphi ]=0$ the conformal
transformation of our 4d dilaton is trivial.
One can integrate by parts the rhs
in (\ref{IIiv21}) and
present the result as a particular case of the theory (\ref{1x}).
As one can show, the divergences of the
conformally-invariant theory (\ref{IIiv21}) appear also in a conformally
invariant form  (up to the total divergence), as it should be.
The functions $A(\varphi)$, $B(\varphi)$, 
$C(\varphi)$ obey the same conformal constraints with some
$F(\varphi)$, $Q(\varphi)$, $P(\varphi)$ instead of 
$f(\varphi)$, $q(\varphi)$, $p(\varphi)$. 
Hence,  the conformal invariant, higher-derivative
scalar theory considered here is renormalizable at the one-loop level
in a conformally invariant way, and therefore it is multiplicatively
renormalizable at one-loop. 

Using explicit one-loop effective action one can search for finite and 
anomaly free models (via condition that one-loop EA is zero).
One can also study renormalization group equations:  
 theory under consideration possesses 
finite fixed points and interesting flows between these fixed 
points \cite{EJOS}. We do not give this analysis here (while it 
may have some applications for construction of infrared sector 
of quantum gravity \cite{AM1, Odn1, hamada}). 
Thus, to conclude one sees that perturbative renormalization of 
4d dilatonic gravity is possible but its detailed analysis looks 
much more complicated than 2d case.

In summary, in this chapter we discussed the questions related with the 
calculation of one-loop effective action in 2d and 4d
dilatonic gravity . It is also applied to study the model
under discussion 
near two dimensions via epsilon-expansion. One can also consider 
applications of these effective actions for cosmology and observable 
consequences of gravity, like dilatonic contributions to Newton 
potential, construction of non-singular Universes (see chapter 5), etc.

\def\cL{{\cal L}}
\def\be{\begin{equation}}
\def\ee{\end{equation}}
\def\bea{\begin{eqnarray}}
\def\eea{\end{eqnarray}}
\def\tr{{\rm tr}\, }
\def\nn{\nonumber \\}
\def\gd{g^\dagger}
\def\e{{\rm e}}
\def\cF{{\cal F}}
\def\det{{\rm det\,}}
\def\Tr{{\rm Tr\,}}
\def\tr{{\rm tr\,}}
\def\e{{\rm e}}
\def\etal{{\it et al.}}
\def\erp2{{\rm e}^{2\rho}}
\def\erm2{{\rm e}^{-2\rho}}
\def\er4{{\rm e}^{4\rho}}

\section{Conformal anomaly and induced effective action for dilaton 
coupled theories}

In the present chapter we will discuss in detail conformal anomaly 
and induced EA for dilaton coupled theories in two and four 
dimensions. The remarkable property of such EA is the fact that 
it will be given for arbitrary dilaton-gravitational background. 
As a result, it will be possible to apply it in different problems 
of BH physics and early Universe.
 
\subsection{2d dilaton coupled scalar and quantum dilaton: conformal anomaly 
and induced action}

In this section 
following to refs.\cite{NO5,NO6},
 we show how to get conformal anomaly and anomaly 
induced EA for 2d dilaton coupled scalar 
with the help of results reported in second 
chapter. 

Let us start from the following action:
\be
\label{I}
S=\int d^2x \sqrt g \left\{ -{1 \over 2}Z(\phi) 
g^{\mu\nu}\partial_\mu\phi\partial_\nu\phi 
+ {1 \over 2}
f(\phi) g^{\mu\nu}\partial_\mu\chi_i
\partial_\nu\chi_i
\right\}
\ee
where $\phi$ is dilaton, $Z$, $f$ are the arbitrary 
dilaton functions, $\chi$ is scalar field, 
$i=1,\cdots,N$.
Here dilaton and scalars are quantum fields and 
gravitational field is considered to be external field.
It is easy to show that such theory is conformally invariant 
in two dimensions.

We work  with non-zero background 
dilaton and non-zero background scalar. 
The calculation of the one-loop effective action for 
such system has been done in 
all details in previous chapter.
The result looks as follows (\ref{IIIxxxxiv}) (see ref.\cite{ENO2}):
\bea
\label{eno-result}
\Gamma_{{\rm div}}
&=&- {1 \over 4\pi(n-2)}\int d^2x \sqrt g 
\Bigl\{ -{N+1 \over 6}R + 
\left({{f'}^2 \over 2fZ}-{f'' \over 2Z}\right) 
(\nabla^\lambda \chi_i)(\nabla_\lambda \chi_i) \nn
&& + \left({N f'' \over 2f}-{N{f'}^2 \over 4f^2} 
  - {{Z'}^2 \over 4 Z^2} \right)
(\nabla^\lambda \phi)(\nabla_\lambda \phi) 
+\left({Nf' \over 2f}-{Z' \over 2Z}\right)
\Delta \phi \Bigr\}
\eea
Here $\phi$, $\chi_i$ denote the dilaton and scalar 
background correspondently.

Let us give very brief introduction to the 
definition of conformal anomaly in two dimensions. It appears 
as result of quantum breaking (due to regularization) of
classical conformal invariance. Hence, it is
induced only by quantum effective action. From 
the definition of quantum energy-momentum tensor for the 
system which is classically conformally 
invariant one gets the following:
\be
\label{IIn1}
T={1 \over \sqrt{g}}g^{\mu\nu}{\delta \Gamma \over 
\delta g^{\mu\nu}}\ .
\ee
This defines conformal anomaly (expectation value of 
trace of stress-energy tensor). Making the 
conformal transformation
\be
\label{IIn2}
g_{\mu\nu}\rightarrow \Omega^2 g_{\mu\nu}\equiv \bar g_{\mu\nu}
\ee
one can rewrite the conformal anomaly as 
\be
\label{IIn3}
T={\Omega \over \sqrt{g}}\left.{\delta\Gamma \over \delta 
\Omega}\right|_{\Omega=1}\ .
\ee
Working in the one-loop approximation and reminding that 
classical action does not give the contribution 
to conformal anomaly (\ref{IIn3}) one understands 
that whole contribution to Eq.(\ref{IIn3}) is due to 
$\Tr\log$ terms. From Eq.(\ref{In4}) one knows that 
\be
\label{Inn4}
\bar\Gamma^{(1)}={i \over 2}\Tr\ln\left\{\mu^2 S_2\right\}
\ee
where dimensional parameter $\mu^2$ is explicitly introduced to 
make $S_2$ dimensionless. 
Under the conformal transformation the one-loop effective 
action becomes (the case of conformal scalar 
is considered for simplicity): 
\be
\label{Inn5}
\bar\Gamma^{(1)}\left[\bar g_{\mu\nu}\right]
={i \over 2}\Tr\log \left\{\mu^2\Omega^2 S_2\right\}\ .
\ee
Then 
\be
\label{IIn6}
T={\Omega \over \sqrt{g}}\left.{\delta\bar\Gamma^{(1)} 
\over \delta \Omega}\right|_{\Omega=1}
={\mu(x) \over \sqrt{g}}\left.{\delta\bar\Gamma^{(1)} 
\over \delta \mu(x)}\right|_{\mu=1}\ .
\ee
Let us take into account that $\mu$-dependent part of 
$\bar\Gamma^{(1)}$ may be presented as follows (see 
\cite{BrDv})
\be
\label{IIn7}
\bar\Gamma^{(1)}=\int d^2x \sqrt{g} a_1(x){1 \over 2}
\ln \mu^2 + \cdots
\ee
where as $\cdots$, we denote the dimensional parameter 
$\mu$-independent terms. 
Then Eq.(\ref{IIn7}) is actually zeta-regularization 
representation of the one-loop effective action. Hence,
\be
\label{Inn8}
T=a_1(x)\ .
\ee
This $a_1(x)$-coefficient of Schwinger-De Witt expansion 
appears also in the calculation of one-loop divergent 
part of effective action in dimensional regularization. 
Hence, multiplying $\Gamma_{div}$ to $n-2$ and dropping integral 
one gets two-dimensional conformal anomaly. 
Similarly, in four dimensions $T$ is defined by 
$a_2$-coefficient of Schwinger-De Witt expansion. 
Note that in such representation of 
conformal anomaly there may be some ambiguity due to the 
fact that Schwinger-De Witt coefficients $a_1$, $a_2$, $\cdots$ 
are defined up to the total derivatives terms. 
(To solve this problem one should not use 
the integration by parts in the derivation of 
$\Gamma_{div}$ in Schwinger-De Witt algorithm. This is the 
case in our calculation!). 
The good algorithm to define the conformal anomaly in universal 
way is given in ref.\cite{IcIk}, where also the  
problems with total derivative terms in conformal anomaly
do not appear.


Thus, one can easily get the 
conformal anomaly of conformally invariant system just 
by calculating the corresponding divergent 
EA multiplied to $(n-2)$ in 2 
dimensions (in other words, as coefficient of the pole). 
Hence, the conformal anomaly of the system under 
discussion is given by \cite{NO5}
\bea
\label{trace}
T&=&{1 \over 24\pi}\Bigl\{ (N+1)R
  -3\left({{f'}^2 \over fZ}-{f'' \over Z}\right) 
(\nabla^\lambda \chi_i)(\nabla_\lambda \chi_i) \nn
&& - 3\left({N f'' \over f}-{N{f'}^2 \over 2f^2} 
  - {{Z'}^2 \over 2 Z^2} \right)
(\nabla^\lambda \phi)(\nabla_\lambda \phi) 
  - 3\left({Nf' \over f}-{Z' \over Z}\right)
\Delta \phi \Bigr\}
\eea
For purely scalar system coupled with dilaton 
(i.e. dilaton is classical) the conformal anomaly is simplified:
\be
\label{trace2}
T={1 \over 24\pi}\Bigl\{ NR
 - 3N\left({ f'' \over f}-{{f'}^2 \over 2f^2} \right)
(\nabla^\lambda \phi)(\nabla_\lambda \phi) 
 - 3{Nf' \over f}\Delta \phi \Bigr\}
\ee
 Anomaly looks very simple in the 
 special case $N=1$ (single 
scalar) .
The calculation of such conformal anomaly has been done 
in the last case, using zeta-regularization 
method in ref.\cite{BH1}.
 Note that the regularization dependence  
of total derivative terms in conformal anomaly is possible.
Thus we got the conformal anomaly for most general 2d 
dilaton coupled scalar-dilaton system with 
arbitrary dilaton couplings. 
Similar calculations have been also made in 
refs.\cite{CS1,Ich2,MR1,KLV1,Dwk1,IO1} 
for dilaton coupled scalars.
Using the conformal anomaly (see eq.(\ref{trace})),  
we may calculate the induced non-local EA.

Let us write the general form of conformal anomaly as following
\be
\label{trace4}
T=cR + F_1(\phi)
(\nabla^\lambda \chi_i)(\nabla_\lambda \chi_i) 
+ F_2(\phi)(\nabla^\lambda \phi)(\nabla_\lambda \phi) 
 +F_3(\phi)\Delta \phi 
\ee
where the explicit form of $c$, $F_1$, $F_2$, $F_3$ is 
evident from the comparison with (\ref{trace2}):
\bea
\label{comp}
&& c={N+1 \over 24\pi}\ ,\ \ 
F_1(\phi)=-{1 \over 8\pi}
\left({{f'}^2 \over fZ}-{f'' \over Z}\right), \nn 
&& F_2(\phi)=-{1 \over 8\pi}
\left({N f'' \over f}-{N{f'}^2 \over 2f^2} 
  - {{Z'}^2 \over 2 Z^2} \right)\ ,  \quad 
 F_3(\phi)=-{1 \over 8\pi}
\left({Nf' \over f}-{Z' \over Z}\right)\ .
\eea
In situation when dilaton is purely classical the above functions are
simplified.
The (non-local) effective action induced by the 
conformal anomaly is defined as 
\be
\label{effaction}
\int d^2x\sqrt g T = 2 \left.{dW \over dt}
\right|_{t=1}
\ee
where $\tilde g^{\mu\nu}=t^{-1}g_{\mu\nu}$. 
In other words, the effective action (\ref{effaction}) 
should derive the anomaly $T$.
Integrating anomaly one gets
\bea
\label{indaction2}
W&=&-{1 \over 2}\int d^2x \sqrt g \Bigl[ 
{c \over 2}R{1 \over \Delta}R + F_1(\phi)
(\nabla^\lambda \chi_i)(\nabla_\lambda \chi_i) 
{1 \over \Delta}R \nn
&& + \left(F_2(\phi)- {\partial F_3(\phi) \over 
\partial \phi}\right)\nabla^\lambda \phi
\nabla_\lambda \phi {1 \over \Delta}R 
+ R\int F_3(\phi) d\phi \Bigr]
\eea
The remarkable property of this action is that it is defined 
for arbitrary dilaton-gravitational background. Note, 
however, that such action is not yet full effective action. 
The reason is that it is calculated from Eq.(\ref{effaction}) 
as integral from conformal anomaly. Hence, it is defined up to 
$t$-dependent (conformal factor dependent) constant, or in 
other words, up to conformally invariant functional.

There is no scheme where one can find such conformally 
invariant functional in closed form. At best, one can use some 
kind of approximation  to calculate it. Hence, 
exact and complete expression for one-loop EA looks as 
\be
\label{Gammai}
\Gamma=W+\Gamma\left[1,\tilde g_{\mu\nu}\right]
\ee
where second term is conformally invariant functional. 
We mention briefly how one can get it. Using SD expansion, 
in the case of classical dilaton and only quantized scalars, 
one gets \cite{NO7}
\be
\label{Gammaii}
\Gamma\left[1,\tilde g_{\mu\nu}\right] 
= - {N \over 4\pi}\int d^2x \sqrt{g}
\left(\nabla_\mu\phi\right)\left(\nabla^\mu\phi\right)
\ln\mu^2
\ee
where numerical factor in front of integral does not matter, 
as it may be reabsorbed to definition of $\mu^2$. Other, non-local terms 
of this expansion have been recently calculated in ref.\cite{gusev}.

\subsection{ Dilaton coupled spinor conformal anomaly in two dimensions}

In the present section, we give another (simpler) method \cite{NNO} to 
find conformal anomaly for dilaton coupled spinor. 
Unfortunately, it cannot be applied to other theories. 

Let us start from 2d dilaton coupled spinor Lagrangian:
\be
\label{i}
L=\sqrt{-g}f(\phi)\bar\psi \gamma^\mu \partial_\mu \psi
\ee
where $\psi$ is 2d Majorana spinor.\footnote{As we see below in the present
formalism where only notations for classical external fields are changed 
the dilatonic contribution to conformal anomaly appears. This contribution 
has the form of total derivative term. In ref.\cite{IO1} using
transformations of quantum fields,Fujikawa regularization and the proper
choice of integration measure the calculation 
of dilaton dependent contribution to conformal anomaly gives zero.
This fact maybe regarded as the regularization dependence of conformal
anomaly. It is not strange as it is known for some time that total
derivative terms of conformal anomaly in various dimensions may depend on
the choice of regularization. Moreover, in dilaton dependent SUSY theories
like super YM coupled with conformal supergravity (where dilaton presents) 
the fact of the presence of dilaton dependent terms in conformal anomaly is
proven by various formalisms. It indicates that from SUSY point of view 
the measure of integration in functional space which leads to non-trivial
dilaton dependent term in conformal anomaly looks more physical.}

Let us make now the following classical transformation of background 
field $g_{\mu\nu}$: 
$g_{\mu\nu}\rightarrow f^{-2}(\phi)\tilde g_{\mu\nu}$.
Then it is easy to see that $\gamma^\mu(x)\rightarrow f(\phi)
\tilde\gamma^\mu(x)$ and in terms of new classical metric we obtain usual, 
non-coupled with dilaton (minimal) Lagrangian for 2d spinor:
\be
\label{iii}
L=\sqrt{-\tilde g}\bar\psi\tilde\gamma^\mu\partial_\mu\psi\ .
\ee
The conformal anomaly for Lagrangian (\ref{iii}) is well-known:
\be
\label{iv}
\sqrt{-\tilde g}T={\sqrt{-\tilde g} \over 24\pi}\left\{{1 \over 2}
\tilde R\right\} \ .
\ee
Now, transforming Eq.(\ref{iv}) to original variables:
$\tilde g_{\mu\nu}=f^2(\phi)g_{\mu\nu}$, 
$\tilde R=f^{-2}(\phi)\left(R - 2 \Delta\ln f\right)$, 
we get the following conformal anomaly for dilaton coupled Majorana 
spinor (\ref{i}):
\be
\label{vi}
\sqrt{-g}T = {\sqrt{-g} \over 24\pi}\left[{1 \over 2}R - 
\Delta \ln f \right] 
= {\sqrt{-g} \over 24\pi}\left[{1 \over 2}R - {f' \over f}
\Delta f - {\left(f''f - {f'}^2 \right) \over f^2}
g^{\mu\nu}\partial_\mu\phi\partial_\nu\phi \right]\ .
\ee
For 2d dilaton coupled Dirac spinor the conformal anomaly is 
twice of that for Majorana spinor. Note also that it could be as in case 
of dilaton coupled scalar, conformal anomaly (its total derivative terms) 
depends on the regularization. 

In the conformal anomaly (\ref{vi}) dilaton dependent 
terms appear in the form of total derivative. In principle, 
it means that this term is ambiguous by physical reasons. 
Indeed, in two dimensions the analogue of Einstein action 
looks like:
\be
\label{vii}
S={1 \over G}\int d^2x \sqrt{-g}Rf(\phi)\ .
\ee
Now, there exists the following relation
\be
\label{viii}
g_{\mu\nu}{\delta \over \delta g_{\mu\nu}}\int
d^2x \sqrt{-g}R f(\phi)=\Delta f(\phi)\ .
\ee
In other words, by finite renormalization of gravitational 
action (\ref{vii}), we can always change the coefficient 
of $\Delta f$ term in conformal anomaly. So in 2d gravity this term 
is only fixed by the physical renormalization condition. 
That is also why total derivative term of conformal anomaly 
for dilaton coupled scalar is ambiguous.

Now we discuss the anomaly induced effective action for dilaton coupled 
spinor. The derivation goes in the same way as it was for dilaton 
coupled scalar. Making the conformal transformation of the metric 
$g_{\mu\nu}\rightarrow \e^{2\sigma}g_{\mu\nu}$ in the conformal anomaly, 
and using relation 
\be
\label{AAA}
\sqrt{-g}T={\delta \over \delta\sigma}W[\sigma]
\ee
one can find the anomaly induced effective action $W$. 
In the covariant, non-local form it may be written as following:
\be
\label{x}
W=-{1 \over 2}\int d^2x \sqrt{- g} \left\{ 
{1 \over 96\pi}R{1 \over \Delta}R + \left(F_2(\phi) 
  - {\partial F_3(\phi) \over \partial \phi} \right)
(\nabla^\lambda \phi)(\nabla_\lambda \phi) {1 \over \Delta}R 
+ R \int F_3(\phi) d\phi\right\}
\ee
where $F_2(\phi)= - {f''f - {f'}^2 \over 24\pi f^2}$, 
$F_3(\phi)= - {f' \over 24\pi f}$.
Note that coefficient of second term is actually zero as is easy to 
check. Hence, 
we got anomaly induced effective action for dilaton coupled spinor. 

An interesting thing is that the effective action (\ref{x}) 
exactly reproduces the effective action of RST model \cite{RST1}, 
which is exactly solvable. The RST model is given by adding the 
quantum correction 
\be
\label{RSTq}
W_{RST}=-{1 \over 2}\int d^2x \sqrt{- g} \left\{ 
{N \over 48\pi}R{1 \over \Delta}R 
+ {N \over 24\pi}\phi R\right\}
\ee
to the action of the CGHS model \cite{CGHS1}
\be
\label{CGHS}
S_{CGHS}={1 \over 2\pi}\int d^2x \sqrt{-g}\e^{-2\phi}
\left(R + 4 \nabla_\mu\phi \nabla^\mu \phi + 4\lambda^2\right)
\ .
\ee
In \cite{RST1}, the second term in $W_{RST}$ (\ref{RSTq}) is added 
by hands. 

It is interesting that full EA is again defined with 
accuracy up to conformally invariant functional. 
To conclude this section let us note that in calculation of EA 
for consistent 2d dilatonic gravity with matter one needs also 
quantum gravitational contribution. However, dilatonic gravity 
is not renormalizable in the standard way 
as it was shown in previous chapter. As a 
result finite contributions to effective action will depend  on 
regularization scheme under discussion. 

\subsection{ Induced effective action for 4d dilaton coupled scalar}

It could be interesting to generalize the results of 
previous sections for 4d case. The purpose of the present section 
will be to calculate the non-local effective action for 4d 
dilaton coupled conformal scalar \cite{NO6}.
Let us consider the theory with the following 
Lagrangian in curved spacetime 
\be
\label{v1}
L=\sqrt{-g}\varphi f(\phi) (\Box - \xi R )\varphi
\ee
where $\varphi$ is quantum scalar field, 
$\Box=g^{\mu\nu}\nabla_\mu \nabla_\nu$, 
$\phi$ is an external field (dilaton), 
$f(\phi)$ is an arbitrary function.
It is very easy to check that 
the theory (\ref{v1}) is conformally 
invariant for $\xi={1 \over 6}$.

Let us calculate the divergent part of the effective 
action for the theory (\ref{v1}):
\bea
\label{v3}
\Gamma_{\rm div}&=&-{i \over 2}\Tr\ln
\left\{f(\phi)\left[ \Box - \xi R 
+{\Box f(\phi) \over 2f(\phi)}
+{(\nabla^\mu f(\phi) ) \over f(\phi)}\nabla_\mu 
\right]\right\} \nn
&=& {1 \over (n-4)}\int d^4x \sqrt{-g} 
b_4
\eea
where $b_4$ (or $a_2$) is the $b_4$-coefficient of 
Schwinger-De Witt expansion. 
Applying standard methods, we find $b_4$ explicitly, where
 $f(\phi)$-multiplier in Eq.(\ref{v3}) 
does not contribute.
For $\xi = {1 \over 6}$, we get the conformal anomaly which is equal to $b_4$
\be
\label{v6}
T={1 \over (4\pi)^2}
\Bigl\{ {1 \over 32}{[(\nabla f) (\nabla f)]^2 \over f^4}
+{1 \over 24}\Box \left(
{(\nabla f) (\nabla f) \over f^2}\right) 
 + {1 \over 180}
\left( R_{\mu\nu\alpha\beta}^2 - R_{\mu\nu}^2 
+ \Box R \right)\Bigr\}
\ee
Here, the last term is the well-known Weyl
anomaly (for a review, see \cite{Dff1,DS1}) for conformally 
invariant scalar. The first two terms in (\ref{v6}) are the 
dilaton contribution to conformal anomaly.

Let us write the Eq.(\ref{v6}) in a slightly 
different form:
\be
\label{v7}
T=\Bigl\{b\left(F+ {2 \over 3}\Box R \right) 
+ b'G + b'' \Box R + a_1 
{[(\nabla f) (\nabla f)]^2 \over f^4}
+a_2\Box \left(
{(\nabla f) (\nabla f) \over f^2}\right) \Bigr\}
\ee
where $F$ is the square of Weyl tensor in four 
dimensions, $G$ is Gauss-Bonnet invariant.
For scalar field, it follows from (\ref{v6}) that 
$b={ 1 \over 120 (4\pi)^2 }$, $b'=-{ 1 \over 360 (4\pi)^2 }$, 
$a_1={ 1 \over 32 (4\pi)^2 }$, $a_2={ 1 \over 24 (4\pi)^2 }$ 
and in principle $b''$ is an arbitrary parameter 
(it may be changed by the finite renormalization 
of local counterterm).

The non-local effective action induced by the 
conformal anomaly (without dilaton) has been 
calculated sometime ago \cite{Rgr}.
Using Eq.(\ref{AAA}) and 
integrating it, one can restore the non-local 
effective action $W$ induced by the conformal 
anomaly:
\bea
\label{v10}
&& W=b\int d^4x \sqrt{-g} F\sigma 
 +b'\int d^4x \sqrt{-g} \Bigl\{\sigma\left[
2\Box^2 + 4 R^{\mu\nu}\nabla_\mu\nabla_\nu 
  - {4 \over 3}R\Box + {2 \over 3}(\nabla^\mu R)\nabla_\mu 
\right]\sigma \nn
&& \quad + \left(G-{2 \over 3}\Box R\right)\sigma \Bigr\} 
 -{1 \over 12}\left(b'' + {2 \over 3}(b + b')\right)
\int d^4x \sqrt{-g}\left[R - 6 \Box \sigma 
  - 6(\nabla \sigma)(\nabla \sigma) \right]^2 \nn
&& \quad + \int d^4x \sqrt{-g} \Bigl\{ 
a_1 {[(\nabla f) (\nabla f)]^2 \over f^4}\sigma 
+a_2\Box \left({(\nabla f) (\nabla f) \over f^2} 
\right)\sigma \nn
&& \qquad +a_2{(\nabla f) (\nabla f) \over f^2} [(\nabla \sigma) 
(\nabla \sigma )]\Bigr\}
\eea
Here the $\sigma$-independent term is dropped, last terms 
represent the contribution from the dilaton dependent 
 terms in conformal anomaly. It is interesting that dilatonic contribution 
may be important in the study of quantum irreversibility \cite{anselmi}.

\subsection{Conformal and chiral anomaly for 4d dilaton coupled spinor}

We consider now 4d dilaton coupled fermion which may appear 
as the result of spherical reduction of higher dimensional 
minimal spinor (we follow to ref.\cite{NNO}). 
The corresponding conformally invariant Lagrangian may be taken as follows
\footnote{As we mentioned in section 3.2 in the present formalizm 
where no transformations of quantum fields is done we get the dilaton
dependent contribution to conformal anomaly. Such dilaton dependent
contribution
to conformal anomaly looks very natural from SUSY point of view. In
ref.\cite{IO1} in terms of Fujikawa regularization the other answer has
been obtained. The difference in two answers maybe regarded as
regularization dependence of conformal anomaly.} :
\be
\label{Vi}
L=\sqrt{-g}f(\phi)\bar\psi \gamma^\mu(x) \nabla_\mu \psi
\ee
where $\nabla_\mu=\partial_\mu + {1 \over 2}\omega^{ab}_\mu\sigma_{ab}$.

Let us make the following transformation of background gravitational field:
\be
\label{Vii}
g_{\mu\nu}=\e^{2\sigma(x)}\tilde g_{\mu\nu}\ ,\ \ 
\gamma^\mu(x)=\e^{-\sigma(x)}\tilde\gamma^\mu(x)\ ,\ \ 
\sqrt{-g}=\e^{4\sigma(x)}\sqrt{-\tilde g}\ .
\ee
where $\sigma(x)$ satisfies $\e^{3\sigma(x)}f(\phi)=1$.
Then, Lagrangian (\ref{Vi}) after transformation (\ref{Vii}) 
takes the form:
\be
\label{Vv}
L=\sqrt{-\tilde g}\left[\bar\psi\tilde\gamma^\mu(x)
\tilde\nabla_\mu\psi 
+ \bar\psi\tilde\gamma^\mu\tilde A_\mu\psi \right]
\ee
where we used $\partial_\mu\equiv \tilde\partial_\mu$ and 
$\tilde A_\mu = {3 \over 2}\partial_\mu\sigma(x)$. Note that 
field strength for above vector field is equal to zero, that is 
why no terms of the sort : square of field strength for above 
vector appear in conformal anomaly. 
Hence, the calculation of $a_2$ Schwinger-De Witt coefficient in theory 
(\ref{Vi}) in curved spacetime with nontrivial dilaton is equivalent to 
the calculation of $a_2$ in an external gravitational field 
$\tilde g_{\mu\nu}$ (but no dilaton) and external vector field 
$\tilde A_\mu$.
 Conformal anomaly for such quantum (Dirac) fermion is 
well-known:
\be
\label{Vvii}
{\sqrt{-\tilde g} \over (4\pi)^2}\left\{{1 \over 20}\left( \tilde F 
+ {2 \over 3}\tilde\Box\tilde R \right) - {11 \over 360}\tilde G
\right\}
\ee
where
$\tilde F = \tilde R_{\mu\nu\alpha\beta}
\tilde R^{\mu\nu\alpha\beta}-2\tilde R_{\mu\nu}\tilde R^{\mu\nu}
+ {1 \over 3}\tilde R^2$, 
$\tilde G = \tilde R_{\mu\nu\alpha\beta}
\tilde R^{\mu\nu\alpha\beta}-4\tilde R_{\mu\nu}\tilde R^{\mu\nu}
+ \tilde R^2$. One can also present coefficients of conformal 
anomaly as 
 $b={1 \over 20(4\pi)^2}$ and $b'=-{11 \over 360(4\pi)^2}$. 
Now, one can transform the relation (\ref{Vvii}) back to original 
metric tensor. 
It is easy to see that there are dilaton dependent 
contributions to conformal anomaly. One can also find 
anomaly induced effective action.
Starting from Eq.(\ref{Vvii}) for conformal anomaly in terms of 
tilded metric the corresponding anomaly induced action in 
this case is quite known (see \cite{Rgr,FT1,BOS2,AM1,Odn1}): 
\bea
\label{Vx}
W&=&-{1 \over 4b'}\int d^4x \sqrt{-\tilde g}
\int d^4x' \sqrt{-\tilde g'}\left[b\tilde F 
+ b'\left(\tilde G - {2 \over 3} \tilde \Box \tilde R\right) \right]_x \nn
&& \times \left[ 2\tilde \Box^2 + 4 \tilde R^{\mu\nu}
\tilde\nabla_\mu\tilde\nabla_\nu - {4 \over 3}\tilde R \tilde \Box 
+ {2 \over 3}\left(\tilde\nabla^\mu\tilde R\right)\tilde\nabla_\mu
\right]^{-1}_{xx'} \nn
&& \times \left[b\tilde F 
+ b'\left(\tilde G
 - {2 \over 3} \tilde \Box \tilde R\right) \right]_{x'} 
 - {1 \over 18}(b+b')\int d^4x \sqrt{-\tilde g}\tilde R^2 \ .
\eea
It is trivial to substitute metric $\tilde g_{\mu\nu}=\e^{-2\sigma}
g_{\mu\nu}$, $\sigma=-{1 \over 3}\ln f(\phi)$ and 
rewrite above equation in terms of original metric 
and dilaton function but the final result is a bit complicated. 

As final remark, we note that similarly one can calculate the chiral 
anomaly for dilaton coupled spinor. Chiral anomaly for theory 
(\ref{Vv}) is known \cite{EF1,Nwn1,ET}:
\bea
\label{Vxi}
A_{1 \over 2}&=&{2i \over (4\pi)^2}\left[-{1 \over 48}
\epsilon^{\mu\nu\rho\sigma}{\tilde R^\xi}_{\zeta\mu\nu}
{\tilde R^\zeta}_{\xi\rho\sigma}\right] \nn
&=& {2i \over (4\pi)^2}\left[-{1 \over 48}
\epsilon^{\mu\nu\rho\sigma}\left\{{R^\xi}_{\zeta\mu\nu}
{R^\zeta}_{\xi\rho\sigma}
  -16 {R^\xi}_{\rho\mu\nu}\nabla_\sigma\nabla_\xi\sigma \right. 
\right. \nn
&& \left.\left. -16 {R^\xi}_{\rho\mu\nu}\nabla_\sigma\sigma\nabla_\xi
\sigma 
  -8 R_{\mu\nu\rho\sigma}\nabla_\alpha\nabla^\alpha \sigma
\right\}\right] 
\ .
\eea
Hence we found explicitly dilaton dependent corrections 
for chiral and conformal anomaly in the theory of 4d dilaton coupled spinor.

\subsection{Conformal anomaly for 4d dilaton coupled vector field}

In the study of string theory the low energy 
4d string effective action  may be presented as following
\be
\label{I1}
S=\int d^4x \sqrt{-g}\left[ R + 4 (\nabla \phi)^2 + 
F_{\mu\nu}^2\right]\e^{-2\phi}
\ee
where $\phi$ is dilaton, $F_{\mu\nu}=\nabla_\mu A_\nu 
 - \nabla_\nu A_\mu$, $A_\mu$ is electromagnetic field. 
 Investigating of string gravity (\ref{I1}) the quantum effects 
of electromagnetic field with the lagrangian:
\be
\label{I2}
L=-{1 \over 4}f(\phi)F_{\mu\nu}F^{\mu\nu}
\ee
may be dominant in some regions,especially if we consider 
generalization of above model with $N$ vectors and apply 
large $N$ expansion. 
 From another point if we 
start from usual Einstein-Maxwell gravity in $d$-dimensions,
we can do spherical reduction to the space 
$R_4\times S_{d-4}$ where $R_4$ is an arbitrary curved space.
Then the reduced action becomes again of the form (\ref{I1})
(maybe with change of some terms and some 
coefficients) where the radius of 
$S_{d-4}$ plays the role of dilatonic function. 

Hence, the study of dilaton coupled electromagnetic field 
with the Lagrangian (\ref{I1}) which describes conformally 
invariant theory may be of interest in different respects.
In the present section, we study conformal anomaly in the 
theory (\ref{I1}) in four dimensions (in two dimensions theory is 
not conformally invariant). 

Adding to the Lagrangian (\ref{I2}) the gauge-fixing 
Lagrangian $L_{gf}$:
\be
\label{II}
L_{gf}=-{1 \over 2}f(\phi)(\nabla_\nu A^\nu)^2\ ,
\ee
one can apply the standard algorithm to get the effective action.
For ghosts it is the same as in the theory with no dilaton 
\bea
\label{XI}
&& \Gamma^{(1)}_{ghost}
={1 \over (4\pi)^2 (n-4)}
\int d^4x \sqrt{-g}\left\{-{1 \over 90}R_{\mu\nu\alpha\beta}^2
+{1 \over 90}R_{\mu\nu}^2 
 -{R^2 \over 36} -{1 \over 15}\Box R\right\}  \nn
&& \quad ={1 \over (n-4)}\int d^4 x \sqrt{-g}b_4^{ghost} \ .
\eea

The one-loop effective action due to vectors may be found as
\cite{NO1,IO1}
\bea
\label{XIII}
\Gamma^{(1)}_{A_\mu}&=&{1 \over (4\pi)^2 (n-4)}
\int d^4x \sqrt{-g}\left\{
  -{11 \over 180}R_{\mu\nu\alpha\beta}^2
+{43 \over 90}R_{\mu\nu}^2 -{1 \over 9}R^2 \right. \nn
&& - {1 \over 30}\Box R -{1 \over 3}\Box\left({1 \over f}
\Box f\right) + {5 \over 12}\Box \left[{1 \over f^2} 
(\nabla_\alpha f)(\nabla^\alpha f) 
\right]
\nn
&& -{1 \over 3f^2}
R_{\mu\nu}(\nabla^\mu f)(\nabla^\nu f) 
 -{1 \over 3f}R_{\mu\nu}(\nabla^\mu\nabla^\nu f) 
+{R \over 6f}(\Box f) \nn
&& +{9 \over 16f^4} (\nabla_\mu f)(\nabla^\mu f) 
(\nabla_\alpha f)(\nabla^\alpha f) 
+{1 \over 6f^3}(\nabla^\beta f)(\nabla^\nu f) 
(\nabla_\beta\nabla_\nu f) \nn
&& - {11 \over 12f^3}(\nabla^\mu f)(\nabla_\mu f)(\Box f)
 -{1 \over 6f^2}(\nabla_\alpha\nabla_\beta f)
(\nabla^\alpha\nabla^\beta f) 
 \left. +{5 \over 12f^2}(\Box f)(\Box f)\right\} \nn
&=&{1 \over (n-4)}\int d^4 x \sqrt{-g}b_4^{vector}  \ .
\eea
The total one-loop effective action $\Gamma^{(1)}$ is 
given by sum of (\ref{XI}) and (\ref{XIII}).
 From the expression for $\Gamma^{(1)}$ one can easily 
get conformal anomaly:
$T=b_4 =b_4^{ghost}+b_4^{vector}$. 
The first four terms give the well-known conformal anomaly 
of electromagnetic  field \cite{BC1,CD1}.
We have to note that using another regularization (like 
zeta-regularization) may slightly alter the coefficients 
of total derivative terms, like already 
happened in this case in the absence of dilaton 
\cite{BC1,CD1}.

We have to note, however, that there are some unclear points. 
First, it was shown \cite{NO1,IO1} that such conformal anomaly 
cannot be integrated and anomaly induced action cannot be 
constructed. That indicates that convenient approach (as 
applied above) fails to reproduce correct conformal anomaly. 
It is known \cite{DS1} that conformal anomaly consists from 
only trivial (total derivative) terms and conformally 
invariant terms. Above conformal anomaly does not satisfy 
this property as it was shown in \cite{IO1} via classification 
of terms entering to conformal anomaly. It may be that BRST 
approach (for a recent review, see \cite{Lavrov}) in 
combination with some convenient regularization may help 
in resolution of this problem as it was discussed in \cite{IO1}. 
In the last chapter, we will show that correct conformal anomaly 
for this case may be reproduced from AdS/CFT correspondence. 
Nevertheless, it would be interesting to understand why conventional approach 
to conformal anomaly as in this section fails.

\subsection{Conformal anomaly and induced action in 
2d dilatonic supergravity with dilaton coupled 
Wess-Zumino matter}

\newcommand{\inv}[1]{\left[#1\right]_{\mbox{inv}}}
\newcommand\DS{D \hskip -3mm / \ }

Based on \cite{NO3}, in this section, we construct 
the theory of 2d dilatonic 
supergravity(SG) with matter and dilaton supermultiplets coupled to
dilaton functions. The 2d model also appears by the spherical 
reduction from 4d supergravity theories \cite{NO3}.  
Conformal anomaly and induced effective action
for matter supermultiplet are calculated, which gives also
large-$N$ effective action for dilatonic SG.

At first we construct the action of 2d dilatonic supergravity 
with dilaton supermultiplet and with matter supermultiplet.
In order to find the Lagrangian of 2d dilatonic
supergravity, we use the component formulation of ref.\cite{HUY1,
Umt1,Umt2} (for introduction, see book\cite{GSRG}).
Here, all the scalar fields are real and
all the spinor fields are Majorana spinors.
We introduce dilaton multiplet $\Phi=(\phi,\chi,F)$
and matter multiplet $\Sigma_i=(a_i,\chi_i,G_i)$,
which has the conformal weight $\lambda=0$, and
the curvature multiplet $W$
\be
\label{W}
W=\left(S,\eta,-S^2-{1 \over 2}R-{1 \over 2}
\bar\psi^\mu\gamma^\nu\psi_{\mu\nu}
+{1 \over 4}\bar\psi^\mu\psi_\mu\right)\ .
\ee
Here $R$ is the scalar curvature and 
\bea
\label{eta}
&& \eta=-{1 \over 2}S\gamma^\mu\psi_\mu 
+{i \over 2}e^{-1}\epsilon^{\mu\nu}\gamma_5
\psi_{\mu\nu} \ , \quad 
\psi_{\mu\nu}=D_\mu\psi_\nu - D_\nu\psi_\mu \ , \nn
&& D_\mu\psi_\nu=\left(\partial_\mu 
  -{1 \over 2}\omega_\mu\gamma_5\right)\psi_\nu \ , \quad
\omega_\mu=-ie^{-1}e_{a\mu}\epsilon^{\lambda\nu}
\partial_\lambda e^a_\nu
  -{1 \over 2}\bar\psi_\mu\gamma_5\gamma^\lambda
\psi_\lambda\ .
\eea
The curvature multiplet has the conformal 
weight $\lambda=1$.

Then the general action of 2d dilatonic supergravity is given 
in terms of general functions of the dilaton $C(\phi)$, $Z(\phi)$, 
$f(\phi)$ and $V(\phi)$ as follows
\bea
\label{lag}
&&{\cal L}=-\inv{C(\Phi)\otimes W} + \inv{V(\Phi)} 
 +{1 \over 2} \inv{\Phi\otimes\Phi\otimes T_P(Z(\Phi))}
  -\inv{Z(\Phi)\otimes\Phi \otimes T_P(\Phi)} \nn
&& +\sum_{i=1}^N \left\{{1 \over 2}
\inv{\Sigma_i\otimes\Sigma_i\otimes T_P(f(\Phi))}
  -\inv{f(\Phi)\otimes \Sigma_i\otimes T_P(\Sigma_i)}
\right\}\ .
\eea
$T_P(Z)$ is called the kinetic multiplet for 
the multiplet $Z=(\varphi, \zeta, H)$ and 
when the multiplet $Z$ has the comformal weight 
$\lambda=0$, $T_P(Z)$ has the following form
\be
\label{kin}
T_P(Z)=(H, \DS\zeta, \Box\varphi)\ .
\ee
The kinetic multiplet 
$T_P(Z)$ has conformal weight $\lambda=1$.
The product of two multiplets 
$Z_k=(\varphi_k, \zeta_k, H_k)$ $(k=1,2)$ 
with the conformal weight $\lambda_k$ is defined by 
\be
\label{product}
Z_1\otimes Z_2 =(\varphi_1\varphi_2, 
\varphi_1\zeta_2 + \varphi_2\zeta_1, 
\varphi_1 H_2 + \varphi_2 H_1 - \bar\zeta_1\zeta_2)\ .
\ee
The invariant Lagrangian $\inv{Z}$ for multiplet 
$Z$ is defined by
\be
\label{inv}
\inv{Z}=e\left[F+{1 \over 2}\bar\psi_\mu\gamma^\mu \zeta
+{1 \over 2}\varphi\bar\psi_\mu\sigma^{\mu\nu}\psi_\nu 
+S\varphi\right]\ .
\ee
Then if necessary one can rewrite the Lagrangian (\ref{lag}) in lengthy
component form.
The covariant derivatives for the multiplet 
$Z=(\varphi, \zeta, H)$ with $\lambda=0$ are 
defined as
\bea
\label{covdrv}
&& D_\mu \varphi=\partial_\mu \varphi 
  -{1 \over 2}\bar\psi_\mu\zeta \ , \quad
D_\mu \zeta = \left( \partial_\mu + 
{1 \over 2}\omega_\mu \gamma_5\right)\zeta
  -{1 \over 2}D_\nu \varphi \gamma^\nu \psi_\mu 
  - {1 \over 2}H\psi_\mu \ , \\
&& \Box \varphi = e^{-1}\left\{
\partial_\nu(e g^{\mu\nu}D_\mu \varphi) 
+ {i \over 4}\bar\zeta\gamma_5\psi_{\mu\nu}
\epsilon^{\mu\nu}
  -{1 \over 2}\bar\psi^\mu D_\mu\zeta
  -{1 \over 2}\bar\psi^\mu\gamma^\nu\psi_\nu 
D_\mu \varphi\right\}\ . \nonumber
\eea
\def\half{{1\over 2}}
\def\zebar{\bar\zeta}
\def\psbar{\bar\psi}
The action (\ref{lag})  is, by construction, 
invariant under the following local supersymmetry transformation,
\bea
\label{str}
&& \delta e_\mu^a= \bar\epsilon\gamma^a\psi_\mu ,\ 
\delta \psi_\mu=2\left(\partial_\mu+\half\omega_\mu\gamma_5
+\half\gamma_\mu S\right) \epsilon ,\ 
\delta S=-\half S\bar\epsilon\gamma^\mu\psi_\mu
+\half i e^{-1}\epsilon^{\mu\nu}\bar\epsilon\gamma_5\psi_{\mu\nu} \nn
&& \delta\phi=\bar\epsilon\chi \ , \quad 
\delta\chi=\Bigl\{F+\gamma^\mu(\partial_\mu\phi
  -\half\psbar_\mu\chi)\Bigr\}\epsilon \nn
&& \delta F=\bar\epsilon\gamma^\mu\Bigl\{\Bigl(\partial_\mu
+\half\omega_\mu\gamma_5\Bigr)\chi 
 -\half\gamma^\nu\Bigl(\partial_\nu A
  -\half\psbar_\nu\chi\Bigr)\psi_\mu-\half F\psi_\mu\Bigr\} \nn
&& \delta a_i=\bar\epsilon\xi_i \ ,\quad 
\delta\xi_i=\Bigl\{{G}_i+\gamma^\mu(\partial_\mu a_i
  -\half\psbar_\mu\xi_i)\Bigr\}\epsilon \nn
&& \delta {G}_i=\bar\epsilon\gamma^\mu\Bigl\{\left(\partial_\mu
+\half\omega_\mu\gamma_5\right)\xi_i 
 -\half\gamma^\nu\left(\partial_\nu \phi
  -\half\psbar_\nu\xi_i\right)\psi_\mu-\half {G}_i\psi_\mu\Bigr\}\ .
\eea
Here $\epsilon$ is an anti-commuting spinor parameter of local 
supersymmetry transformation and $\omega_\mu$ is the spin connection 
and given by 
$\omega_\mu=-ie^{-1}e_{a\mu}\epsilon^{\lambda\nu}
\partial_\lambda e^a_\nu
  -\half\psbar_\mu\gamma_5\gamma^\lambda\psi_\lambda$.
Hence, we constructed the classical action for 2d 
dilatonic supergravity with dilaton and matter supermultiplets.
The extended versions of 2d dilatonic supergravity have been 
constructed in refs.\cite{GKNO1,wit}.

We now study the conformal anomaly and effective action in 
large-$N$ approximation for the 2d dilatonic supergravity.
We consider only bosonic background below as it will be 
sufficient for the study of black hole or cosmological 
applications. Then, the dilatino $\chi$ and 
the Rarita-Schwinger fields vanish, and one can show that 
 gravity and dilaton part of the Lagrangian has 
the following form:
\bea
\label{gdlag}
&&\inv{C(\Phi)\otimes W} 
= e\left[-C(\phi)\left(S^2 +{1 \over 2}R\right)
  -C'(\phi)FS\right] \ ,\nn
&& \inv{\Phi\otimes\Phi\otimes T_P(Z(\Phi))}
=e\left[\phi^2\tilde\Box(Z(\phi)) + 2Z'(\phi)\phi F^2\right] \ ,\nn
&& \inv{Z(\Phi)\otimes\Phi \otimes T_P(\Phi)}
=e\left[Z(\phi)\phi\tilde\Box\phi
+ Z'(\phi)\phi F^2 + Z(\phi)F^2 \right] \ ,\nn
&& \inv{V(\Phi)} = e\left[ V'(\phi)F  + SV(\phi)\right] \ .
\eea
Using the equations of motion with respect to 
the auxiliary fields $S$, $F$, $G_i$, 
on the bosonic background one can show that
\be
\label{bgaf}
S={C'(\phi)V'(\phi) - 2V(\phi)Z(\phi)
\over {C'}^2(\phi) + 4C(\phi) Z (\phi)} , \ 
F={C'(\phi) V(\phi)+ 2 C(\phi)V'(\phi)
\over {C'}^2(\phi) + 4C(\phi) Z (\phi)} , \ 
G_i=0.
\ee
For the matter on bosonic background
\bea
\label{lag2}
&& \sum_{i=1}^N \left\{{1 \over 2}
\inv{\Sigma_i\otimes\Sigma_i\otimes T_P(f(\Phi))}
  -\inv{f(\Phi)\otimes \Sigma_i\otimes T_P(\Sigma_i)}\right\} \\
&&=ef(\phi)\sum_{i=1}^N
(g^{\mu\nu}\partial_\mu a_i \partial_\nu a_i
+\bar\xi_i\gamma^\mu\partial_\mu\xi_i -f(\phi)G_i^2) + \left(
\begin{array}{c}\mbox{\small total divergence} \\
\mbox{\small terms} \end{array}\right) \nonumber
\eea
Here we have used the fact that $\bar\xi_i \gamma_5 \xi=0$
for the Majorana spinor $\xi_i$. 

Let us write down the effective action 
in above theory. It is clearly seen that theory 
(\ref{lag2}) is conformally invariant on the 
gravitational background under discussion. 
Then using standard methods, we can prove that theory 
with matter multiplet $\Sigma_i$ is superconformally 
invariant theory. 
The anomaly induced effective action due to $N$ scalars and 
spinors may be derived from Eqs.(\ref{indaction2}) and (\ref{x})
as the following:
\be
\label{xv}
W=-{1 \over 2}\int d^2x \sqrt{- g} \left\{ 
{N \over 32\pi}R{1 \over \Delta}R 
 - {N \over 16\pi}{{f'}^2 \over f^2} 
(\nabla^\lambda \phi)(\nabla_\lambda \phi) {1 \over \Delta}R 
  - {N \over 6\pi}R\ln f\right\}\ .
\ee

Thus, using results of first sections of this chapter we were able 
to get the effective action for dilatonic supergravity with matter.
It is interesting to note that similarly one can get large N, 
anomaly induced effective action for dilatonic supergravity 
(with matter) in other dimensions. 
In the next chapter we discuss applications of induced effective action 
in black holes thermodynamics with account of quantum corrections 
and in chapter 5- in quantum cosmology.

\section{Thermodynamics of quantum dilatonic black holes}

Our purpose in this chapter will be to apply induced effective action in 
the study of quantum aspects of black holes thermodynamics. Let us first  
make few general remarks.

When we try to construct the quantum gravity theory, black hole 
evaporation provides a serious problem. A quantum mechanically 
pure state which describes gravitationally collapsing matter to 
form a black hole evolves into a mixed quantum state, which 
describes Hawking radiation \cite{Hwk1}. There are several 
scenarios to solve this problem of the loss of quantum information. 
Hawking gave a most radical proposal that the black hole completely 
evaporates and the quantum coherence is lost in the gravitational 
collapse. On the other hand, 't Hooft has proposed that Hawking 
radiation carries off information about the quantum states of 
the black holes \cite{Hft1}. Another conservative proposal is 
that the process of the collapse and the radiation leaves a stable 
remnant which carries the information of the initial configuration 
of the system \cite{ACN1}. However, if the remnant has the mass of 
order the Planck scale, gravitational effects would produce the 
light remnants in pairs and the lifetime of stars would be shorter 
than observed. Therefore the mass of the remnants should be 
macroscopic. A candidate of such remnants is the extremal black 
hole of Reissner-Nordstr\o m type \cite{HW1}. The extremal 
black holes have vanishing temperature and the space-time metric 
is not singular everywhere.

\subsection{2d dilatonic black holes and Hawking radiation}

\def\pint{{1 \over 2\pi}\int d^2x\,}
\def\sqg{\sqrt{-g}}
\def\ephi{\e^{-2\phi}}
\def\erho{\e^{-2\rho}}
\def\epphi{\e^{2\phi}}
\def\eprho{\e^{2\rho}}
\def\ephirho{\e^{2(\phi-\rho)}}
Let us turn now to overview of the 2d dilatonic black holes properties.
Callan, Giddings, Harvey and Strominger \cite{CGHS1} 
have investigated an interesting toy model of two dimensional 
gravity, which contains Hawking radiation and 
which is now called CGHS model. 
The action of the model is given by adding the action of 
$N$ minimal scalar fields $a_i$ ($i=1,\cdots,N$) 
\be
\label{mnsclr}
S_a=-\pint\sqg\sum_{j=1}^N \partial_\mu a_j \partial^\mu a_j
\ee
to the action  (\ref{CGHS}) .
When we fix the gauge freedom of reparametrization 
invariance by the conformal gauge,
\be
\label{IIIiib}
g_{\mp\pm}=-\half\eprho\ , \quad g_{\pm\pm}=0\ ,
\ee
the general solutions of eqs. of motion are given by,
\bea
\label{IIIviii}
&& a_j=a^+_j(x^+)+a^-_j(x^-) \ , \nn
&& \ephi=\erho={M \over \lambda}-\lambda^2x^+x^-
+ \int^{x^+}dy^+ \int^{y^+}dz^+\sum_{j=1}^N
\partial_+a^+(z^+)\partial_+a^+(z^+) \nn
&& \quad + \int^{x^-}dy^- \int^{y^-}dz^+\sum_{j=1}^N
\partial_-a^-(z^-)\partial_-a^-(z^-) \ .
\eea
Here we have fixed the residual gauge symmetry of 
reparametrization by choosing the condition $\phi-\rho=0$, 
which is given by a sum of holomorphic function $w^+(x^+)$ 
and anti-holomorphic function $w^-(x^-)$ in general :
$\phi(x^+,x^-)-\rho(x^+,x^-)=w^+(x^+)+w^-(x^-)$.

When we consider solutions with $a_j=0$, 
a special solution describing a dilaton vacuum is given by,
\be
\label{IIIix}
\ephi=\erho=-\lambda^2x^+x^-\ .
\ee
One may get more general solutions,
\be
\label{IIIx}
\ephi=\erho={M \over \lambda}-\lambda^2x^+x^-\ .
\ee
Here ${M \over \lambda}$ is an integration constant. 
The solutions (\ref{IIIx}) tell that the metric has a singularity 
when ${M \over \lambda} - \lambda^2x^+x^-=0$ and there
appear horizons $x^+x^-=0$. Therefore, the solutions (\ref{IIIx}) 
describe black holes. The structure of space-time is similar
to that of the 4d Schwarzschild black holes. 

The solutions (\ref{IIIviii}) can also describe matter 
collapsing into black holes. As an example, we can consider a 
shock wave which is given by 
$\sum_{j=1}^N\partial_\pm a_j\partial_\pm a_j =a\delta(x^+-x^+_0)$. 
Then the solution is given by,
\bea
\label{IIIxii}
&x^+<x^+_0\ , \quad \ephi&=\erho={M \over \lambda}
  -\lambda^2x^+x^-\ , \nn
&x^+>x^+_0\ , \quad \ephi&=\erho={M \over \lambda}+ax^+_0
  - \lambda^2 x^+ (x^- +{a \over \lambda^2})\ .
\eea
The event horizon when $x^+>x^+_0$ is given by 
$x^+ \left(x^-+{a \over \lambda^2}\right)=0$.
Note that the past horizon, where
$x^+=-{a \over \lambda^2}$, is shifted by the shock wave. 

In the original paper by CGHS, they found that the Hawking 
radiation and the back reaction of the metric can be described 
by adding correction terms to the classical action.
These terms come from the minimal scalar conformal anomaly 
(first term in (\ref{indaction2})).
\be
\label{IIIxvi}
W=-{1 \over 2}\int d^2x \sqrt{- g} 
{N \over 48\pi}R{1 \over \Delta}R 
\ee
However, they break the solvability in the classical CGHS 
model. Subsequently, Russo, Susskind and Thorlacius \cite{RST2} 
added a dilaton dependent term in (\ref{RSTq}) by hands. 
That  
makes the semi-classical theory solvable. As shown in Eq. 
(\ref{indaction2}) \cite{NNO}, such term as in RST model 
can arise from the conformal 
anomaly for dilaton coupled spinors. CGHS and RST models 
initiated a lot of activity. Some other models which 
mainly represent their modifications and quantization of such models 
have been discussed in refs.\cite{dAl1,BCl1,Mnn1,LK1,PS1,KS1,BPP1,
CFNNS} 
(for a review and more complete list of refs., see \cite{Str1}).             

In \cite{NOd2}, another solvable models which 
also
represent some modification of CGHS theory are proposed. 
The starting classical action is given by
\be
\label{IIIxviii}
S_c=\pint\sqg\left\{\ephi(R+4(\nabla \phi)^2+4\lambda^2)
  -{\e^{a\phi} \over g_A^2}F^2
  -\sum_{j=1}^N i\bar\Psi_j\gamma^\mu(D_\mu-iA_\mu)\Psi_j\right\}\ .
\ee
Here $\phi$ is a dilaton field and $\Psi_j={\psi_j \choose 0}$'s are
$N$ left-handed complex fermions. $g_A$ is a $U(1)$ electromagnetic 
gauge coupling constant and $D_\mu$ is a covariant derivative.
If we fix $a=-2$, this action, except the fermion part, describes
an effective field theory derived from string theory \cite{MNY1}. 
In ref.\cite{NOd2}, it was chosen  $a=2$, which allows 
us to exactly solve this model classically. In fact, when $a=-2$, 
the general solutions of effective equations are given by, 
\bea
\label{IIIxix}
&& \psi_j=\psi_j(x^+)\ ,\quad 
A_+=-{g_A^2 \over 4}x^-\int^{x^+}dy^+\sum_{j=1}^N
\psi_j^*(y^+)\psi_j(y^+)\ ,\nn
&& \ephi=\erho={M \over \lambda}-\lambda^2x^+x^-
+{g_A^2 \over 8}x^- \int^{x^+}dy^+(\int^{y^+}dz^
+\sum_{j=1}^N\psi_j^*(z^+)\psi_j(z^+))^2 \nn
&& \quad -{i \over 4}\int^{x^+}dy^+
\int^{y^+}dz^+\sum_{j=1}^N(\psi_j^*(z^+)\partial_+\psi_j(z^+)
  -\partial_+\psi_j^*(z^+)\psi_j(z^+))\ .
\eea
A special solution with $\psi_j=0$ describes the dilaton vacuum 
in (\ref{IIIix}). And as more general solutions, we find 
\be
\label{IIIxxb}
F_{+-}=C \quad (\mbox{constant})\ ,\quad 
\ephi=\erho={M \over \lambda}-\tilde\lambda^2x^+x^-\ .
\ee
Here ${M \over \lambda}$ is an integration constant
and $\tilde\lambda$ is defined by 
$\tilde\lambda^2\equiv \lambda^2-{2C^2 \over g_A^2}$. 
The metric has a singularity when
${M \over \lambda}-\tilde\lambda^2x^+x^-=0$ and there
appear horizons $x^+x^-=0$. Therefore, if $C\neq 0$, the solutions 
describe charged black holes. The structure of space-time is similar
to that of the Schwarzschild black holes and simple in contrast to
the Reissner-Nordstr\o m black hole solutions in four dimensions.
Note that the singularity vanishes when $\tilde\lambda^2=0$ 
when we fix $M$ to be finite. 
This solution could be a natural analogue of the extremal
Reissner-Nordstr\o m black hole solution. 
In fact, as the temperature $T$ is given by,
\be
\label{IIIxxii}
T={\tilde\lambda \over 2\pi}\ ,
\ee
the extremal solutions, which correspond to $\tilde\lambda^2=0$, 
have a vanishing temperature. 

The solutions (\ref{IIIxix}) can also describe charged matter 
(chiral fermions) collapsing into black holes. As an example, 
we can consider a charged shock wave which is given by,
${i \over 4}\sum_{j=1}^N(\psi_j^*(x^+)\partial_+\psi_j(x^+)
  -\partial_+\psi_j^*(x^+)\psi_j(x^+))=a\delta(x^+-x^+_0)$ and 
${g_A^2 \over 4}\sum_{j=1}^N\psi_j^*(x^+)\psi_j(x^+)=
b\delta(x^+-x^+_0)$. 
Then in the solution, not only future horizon corresponding to
$x^-=-{2Bx^+_0 \over g_A^2\tilde\lambda'^2}$, but past horizon, 
where $x^+=-{a \over \tilde\lambda'^2}$, is shifted by the
shock wave. Here $\tilde\lambda'$ and $B$ are defined by 
$\tilde\lambda'^2\equiv\lambda^2-{2(C+b)^2 \over g_A^2}$ and 
$B\equiv (2Cb+b^2)$. 
In the original paper by 't Hooft \cite{Hft1}, 
the shift of only future horizon was discussed in four dimensional 
black holes. The shift of the past horizon 
would suggest that charged particles collapsing into 
Reissner-Nordstr\o m black hole in four dimensions could shift 
the past horizon.

Now, since we have massless fermions, the terms which come from
the chiral anomaly should be also added to the quantum 
correction besides the terms coming from the conformal anomaly. 
Furthermore, by adding the contribution which corresponds to the term 
in RST model \cite{RST2}, we find the following total quantum 
action $S_q$:
\be
\label{IIIxxx}
S_q=S_c-{N \over 96}\pint\sqg \left\{R{1 \over \Delta}R
+\phi R\right\} + {N \over 24}\pint\left(\epsilon^{\mu\nu}F_{\mu\nu}
\right){1 \over \Delta}\left(\epsilon^{\mu\nu}F_{\mu\nu}\right)\ .
\ee
The above model can be also solved in the same way as RST model. 
Of course, there are many more BH models which are not solvable. 
Their study is extremely difficult even technically.

Similarly, one can discuss other 2d and 4d BHs with Hawking radiation 
(for reviews on Hawking radiation see \cite{Str1,KV1}). As the presentation
is more-less standard we finish this section by last example of 
supersymmetric BHs which are not exactly solvable.

In refs. \cite{NO3,NNO}, the Hawking radiation of the 
supersymmetric extension of CGHS model \cite{NOd3} has 
been investigated. 
We discuss the case that dilaton function is given 
by $f(\phi)=\e^{\alpha\phi}$ in (\ref{lag2}) in the previous 
chapter. The Hawking radiation can be obtained by 
substituting the classical black hole solution
which appeared in the original CGHS model \cite{CGHS1} 
\be
\label{sws}
\rho=-{1 \over 2}\ln \left(1 + {M \over \lambda}
\e^{\lambda (\sigma^--\sigma^+ )} \right) \ , \ \ 
\phi=- {1 \over 2}\ln \left( {M \over \lambda}
+ \e^{\lambda(\sigma^+ - \sigma^-)} \right)  
\ee
(where $M$ is the mass of the black hole and we used asymptotic 
flat coordinates) into the quantum part of the energy momentum 
tensor. Then we find that when 
$\sigma^+\rightarrow +\infty$, the energy momentum tensor behaves as
\be
\label{asT}
T^q_{+-}\rightarrow 0 \ ,\quad 
T^q_{\pm\pm}\rightarrow {N\lambda^2 \over 16}
\alpha^2 + t^\pm(\sigma^\pm)\ .
\ee
Here $t^\pm (\sigma^\pm)$ is a function which is determined by the 
boundary condition. In order to evaluate $t^\pm(\sigma^\pm)$, 
we impose a boundary condition that there is no incoming energy.
This condition requires that $T^q_{++}$ should vanish at the past 
null infinity ($\sigma^-\rightarrow -\infty$) and if we assume 
$t^-(\sigma^-)$ is black hole mass independent, $T^q_{--}$ should 
also vanish at the past horizon ($\sigma^+\rightarrow -\infty$) 
after taking $M \rightarrow 0$ limit. Then we find
\be
\label{rad}
T^q_{--}\rightarrow 0
\ee
at the future null infinity ($\sigma^+\rightarrow +\infty$). 
Eqs.(\ref{asT}) and (\ref{rad}) might tell that there is no 
Hawking radiation in the dilatonic supergravity model 
under discussion when quantum back-reaction of 
matter supermultiplet in large-$N$ approach 
is taken into account as in \cite{NO3}. (That indicates that 
above black hole is extremal one). From another side since we 
work in strong coupling regime it could be that new methods to 
study the Hawking radiation should be developed. This follows 
from the fact that despite various efforts there are still many unclear points
related with Hawking radiation \cite{Hwk2,BDDO1,ILS1,CPW1,RST1,NOd3,Alves}. 
Especially puzzling is its relation with information loss paradox. 

\subsection{Thermodynamics of Schwarzschild-de Sitter black holes}

The strong analogy between thermodynamics and BHs is known 
for quite a long time \cite{BCH1,Hwk3}. One of its bright manifestations 
is related with Bekenstein-Hawking entropy \cite{Bkn1,Hwk1} 
which is known to be proportional to the surface area of BH horizon. 
Despite numerous attempts varying from strings \cite{SV1,HS1,
HMS1,HLM1} (for a review and list of refs. see \cite{Pt1}), 
three-dimensional gravity \cite{Crl1,BVZ} including 
higher derivative terms\cite{Soda}, induced gravity \cite{Jcb1}, 
etc. the derivation of Bekenstein-Hawking entropy from statistical 
mechanics is not yet completely understood. Most probably, 
it is expected to get the complete understanding from string or M-theory.
Our purpose in this section will be more modest than
explanation of the origin of BH entropy. We are going to calculate 
quantum corrections to some classes of BHs semiclassically, 
using effective action obtained in previous chapter.

In \cite{NO4}, the quantum corrections to thermodynamics 
(and geometry) of S(A)dS BHs have been discussed by using 
large $N$ anomaly induced effective action for 
dilaton coupled matter (scalars and spinors). It is found 
the temperature, mass and entropy with account of quantum 
effects for multiply horizon SdS BH and SAdS BH. That also 
gives the corresponding expressions for their limits: 
Schwarzschild and de Sitter spaces.  Note that such BHs are not dilatonic 
ones. Nevertheless,  we develop the picture describing 
the same objects as 2d dilatonic BHs. Hence, they again have the relation 
to dilatonic gravity.

We will start from the action of Einstein gravity with
$N$ minimal real scalars $\chi_i$ and $M$ Majorana 
fermions $\psi_i$ :
\be
\label{e1}
S_{4d}=-\frac{1}{16\pi G}\int
d^4x\,\sqrt{-g_{(4)}} \left(R^{(4)}-2\Lambda\right) 
 + \int d^4x\,\sqrt{-g_{(4)}}\,
\left(\frac{1}{2}\sum_{i=1}^Ng^{\alpha\beta}_{(4)}
\partial_\alpha \chi_i \partial_\beta \chi_i + \sum_{i=1}^M 
\bar\psi_i \gamma^\mu\nabla_\mu \psi_i \right)
\ee
where $N$ and $M$ are considered to be large, $N,M \gg 1$, $G$
and $\Lambda$ are gravitational and cosmological constants, 
respectively.
The convenient choice for the spherically symmetric spacetime 
is the following one
\be
\label{e2}
ds^2=g_{\mu\nu}dx^\mu dx^\nu+e^{-2\phi}d\Omega,
\ee
where  $\mu,\nu=0,1$, $g_{\mu\nu}$ and $\phi$ depend only on  $x^0$,
$x^1$ and $d\Omega$ corresponds to the two-dimensional sphere.
Then the action (\ref{e1}), reduced according to (\ref{e2}) takes 
the form
\be
\label{e3}
S_{\rm red}=\int d^2x \sqrt{-g}\e^{-2\phi}
\left[-{1 \over 16\pi G}
\{R + 2(\nabla  \phi)^2 -2\Lambda + 2\e^{2\phi}\} 
 + {1 \over 2}\sum_{i=1}^N(\nabla \chi_i)^2
+ \sum_{i=1}^{2M} \bar\psi_i \gamma^\mu \nabla_\mu \psi_i \right]
\ee
Working in large $N$ and $s$-wave approximation and 
using 2d conformal anomaly for dilaton coupled 
scalar, calculated in \cite{ENO2} (see also
\cite{NO6,CS1}) one can find the anomaly induced
effective action \cite{BH1,NO5, NO3} from Eqs.(\ref{indaction2}) 
and (\ref{x}) by choosing $f(\phi)=\e^{-2\phi}$. 
The conformally invariant functional can be found (see
(\ref{Gammaii})) as some expansion of Schwinger--DeWitt 
type \cite{NO7} keeping only the leading term.
Then, the effective action may be written in the following form
\cite{NO5,NO3,NO7}
\be
\label{e4}
W=-{1 \over 8\pi}\int d^2x \sqrt{-g}\,\left[
{N+M \over 12}R{1 \over \Delta}R
  - N \nabla^\lambda \phi
\nabla_\lambda \phi {1 \over \Delta}R 
 +\left(N + {2M \over 3} \right)\phi R
+2N\ln\mu_0^2 \nabla^\lambda \phi \nabla_\lambda \phi
\right]\ ,
\ee
where $\mu_0^2$ is a dimensional parameter. 

 In the conformal gauge (\ref{IIIiib}),it is convenient to 
change the radial 
coordinate $r$ by the new coordinate $x$
\be
\label{ss1}
x=\e^{-\phi}\ ,
\ee
which corresponds to the usual coordinate choice in the 
Schwarzschild metric:
\be
\label{ss2}
ds^2=-\e^{2\rho}dt^2 + \e^{2\sigma}dx^2 + x^2d\Omega^2 \ , \quad 
\e^{\sigma}=-\e^{\rho + \phi}\left({d\phi \over dr}\right)^{-1}\ . 
\ee

Note that there were also attempts in 
refs.\cite{HB2,Bss1,NO5,NO3,NO7,BRM1} to apply such EA (usually 
without logarithmic term and only for scalars) for quantum 
considerations around BHs (for calculation of 
semiclassical stress tensor with 
dilaton,see refs.\cite{LMR1,BF1}). 

First we consider the case $\Lambda=0$. In the classical limit 
($N\rightarrow 0$), we obtain, of course, the Schwarzschild 
black hole as solution of classical equations of motion:
\be
\label{sb1}
\e^{2\rho}=\e^{2\rho_0}\ ,\quad 
\e^{2\sigma}=\e^{2\sigma_0}\ ,\quad 
\e^{2\rho_0}=\e^{-2\sigma_0}= 1 - {\mu \over x}\ .
\ee
Here $\mu=2GM_{\rm BH}$ and $M_{\rm BH}$ is the black hole mass. 
We now consider the quantum corrections by regarding $GN$ is 
small and assuming
\be
\label{qc1}
\rho=\rho_0 + GN\Delta_\rho\ ,\quad \sigma=\sigma_0
+ GN \Delta_\sigma\ (\sigma_0=-\rho_0 )\ .
\ee
Then from effective equations of motion we find that $\Delta_\sigma$ 
 behaves near the classical horizon $x\sim\mu$, as
\be
\label{sb7}
\Delta_\sigma\sim {1 \over 3(x-\mu)}
\left[\left({A \over 8\mu} + 6t_0\mu\right)\ln (x-\mu) 
+ \mbox{regular terms}\right]\ .
\ee
Here, $t_0$ is a constant which is determined by the initial 
conditions and 
$A\equiv {N+M \over N}$, $B\equiv {N + {2M \over 3} \over N}$. 
The singularity coming from $\ln (x-\mu)$ vanishes if we 
choose (then $t_0$ here corresponds to the constants 
$C$ and $D$ of the integration in \cite{BRM1}) 
\be
\label{sb8}
t_0=-{A \over 48 \mu^2}\ .
\ee
In the choice of (\ref{sb8}), one gets
\bea
\label{qs1}
&& \Delta_\rho + \Delta_\sigma 
=A\left\{{1 \over 8x^2} + {1 \over 6\mu x}
  - {1 \over 12\mu^2}\ln {x \over l} \right\}  \nn
&& \quad + {a+B-1 \over 2x^2} - {1 \over 4x^2} - {1 \over 2\mu x}
+ {1 \over 2}\left({1 \over x^2} - {1 \over \mu^2}\right)
\ln\left( 1 - {\mu \over x}\right) \ ,\nn
\label{qs2}
&& \Delta_\sigma = {1 \over 3(x-\mu)} 
\times \left\{ \Delta_S 
+ A\left( -{1 \over 8\mu}\ln {x \over l}
+ {1 \over 8x} - {7\mu \over 16x^2}\right) 
+ {a-B+1 \over 2}\left( {3 \over x} - {3\mu \over 2x^2} \right)
\right. \nn
&& \quad \left. \ -{3B\mu \over 4x^2} 
  - {3\mu \over 4}\left({1 \over x} - {1 \over \mu}\right)^2
\ln \left( 1 - {\mu \over x}\right)  
  - {3 \over 4x} + {3\mu \over 8x^2} + {3x \over 2\mu^2}\right\}\ .
\eea
Here 
parameters $l$ and $\Delta_S$ coming from the 
constants of the integration. We now assume the radius $L$ of 
the universe is large $L\gg\mu$ but finite and we require
$\Delta_\rho=\Delta_\sigma=0$ when $x=L$. 
Then we find
\be
\label{sb8c}
l=L\ ,\quad \Delta_S=-{3L \over 2\mu^2}\ .
\ee
Near the classical horizon, $\Delta_\rho+\Delta_\sigma$ 
and the scalar curvature are regular.
The horizon defined by $\e^{2\rho}=0$, which corresponds 
to $x=\mu$ in the classical limit is given by 
\be
\label{sb11}
0\sim {1 \over \mu}\left(x - \mu + 2GNC_S \right) \ .
\ee
Then the entropy, which is defined by the area of horizon divided 
by $4G$, is 
\be
\label{sb12}
S\sim{\pi\mu^2 \over G} 
  - \pi N \left(-{4\mu\Delta_S \over 3} + {A \over 6}\ln{\mu \over l}
+{5A \over 12} - a + 2B - {5 \over 2} \right)\ .
\ee 
The second term in (\ref{sb12}) corresponds to the quantum 
correction. From the behavior of the metric 
near the horizon $x= \mu - 2GNC_S$, we find the temperature 
$T$ (in the following, we put 
the Boltzmann constant $k$ to be unity, $k=1$) is given by
\be
\label{qs8}
T\sim{1 \over 4\pi\mu}\left[ 1 + GN\left\{-{2 \over 3\mu}\Delta_S 
+ {A \over 12\mu^2}\left(-1 + 2\ln{\mu \over l}\right) 
  - {a \over \mu^2} + {B \over 2\mu^2}\right\}\right]\ .
\ee
The second term corresponds to the quantum correction.
We now consider the thermodynamical mass $E$, which is defined by
\be
\label{sb14}
dE=TdS\ .
\ee
Using the parameter $\mu$, (\ref{sb14}) can be rewritten as
\be
\label{sb15}
E=\int d\mu T{dS \over d\mu} \nn
={\mu \over 2G} + N\left(-{L \over 2\mu^2} 
  - {A \over 12\mu}\ln{\mu \over L} + {a \over 2\mu} 
+ {B \over 4\mu}\right)\ .
\ee
The first term expresses the usual classical mass 
$M_{\rm BH}$ since $\mu=2GM_{\rm BH}$ and the second term is 
due to the quantum effects. Notice the regularization scheme 
dependence via the presence of parameter a in above expressions. 
It may be fixed by the choice of some physical regularization. 
The qualitative structure of the entropy (\ref{sb12}), the 
temperature (\ref{qs8}) and the energy (mass) (\ref{sb15}) are 
similar to the corresponding quantities found in \cite{BRM1} for
pure scalar.

Let us turn now to 2d formulation of above results.
The classical black hole solution (\ref{sb1}) can be 
regarded as a purely two dimensional object if we start 
with the reduced action (\ref{e3}). The 
thermodynamical quantities as the energy and entropy for 
 two dimensional charged black hole with dilaton 
are evaluated on the classical 
and one-loop levels in \cite{NO7,MK1}. In \cite{MK1}, the boundary 
of the universe is introduced at the radius $r=L$. 
If there is a boundary, we need to add the boundary 
terms to the action in order that the variation with respect to 
the metric should be well-defined. By including the boundary 
term, the formula for the energy (mass) of the black hole with
one-loop quantum correction was derived in \cite{MK1} as follows,
\be
\label{mk1}
E(\mu,L)=-{\e^{\lambda(L)} \over G} g^{1 \over 2}(L) D'(\phi(L)) 
 - {\hbar \over 3}\e^{\lambda_{CL}(L)} g_{CL}^{-{1 \over 2}}(L)
g_{CL}'(L)
  - {c\hbar \over 6}\e^{\lambda_{CL}(L)} g_{CL}^{1 \over 2}(L)
\phi'(L)\ .
\ee
Here $'={d \over dx}$. The quantities appeared in (\ref{mk1}) 
have the following correspondence with the quantities here,
\be
\label{mk2}
g(x)=\e^{2\rho(x)}\ ,\quad \e^{-\lambda(x)}\ , \quad
D(\phi(x))={\e^{-2\phi(x)} \over 2}={x^2 \over 2} \ .
\ee
Since only the quantum correction from the one scalar field was 
evaluated in \cite{MK1}, we need to do the replacement  
$\hbar\rightarrow NA$ and $c\rightarrow {12B \over A}$. 
We should note that we usually choose $\hbar=1$. 
In the expression (\ref{mk1}), the quantities with suffices $CL$ 
correspond to the classical ones. 
If we choose a proper boundary condition, we find 
\be
\label{mk4}
g(L)=g_{CL}(L)=1-{\mu \over L}\ ,\ \ 
\Lambda(L)=\Lambda_{CL}(L)=0\ .
\ee
Therefore 
\be
\label{mk5}
E(\mu, L)=-{L \over G}\left(1-{\mu \over L}\right)^{1 \over 2}
  -{NA\mu \over 3L^2}\left(1-{\mu \over L}\right)^{-{1 \over 2}}
+{2NB \over L}\left(1-{\mu \over L}\right)^{1 \over 2}\ .
\ee
Since the energy (\ref{mk5}) diverges in the limit  
$L\rightarrow +\infty$, we need to subtract $E(0,L)$ before 
taking the limit of $L\rightarrow +\infty$. Then we obtain
\be
\label{mk6}
E_{sub}\equiv\lim_{L\rightarrow +\infty}
\left(E(\mu,L)-E(0,L)\right)={\mu \over 2G}\ ,
\ee
which is nothing but the classical  black hole 
mass. This tells that there is no quantum correction for 
the mass. 
Even if the black hole is purely two dimensional one
(but with dilaton), the definition of the temperature is not 
changed and one gets (\ref{qs8}).
Then using the energy (\ref{mk6}), the temperature (\ref{qs8}) 
and the definition of the entropy (\ref{sb14}), we obtain the 
following expression of the entropy
\bea
\label{mk7}
S&=&\int {dE \over T}=\int {d\mu \over T(\mu)} {dE \over d\mu} \nn
&=&{\pi \mu^2 \over G}- 2\pi N\left\{-{L \over \mu}+ 
\left\{-{A \over 12} - a + {B \over 2}\right)\ln{\mu \over l}
+ {A \over 12}\left(\ln {\mu \over L}\right)^2 + c \right\} \nn
&& + {\cal O}(L^{-1}) + {\cal O}\left(GN)^2\right)\ .
\eea
Here $c$ is the constant of the integration. The classical part 
coincides with the usual Bekenstein-Hawking entropy when we regard 
the black hole as the four dimensional object.

Hence we calculated quantum corrections to simplest black hole 
thermodynamical quantities. The result is actually obtained for
two objects: 4d  BH and the same object described as 2d 
dilatonic BH. That shows remarkable property of $s$-wave EA that 
it could be applied to 4d as well as to 2d geometry (where it 
looks already as complete EA). It is not difficult to extend 
the discussion for other types of BHs. 

We now consider more general Schwarzschild-(anti-)de Sitter black 
holes. Their convenient description is given in next section.
In the classical limit ($N\rightarrow 0$) 
with non-vanishing cosmological constant $\Lambda$, 
we obtain the usual Schwarzschild-(anti)de Sitter as 
solution of equations of motion
\be
\label{ss6}
\e^{2\rho}=\e^{-2\sigma}=\e^{2\rho_0} 
\equiv 1 - {\mu \over x} - {\Lambda \over 3}x^2 
=-{\Lambda \over 3x}\prod_{i=1}^3 (x - x_i)\ .
\ee
Here $\mu$ is a constant of the integration corresponding to 
the black hole mass ($\mu =2GM_{\rm BH}$). 
The parameters $x_i$ ($i=1,2,3$) 
are solutions of the equation $\e^{2\rho_0}=0$. Among $x_i$'s, 
two are real and positive if $\Lambda>0$ and 
$\mu^2 < {4 \over 9\Lambda}$ and they correspond to black hole 
and cosmological horizons in the Schwarzschild-de Sitter black 
hole. On the other hand, only one is real if $\Lambda<0$. 
If we start from the reduced action (\ref{e3}), the classical 
solution (\ref{ss6}) can be regarded to express the 
purely two dimensional Schwarzschild-(anti) de Sitter black 
holes with dilaton. Hence, we again obtain 2d or 4d formulation 
for such object. Note that reduction of SAdS BH may be understood 
as 2d dilatonic AdS BH where quantum effects of dilaton 
were recently discussed in refs.\cite{KO, strominger}.

As in (\ref{qc1}), we consider the quantum corrections  
regarding $GN$ as small. 
In a similar way as in the case of the Schwarzschild black hole, we 
obtain the following expressions for the entropy $S$ and the 
temperature $T$:
\bea
\label{qc10}
S&\sim&\pi x_I^2 - 4\pi GNC_I x_I \nn
T&\sim& \left| {\Lambda \over 12\pi}\left\{Y_I
+\left(- 6 + {2Y_I \over x_I}\right)GN C_I
  -GNY_IB_I \right\}\right| \ .
\eea
The explicit forms of $B_I$ and $C_I$ are given in \cite{NO4} 
and
\be
\label{qc3}
Y_i\equiv {1 \over x_i}\prod_{i,j=1,i\neq j}^3 (x_j - x_i) \ ,
\quad a'\equiv \ln\left(-{\Lambda \over 3}\right) + a \ .
\ee

We now consider de Sitter space as a limit of $\mu\rightarrow 0$.
Let us choose $x_I$ as a cosmological horizon : 
$x_I=h\equiv \sqrt{3 \over \Lambda}$.
Then we find the expressions for the entropy $S$ and the temperature 
$T$ :
\bea
\label{sd8}
S&=&{3\pi \over \Lambda}-2\pi GN 
\left\{\tilde\Delta_1\sqrt{\Lambda \over 3} + {A \over 6} 
+ a + 1 + 2\ln 2 \right\} \nn
T&=&{1 \over 2\pi}\sqrt{\Lambda \over 3} + {GN\Lambda \over 12\pi}
\left\{{\Lambda\tilde\Delta_1 \over 3} 
  - 2\tilde\Delta_0\sqrt{3 \over \Lambda}
+ \left({A \over 2} + 2 + 2\ln 2 - B\right)\sqrt{\Lambda \over 3}
\right\}\ .\nonumber
\eea
Here $\tilde\Delta_0$ and $\tilde\Delta_1$ 
are the constants of the integration. 
It is interesting to note that the expression for $S$ describes 
quantum corrections to the entropy of expanding inflationary Universe 
(as de Sitter space may be considered as such inflationary Universe). 
That gives new terms proportional to particles number as compare with 
classical entropy of expanding Universe discussed 
extensively in refs.\cite{Gbn4,Gbn2,Gbn3}.

Hence we found quantum corrections to the temperature and the entropy 
for 4d  de Sitter space (as the limit of SdS BH) as well as 
for SdS BH and SAdS BH. In the last case of SdS or SAdS BHs 
we also defined the quantum correction to the temperature of
corresponding 2d object (i.e. corresponding BH with dilaton). 
The calculation of 2d quantum entropy is more difficult and 
cannot be done by using only 4d point of view unlike the case 
of Schwarzschild  BH. Similarly, one can calculate quantum 
corrections to other BHs, for example, for charged 4d BHs it 
was done in ref. \cite{BR1}.
Below we consider one more example of 2d charged BHs.

In \cite{MK1}, it was investigated the one-loop quantum 
correction to the thermodynamical quantities in general two- 
dimensional charged black holes, whose special cases correspond 
to the dimensionally reduced higher dimensional black holes like 
4d charged spherically symmetric one or 3d rotating BTZ one. 
The starting classical action is given by
\be
\label{MKi}
W\left[g_{\mu\nu},\phi,A_\mu \right]
={1 \over 2G}\int d^2x \sqrt{-g}\left[ D(\phi) R 
+{1 \over 2}\partial_\alpha\phi \partial_\beta\phi 
 +{1 \over l^2}V(\phi) 
  - {2G \over 4}Y(\phi)F^{\alpha\beta}F_{\alpha\beta}\right]
\ee
where $A_\mu$ is the abelian gauge field and $F_{\mu\nu}
=\partial_\mu A_\nu - \partial_\nu A_\mu$. $D(\phi)$, $V(\phi)$, 
and $Y(\phi)$ are arbitrary dilatonic functions. 
The free energy $F$ can be evaluated from the action $W$ after 
Wick-rotating to Euclidean signature $t\rightarrow i\tau$ by 
$F={1 \over 2\pi \bar\beta}W$ after the subtraction of the 
contribution from the adequate background geometry. 
Here $\bar\beta$ is the inverse of the temperature which is 
red-shifted at the boundary at $x=L$ $\bar\beta=g^{1 \over 2}\beta$. 
In order to well define the quantities, the space-time is assumed to 
be the disc, where the space coordinate $x$, corresponding the radial 
coordinate in the choice of the Schwarzschild like metric
\be
\label{SCW}
ds^2=g(x) d\tau^2 + \e^{-\lambda(x)}g^{-1}(x)dx^2\ ,
\ee
is defined between the outer horizon at $x=x_+$ to the finite 
boundary at $x=L$. 
Other thermodynamical quantities like the entropy $S$ and the energy 
$E$ can be obtained by using the standard definitions in the 
thermodynamics:
\be
\label{SM}
S={\bar\beta}^2{\partial F \over \partial\bar\beta}\ ,\quad 
E={\partial\left(\bar\beta F\right) \over \partial\bar\beta}\ .
\ee
In \cite{MK1}, the quantum corrections were included by adding the 
one-loop effective action $\Gamma$ to the classical action 
(\ref{MKi}).  $\Gamma$ is generated by the 2d conformal anomaly 
as given in Eq.(\ref{indaction2}) or (\ref{x}) : 
\be
\label{MKii}
\Gamma={1 \over 12}\int d^2x \sqrt{-g}\left[aR
{1 \over \triangle}R
+ b\left(\nabla\phi\right)^2 {1 \over \triangle}R + c\phi R\right]\ .
\ee
The parameter $a$ is chosen to be unity $a=1$ and $b$ and $c$ are 
arbitrary.
Then the entropy $S$ is given by
\bea
\label{MKiii}
&& S={2\pi \over G}D\left(\phi_{CL}(x_+)\right) \nn
&& \quad - {2\pi\hbar \over 6}
\left[ 2\psi(x_+) - c\phi_{CL}(x_+) 
 -b\int_{x_+}^Ldx {\e^{-\lambda_{CL}(x)} \over g_{CL}(x)}
\int_x^Ld\bar x \e^{\lambda_{CL}(\bar x)}\phi_{CL}'(\bar x)^2
g_{CL}(\bar x)\right] \ ,
\eea
and the energy $E$ is given by Eq.(\ref{mk1}). 
Here the classical quantities are notated by the suffices 
$CL$ and $\psi$ is defined when we choose the Euclidean metric 
in the following way:
\be
\label{MKiv}
ds^2 = \e^{-\psi}\left(z^2 d\theta^2 + dz^2\right)\ ,
\ee
that is
\be
\label{MKv}
g(x)dt^2=\e^{-\psi}z^2d\theta^2\ ,\quad 
g^{-1}(x)\e^{-2\lambda(x)}dx^2 = \e^{-\psi}dz^2\ .
\ee
We can regard the above 2d model as 4d spherically symmetric 
reduced model if we identify
\be
\label{MKvi}
\phi={\sqrt{2} r \over l}\ ,\quad D(\phi)={r^2 \over 2l^2}\ ,
\quad V(\phi)- {Gl^2 q^2 \over Y(\phi)}= 1 - 
{Q^2 \over r^2}\ .
\ee
Here $q$ corresponds to the charge in 2d model. Then we obtain 
the following expressions for the above thermodynamical 
quantities: 
\bea
\label{MKvii}
&& S\sim S_{CL}+2\pi \beta_{CL}m(r_{+CL}) 
+ {2\pi \over 3}\left(1 - {r_{-CL}^2 \over r_{+CL}^2}\right)
\ln \left({L \over r_{+CL} - r_{+CL}}\right) 
+ {4\pi \over 3}\ln \left({r_{+CL} \over z_0}\right) \nn 
&& E\sim M_{CL}+ {\hbar Gl^2 M_{CL} \over 3\beta^2_{CL}}
\left(2\ln (L) + 1 \right)\ .
\eea
Here we assume that $L$ is large and neglect ${\cal O}(L^{-2})$ 
terms. $m(r)$ expresses the quantum correction to the mass: 
\be
\label{MKviii}
m(r)=M(r)-M_{CL}\sim {\hbar \over 3\beta^2_{CL}}\left[
L + 2M_{CL} Gl^2 \ln (L)\right]
\ee
and $z_0$ is the value of $z$ corresponding to $x=L$.

Thus, we presented the derivation of quantum corrections to thermodynamics
of different 2d and 4d BHs using the effective action derived in previous
chapter. There is no problem to extend such semiclassical investigation 
for other BH models. 
It is also interesting to note that some integrable dilatonic gravity 
models lead to the solutions similar to Nariai space \cite{CFNNS} 
where again quantum corrections may be found.

\subsection{Anti-evaporation of multiply horizon black holes}

In \cite{Hwk1}, Hawking made one of the most striking discoveries 
of the history of black hole physics: the possibility for black 
holes to evaporate, as a result of particle creation. This effect 
 ---which is now called the Hawking radiation process--- produced 
a deep impact in our  understanding of quantum gravity. 
 
There exists, however, an exotic class of black holes 
(see Ref. \cite{Hwk3} for a review) which possesses a multiple horizon 
and for which the {\it opposite} effect might occur. It is interesting to  
investigate this possibility in detail. Consider a nearly 
degenerate Schwarzschild-de Sitter 
black hole (the so-called Nariai black hole \cite{Nr1}).
 The Schwarzschild-de Sitter 
black hole represents the neutral, static, spherically symmetric  
solution of Einstein's theory with a cosmological constant.  
The corresponding metric looks as follows 
\bea
ds^2= -V(\tau )dt^2 + V^{-1}(\tau )d\tau^2+\tau^2 d\Omega^2, 
\qquad
\label{iiV} 
V(\tau ) = 1-\frac{2\mu}{\tau} - \frac{\Lambda}{3} \tau^3, 
\eea 
where $\mu$ is the mass and $\tau$ the radius of the black hole, 
$d\Omega^2$ the metric corresponding to a unit two-sphere,  
and $\Lambda$ is the cosmological constant. It can be easily checked  
that the equation $V(\tau )=0$ 
has two positive roots, $\tau_c$ and $\tau_b$ 
(we set $\tau_c >\tau_b$). Here, $\tau_c$ and $\tau_b$ have the 
meaning of a cosmological and a black-hole horizon radius, 
respectively. In the degenerate case (which corresponds to  
a black hole of maximal mass) both radii coincide and the black hole 
is  in thermal equilibrium. It is called a Nariai black hole. 
If we redefine new time and radial coordinates $ \psi $ and $ \chi $ by
\begin{equation}
\tau = \frac{1}{\epsilon\sqrt{\Lambda}} \psi\ , \quad 
r = \frac{1}{\sqrt{\Lambda}}
\left[1 - \epsilon\cos\chi - \frac{1}{6} \epsilon^2 \right]\ ,
\label{eq-transformations}
\end{equation}
the Nariai limit is given by $\epsilon\rightarrow 0$. 
In the Nariai limit, the space-time has the topology 
of $S^1\times S^2$ and the metric is given by
\be
\label{Nsol}
ds^2={1 \over \Lambda}\left( \sin^2 \chi d\psi^2 - d\chi^2 - d\Omega^2
\right)\ .
\ee
Here coordinate $\chi$ has a period $\pi$. 
If we change the coordinates 
variables by
\be
\label{cv}
r=\ln\,\tan{\chi \over 2}\ ,\ \ \ t={\psi \over 4}\ ,
\ee
we obtain 
\be
\label{Nsol2}
ds^2= {1 \over \Lambda\cosh^2 r}\left(-dt^2 + dr^2\right) 
+ {1 \over \Lambda}d\Omega\ .
\ee
This form corresponds to the conformal gauge in two dimensions. 

There are in fact two opposite sources contributing to 
the thermal equilibrium, namely a radiation flux coming from the 
cosmological horizon and the Hawking evaporation originated at 
the black hole horizon. It is plausible that such a state might 
be unstable, since it could be affected by small perturbations 
of the geometry.  
 
It was demonstrated in by Bousso and Hawking 
\cite{HB2}, that a nearly maximal (or nearly degenerated) 
Nariai black hole may not only evaporate ---as shown in 
Ref. \cite{Hwk1}--- but also anti-evaporate \cite{HB2}. 
In other words, Nariai black holes actually develop two 
perturbative modes: an evaporating one and an anti-evaporating 
one. The realization of this quantum process, 
carried out in Ref.\cite{HB2}, is based on the $s$-wave approximation 
by the spherical reduction and the 2d conformal anomaly induced 
effective action due to N dilaton coupled scalars. 

In the model of \cite{HB2}, the possibility of evaporation or 
anti-evaporation (i.e., increase in size) of a black hole of 
this kind is connected with the initial conditions chosen for 
the perturbations. The use of the commonly employed Hartle-Hawking 
no-boundary conditions \cite{HH1} shows that such black holes 
will most likely evaporate. 

The Nariai black holes are not asymptotically flat and they 
will never appear in the process of star collapse. Nevertheless, they  
could actually be present in the early inflationary universe. 
There may be at least two mechanisms to produce such multiple 
horizon BHs. First of all, they may appear via black holes 
pair creation \cite{HB3} in the inflationary Universe. Second, 
it is possible the direct quantum generation of such objects \cite{BNS1} 
in effective dilatonic gravity models. 
In the model in Ref.\cite{HB2}, however, the primordial  
multiple horizon black holes should quickly evaporate, and it is very  
unlikely that they could be detected in the present universe. 
 
However, a different model (with another matter content) 
for anti-evaporation of black holes has  
been proposed in Refs. \cite{NO8,ENjO1}, in which these difficulties 
may be solved. In that model, the quantum effects of the 
conformally invariant matter have been taken into account. 
What is more interesting, this theory allows for the possibility 
of including not only scalar fields, but also fermionic and vector 
fields (typical of all grand unified theories, GUTs), whose 
classical action is  
\def\D{{D \hskip -3mm /\,}} 
\be
\label{OI} 
S=\int d^4x \sqrt{-g_{(4)}}\left\{{1 \over 2}\sum_{i=1}^N  
\left(g_{(4)}^{\alpha\beta}\partial_\alpha\chi_i 
\partial_\beta\chi_i + {1 \over 6}R^{(4)}\chi_i^2 \right) 
  -{1 \over 4}\sum_{j=1}^{N_1}F_{j\,\mu\nu}F_j^{\mu\nu} 
+\sum_{k=1}^{N_{1/2}}\bar\psi_k\D\psi_k \right\}, 
\ee 
and its quantum correction $\Gamma$ is the sum of the conformal  
anomaly induced action $W$, as given by Eq.(\ref{v10}) ($f=1$) 
and one corresponding to (\ref{Vx}), 
and the action $\Gamma'$ deduced from the  
Schwinger-de Witt type expansion : 
\bea 
\label{OVIII} 
&& \Gamma=W+\Gamma', \nn 
&& W=b\int d^4x \sqrt{-g} F\sigma  
+b'\int d^4x \sqrt{-g} \Bigl\{\sigma\left[ 
2\Box^2 + 4 R^{\mu\nu}\nabla_\mu\nabla_\nu 
 - {4 \over 3}R\Box + {2 \over 3}(\nabla^\mu R)\nabla_\mu  
\right]\sigma \nn 
&& \quad + \left(G-{2 \over 3}\Box R\right)\sigma \Bigr\} 
 -{b + b' \over 18} 
\int d^4x \sqrt{-g}\left[R - 6 \Box \sigma  
  - 6(\nabla \sigma)(\nabla \sigma) \right]^2,  \nn
&& \Gamma'=\int d^4x  
\sqrt{-g}\left\{\left[{b}F+  {b'}G 
+ { 2b\over 3}\Box R\right]\ln { R \over \mu^2}\right\} 
+{\cal O}(R^3). 
\eea 
Here $b={(N +6N_{1/2}+12N_1)\over 120(4\pi)^2}$,  
$b'=-{(N+11N_{1/2}+62N_1) \over 360(4\pi)^2}$,  
$\mu$ is a mass-dimensional constant parameter, and we choose 
the spherically symmetric metric as  
\be
\label{at3}
ds^2=\e^{-2\sigma} 
\left(\sum_{\alpha,\beta=0,1}g_{\alpha\beta}dx^\alpha dx^\beta  
+ r_0^2 d\Omega^2\right)\ ,
\ee
where $d\Omega^2$ is the metric on the unit two-sphere. 
The parameter $r_0$ is introduced by hands. Note that the 
Schwinger-De Witt expansion is the corresponding power expansion 
with respect to the curvature. Having introduced the parameter 
$r_0$, the scalar curvature given by the metric $g_{\mu\nu}$  
is of the order of $1/r_0^2$ if there is no singularity, as in 
the Nariai black hole. Therefore, if we choose $r_0$ to be large, 
the Schwinger-de Witt type expansion becomes exact. 

Solving the quantum effective equations of motion derived from 
(\ref{OI}) and (\ref{OVIII}), we can find the quantum analogue of 
the Nariai black hole, which has constant scalar curvature $R=R_0$ 
and radius $\e^\sigma=\e^{\sigma_0}$:  
\bea 
\label{solrl} 
R_0&=&\left\{2+\left( \ln (\mu r_0) \right)^{-1} 
\left({2b+3b' \over b}+{9 \over 512\pi^2 b G\Lambda}\right) \right\}  
+ {\cal O}\left( \left( \ln (\mu r_0) \right)^{-1}\right), \nn 
\sigma_0&=&- \ln (\mu r_0) + {1 \over 2}\ln \left( 
{3\mu^2 \over 2\Lambda}\right) \nn && +\left( \ln (\mu r_0) \right)^{-1} 
{\mu^2 \over 8\Lambda}\left({2b+3b' \over b}+{9 \over 512\pi^2 b G
\Lambda} \right) 
+ {\cal O}\left( \left( \ln (\mu r_0) \right)^{-1}\right).
\eea 
Here we assume $r_0$ to be large. 
Then the metric in (\ref{at3}) has the following form: 
\be
\label{at3b}
ds^2=\e^{2\sigma_0+2\rho} 
\left(-dt^2 + dr^2 + r_0^2 d\Omega^2\right)\ ,\quad 
\e^{2\rho}=\e^{2\rho_0}
\equiv{2C \over R_0}\cdot{1 \over \cosh^2 \left(r\sqrt{C}\right)} .
\ee
Here $C$ is a constant of integration. 

Furthermore, we can find the  
perturbation around the solution: 
\be
\label{pert}
\rho=\rho_0 + \epsilon R(t,r)\ ,\ \ \ 
\sigma=\sigma_0 + \epsilon S(t,r) \ .
\ee
If we assume $R$ and $S$ are the eigenfunctions with eigenvalue 
$A$ of the Laplacian $\Delta$ in the two-dimensional hyperboloid, 
which corresponds to the subspace in the Nariai black hole, given 
by radial and time coordinates
\be
\label{Delta}
\Delta=\cosh^2\left(r\sqrt{C}\right)\partial_+\partial_-\ ,
\ee
the perturbative equations of motion become two linear algebraic 
equations. From the condition that the two equations have non-trivial 
solutions,  the eigenvalue $A$ can be determined. 

The fate of the perturbed black hole is governed by 
the eigenvalue $A$ of the Laplacian $\Delta$. 
We now consider the following function as an eigenfunction 
of $\Delta$ in (\ref{Delta}):
\be
\label{cshS}
f_A(t,r)=\cosh t\alpha\sqrt{C}\cosh^\alpha r\sqrt{C} \ ,\quad 
A\equiv {\alpha (\alpha - 1) C \over 4}\ .
\ee
Note that there is one to one correspondence between $A$ and $\alpha$ if 
we restrict $A>0$ and $\alpha<0$. The horizon is given by 
the condition
\be
\label{horizon}
\nabla\sigma\cdot\nabla\sigma=0\ .
\ee
Substituting (\ref{cshS}) into (\ref{horizon}), 
we find the horizon is given by $r=\alpha t$.
Therefore on the horizon, we obtain
\be
\label{atv} 
S(t,r(t))=Q\cosh^{1+\alpha} t\alpha\sqrt{C}\ .
\ee
This tells that the system is unstable if there is a solution 
$0>\alpha >-1$, i.e., $0<A<{C \over 2}$. On the other hand, the 
perturbation becomes stable if there is a solution where 
$\alpha<-1$, i.e., $A>{C \over 2}$. The radius of the horizon 
$r_h$ is given by 
\be
\label{atvi}
r_h=\e^\sigma =\e^{\sigma_0 + \epsilon S(t,r(t))}\ .
\ee
Let the initial perturbation is negative $Q<0$. Then the radius 
shrinks monotonically, i.e., the black hole evaporates in case of 
$0>\alpha>-1$. On the other hand, the radius increases in time 
and approaches to the Nariai limit asymptotically 
$S(t,r(t))\rightarrow Q\e^{(1+\alpha) t |\alpha|\sqrt{C}}$ 
in case of $\alpha<-1$. The latter case corresponds to the 
anti-evaporation of black hole observed by Bousso and 
Hawking \cite{HB2}.
We should be more careful in the case of $A={C \over 2}$. When 
$A={C \over 2}$, $f_A(r,t)$ is, in general, given by
\be
\label{solS}
f_A(r,t)= \left\{{\cosh\left((t+a)\sqrt{C}\right) \over 
\cosh \left(r\sqrt{C}\right)}
+\sinh \left(b\sqrt{C}\right)  \tanh \left(r\sqrt{C}\right)\right\}\ .
\ee
Then the condition (\ref{horizon}) gives $t+a=\mp (r-b)$.
Therefore on the horizon, we obtain $S(t,r(t))= Q\cosh b$.
This is a constant, that is, there does not occur evaporation nor 
anti-evaporation. The radius of the horizon does not develop in time.
In classical case, we find $A={C \over 2}$. Therefore the horizon 
does not develop in time and the black hole does not evaporate 
nor anti-evaporate. The result is, of course, consistent with 
that of \cite{HB2}. 

The eigenvalue $A$ is now explicitly given by
\be
\label{atvii}
A={C \over 2}+\left( \ln (\mu r_0) \right)^{-1}a_1\ ,
\quad a_1=0\ ,\ \mbox{or}\ - {(b+b')C \over 8b'}
\ee
Here we neglect the term of ${\cal O}\left(\left( 
\ln (\mu r_0) \right)^{-2} \right)$ since we assume $r_0$ 
is large. In the first solution for $a_1$, the horizon does 
not develop in time and we would need the analysis of the higher 
order of $\left( \ln (\mu r_0) \right)^{-1}$. An important thing 
is the second solution is positive 
when 
\be
\label{Ns} 
2N+7N_{1/2}>26 N_1 . 
\ee
When $a_1$ is positive, $A>{C \over 2}$, i.e., there occurs 
anti-evaporation. 
Since the equations of motion in the model by Bousso and Hawking 
contain only second order derivatives, anti-evaporation is 
excluded there by the no-boundary condition of Hartle and Hawking. 
The quantum effective  equations of motion given by (\ref{OI}) 
and (\ref{OVIII}), however, contain fourth order derivatives 
and there the anti-evaporation phenomenon can be 
consistent with the  no-boundary condition.  

For the usual SO(10) GUT, which would be a typical elementary 
particle physics model in 
the early universe , it is shown that 
anti-evaporation may occur there only in its SUSY 
version \cite{ENjO1}. One may conclude that the existence 
of primordial black holes in the present universe might be 
considered as an indirect evidence for supersymmetry.

It could be really interesting to investigate the phenomena of the 
anti-evaporation in other types of BHs. The natural candidate to 
think about is charged BH. Working in large $N$ approach for dilaton 
coupled quantum scalars we present below some quantum properties of 2d
charged BHs. The corresponding classical solutions have been found in 
\cite{MNY1} (see also ref.\cite{GP1}). They may be considered as 
some compactifications of Type II string solutions. Moreover, 
very naturally they may be considered as 2d analog of 
Reissner-Nordstr\" om (RN) charged 4d BH.

We start from the action which has been considered by McGuigan, Nappi 
and Yost in ref.\cite{MNY1}.  
This action follows from compactification of heterotic string theory:
\be
\label{cMNY}
S={1 \over 16\pi G}\int d^2x \sqrt{-g}\e^{-2\phi}
\left( R + 4 (\nabla \phi)^2 + 4\lambda^2 - 4 F_{\mu\nu}^2 \right)\ .
\ee
 Note that above action has the form typical for 4d or 5d 
Einstein-Maxwell theory spherically reduced to two dimensions. 
Hence, it can be considered 
as toy model to describe 4d or 5d BH with spherical reduction.

It is remarkable that action (\ref{cMNY}) has the classical solutions which 
correspond to the 2d charged black hole (2d analogue of 
Reissner-Nordstr\" om black hole) with multiple horizon.
In the explicit form its metric and dilaton look like \cite{MNY1,GP1}
\bea
\label{MNYmetric}
&& ds^2=-\left( 1-2m \e^{-2\lambda x} 
+ q^2 \e^{-4\lambda x}\right)dt^2 
+\left( 1-2m \e^{-2\lambda x} 
+ q^2 \e^{-4\lambda x}\right)^{-1}dx^2 \nn
&& \e^{-2(\phi-\phi_0)}=\e^{2\lambda x}\ .
\eea
Here $q$ and $m$ are parameters related to the
charge and the mass of BH, respectively. 
The extremal solution is given by putting $q^2=m^2$. 
If we define new coordinates $r$ and $\tau$ by
\be
\label{r}
\e^{-2\lambda x}={m+ \epsilon \tanh 2\lambda r \over q^2} \ ,
\quad t={q \over \epsilon}\tau \ ,\quad 
\epsilon^2 \equiv m^2 - q^2\ ,
\ee
$r\rightarrow +\infty$ corresponds to outer horizon and 
$r\rightarrow -\infty$ corresponds to inner horizon.
Taking the limit $\epsilon\rightarrow 0$, we obtain
\be
\label{exmetric}
ds^2={1 \over \cosh^2 \lambda r}\left(d\tau^2 - dr^2\right) \ ,\quad 
\e^{-2(\phi-\phi_0)}={1 \over q}\ .
\ee
Note that the dilaton field becomes a constant in the limit.

We will discuss now the quantum corrections induced by $N$ free 
conformally invariant dilaton coupled scalars $f_i$ (no 
background scalars $f_i$):
\be
\label{sS}
S^f=-{1 \over 2}\int d^2x \sqrt{-g}\e^{-2\phi}
 \sum_{i=1}^N(\nabla f_i)^2\ .
\ee
Using the effective action induced by the conformal anomaly plus 
conformally invariant part calculated in the 
Schwinger-De Witt type expansion, we find the quantum analogue of 
the extremal solution in (\ref{exmetric}), where $\phi$ and the 
scalar curvature are constant:
\be
\label{MNYr1}
\phi=\phi_0 \ ,\quad 
R=-2\e^{2\rho}\partial_r^2\rho=R_0
\equiv -{8\lambda^2 \e^{-2\phi_0} \over 
\e^{-2\phi_0}-{2GN \over 3} }
\ee
and the field strength of the gauge field is given in 
the conformal gauge by 
\be
\label{MNYr2}
F_{+-}=B\e^{2\phi_0+2\rho_0} \ ,\quad 
B^2={\lambda^2\e^{-4\phi_0}\left(\e^{-2\phi_0}-{4GN \over 3} 
\right) \over \e^{-2\phi_0}-{2GN \over 3} } \ ,\quad
 \e^{2\rho_0}= {2C \over R_0}
\cdot {1 \over \cosh^2 \left(r\sqrt{C} \right)}\ .
\ee
As in the case of the Nariai black hole, we can investigate the 
(in)stability of the above solution by using the eigenvalue $A$ 
of the Laplacian in the hyperboloid and find
\be
\label{Asol}
{A \over C}
\sim{1 \over 2} \pm {1 \over 8} g^{1 \over 2} + {\cal O}(g) 
\ ,\quad g\equiv 8GN\e^{2\phi_0} \ .
\ee
In the classical limit $g=0$, 
 there does not occur any kind of the radiation 
in the solution.  Near the classical limit $g\sim 0$, there are solutions 
corresponding to both of stable and instable ones. 
It might be surprising that there is an instable mode since the 
extremal solution is usually believed to be stable.

Hence, we presented the evidence \cite{NO7,HB2,ENjO1} that 
anti-evaporation may be natural phenomenon for different types of 
multiple horizon BHs. It may occur also for BTZ BH \cite{BTZ1} 
as it was shown in ref. \cite{NO7}. Note once more that multiple horizon 
BHs are quite exotic objects. Even if they could present in our Universe 
\cite{HB3,BNS1} they should be rather rare in the present epoch. 
Nevertheless, even so there are indications \cite{Bss1,BNS1}   
that instabilities in such BHs( proliferation of de Sitter space) 
may become extremely important in the early Universe. 
In fact, their instability ( in particular, anti-evaporation) could lead to
kind 
of topological phase transition which may represent the mechanism 
of inflationary Universe creation!

\section{Quantum cosmology}

In the present chapter we discuss the applications of the effective action 
found in chapter 3 to dilatonic cosmology. We use two forms of effective
action: s-wave effective action (obtained as 2d anomaly induced action 
after spherical reduction) and 4d anomaly induced effective action. It is 
known that these two approximations are different \cite{FSZ1, NOdi}, 
generally speaking. They lead to qualitatively the same results only 
for sufficiently smooth backgrounds like black holes or not very quickly 
oscillating cosmologies. We mainly use 4d anomaly induced effective action 
except first section where s-wave approximation is taken. As one can see 
from results of this chapter it is possible to create quantum inflationary
Universe 
(or more generally, non-singular Universes) even in the presence of 
time-dependent dilaton.

\subsection{(Non)-singular Kantowski-Sachs Universe}

It is quite common belief that two-dimensional dilatonic 
gravity may be useful only as toy model for the study of realistic 
4d gravity, especially in quantum regime. However, it is 
quite well-known (for example, see refs.\cite{Hjc1,TIH1}) that spherical 
reduction of Einstein gravity leads to some specific 
dilatonic gravity.
On the same time, spherical reduction of minimal 4d matter leads 
to 2d dilaton coupled matter. 

The conformal anomaly for 2d 
conformally invariant, dilaton coupled matter  
and the correspondent anomaly induced 
effective action have been extensively discussed in chapter 3. Using such
anomaly induced effective 
action (i.e. working in $s$-wave and large $N$ approximation) 
and adding it to reduced Einstein action one may study FRW or  
four-dimensional Kantowski-Sachs (KS) quantum cosmology \cite{KS1c} 
in consistent way 
as it was done in refs.\cite{GKNO1,NOOO} (for a discussion 
of 2d dilatonic 
quantum cosmology, see for example 
\cite{MzR1,FR1,GV1,KY1,GKNO1} and references therein). 

In the (mainly numerical) study of refs.\cite{GKNO1} 
it was found 
that most of KS cosmologies under investigation are singular 
at the initial stage of the Universe evolution. The interesting 
question is : can we construct (non) singular KS quantum cosmologies 
using purely analytical methods?

In the first part of this section, we answer to this question following 
to study of ref. 
\cite{NOOO}. 
Using the analogy between KS cosmology
 and Schwarzschild BH 
(or its generalizations) after the 
interchange of time and radial coordinates (see \cite {KS1c,OR1})
we present the particular solution of quantum equations of motion 
. This solution represents non-singular KS cosmology 
(expanding Universe with always non-zero radius)  
which appears from Schwarzschild-de Sitter (or -anti de Sitter) BH 
after interchange of time and radial coordinates.
For purely induced gravity (when cosmology is defined completely 
by quantum effects of matter) we present general analytical solution
of quantum equations of motion. Unfortunately, in situation under discussion
all given examples of KS quantum cosmologies are singular. 
  
We start from the action of Einstein gravity with
$N$ minimal real scalars and $M$ Majorana fermions. 
In order to apply large $N$ approach $N$ and $M$ are
considered to be large, $N,M \gg 1$, $G$.
We now assume the spherically symmetric spacetime as in (\ref{e2}). 
Working in large $N$ and $s$-wave approximation one can calculate the
quantum correction to $S_{red}$ (effective action). Using 2d conformal
anomaly for dilaton coupled scalar and dilaton coupled spinor, 
one can find the anomaly induced effective action. 
In the case of dilaton absence such induced effective action gives 
the complete effective action which is valid for an arbitrary 
two-dimensional background. In the presence of dilaton as above 
the complete effective action consists of two pieces. First one is 
induced effective action which is given actually for any background 
but with accuracy up to conformally invariant functional.
The second piece of it, i.e. conformally invariant functional
can not be found in closed form. We use Schwinger-De Witt technique 
(curvature expansion) to
calculate it. We keep only leading part of such expansion,
for more detail and explanation, see refs.\cite{NO7,NNO}.

We choose the conformal gauge (\ref{IIIiib}), 
($x^\pm\equiv t\pm r$) but it is often convenient to use 
the cosmological time $\tau$ instead of $t$, where the metric 
is given by
\be
\label{e10}
ds^2=-d\tau^2 + \e^{2\rho}dr^2 + \e^{-2\phi}d\Omega^2 .
\ee
Since we have $d\tau=\e^\rho dt$, 
we obtain $\partial_t = \e^\rho\partial_\tau$ and 
$\partial_t^2 = \e^{2\rho}\left(\partial_\tau^2
+ \partial_\tau\rho\partial_\tau \right)$.

We now consider a special solution 
corresponding to the (Wick-rotated) Nariai 
solution \cite{Nr1},
where $\phi$ is a constant : $\phi=\phi_0$.
Then from the effective equations written explicitly in \cite{NOOO},
 we find the solution 
\be
\label{Niii}
\e^{-2\phi_0}={(2N+M)G \over 6} + {1 \over 2\Lambda}
\pm {1 \over 2}\sqrt{{(2N+M)^2G^2 \over 9}+{1 \over \Lambda^2}
  - {(8N + 6M)G \over 3\Lambda}}
\ee
The sign $\pm$ in (\ref{Niii}) should be $+$ if we require the
solution coincides with the classical one
$\e^{-2\phi_0}={1 \over \Lambda}$ in the
classical limit of $N,M\rightarrow 0$.
On the other hand, in the solution with the $-$ sign, 
we have $\e^{-2\phi_0}\sim {(3N+2M)G \over 3}\rightarrow 0$
in the classical limit. Therefore the second solution 
does not correspond to any classical solution but the solution 
is generated by the quantum effects. 
The solution for $\e^{2\rho}$ is given by
\be
\label{Niv}
\e^{2\rho}=\left\{\begin{array}{ll}
{2C \over R_0}{1 \over \cos^2\left(t\sqrt{C}\right)}
\ \ \ \ &\mbox{when}\ R_0>0 \\
  -{2C \over R_0}{1 \over \cosh^2\left(t\sqrt{C}\right)}
\ \ \ \ &\mbox{when}\ R_0<0 \\
\end{array}\right. \ .
\ee
Here $C>0$ is a constant of the integration and $R_0$ is 2d scalar
curvature, which is given by 
\bea
\label{Nv}
R_0=2\e^{-2\rho}\partial_t^2\rho 
&=&-{3\Lambda \over (N+M)G}\left({(2N+M)G \over 3} 
  - {1 \over \Lambda}
\right. \nn && \left.
\pm \sqrt{{(2N+M)^2G^2 \over 9}+{1 \over \Lambda^2}
  - {(8N + 6M)G \over 3\Lambda}}\right)\ .
\eea
Note that 4d curvature $R_4=R_0 + 2\e^{2\phi_0}$ becomes a
constant. It should be also noted that the solution exists for 
the both cases: of positive $\Lambda$ and negative $\Lambda$.
In Eq.(\ref{Nv}), the $+$ sign corresponds to the classical limit 
($N$, $M\rightarrow 0$). In the limit, we obtain 
$R_0\rightarrow 2\Lambda$ ($R_4\rightarrow 4\Lambda$). 
On the other hand, the $-$ sign in Eq.(\ref{Nv}) corresponds 
to the solution with $-$ sign in (\ref{Niii}) generated by 
the quantum effect. In the classical 
limit for the solution, the curvature $R_0$ in (\ref{Nv}) 
diverges as $R_0\sim {3 \over 2(N+M)G}\rightarrow +\infty$.
Therefore from (\ref{Niv}), we find that $\e^{2\rho}$ vanish.
(Note that $R_0>0$ in the limit) : 
$\e^{2\rho}={4(N+M)GC \over 3\cos^2\left(t\sqrt{C}\right)}
\rightarrow 0$.
Therefore by using (\ref{e2}), we obtain the following 
metric near the classical limit: 
\be
\label{q4}
ds^2 = {4(N+M)GC \over 3\cos^2\left(t\sqrt{C}\right)}
\left(-dt^2 + dr^2\right) + {(3N+2M)G \over 3}d\Omega^2\ .
\ee
This is non-singular metric for fixed $N$, $M$.

It should be interesting to consider the limit 
$\Lambda\rightarrow 0$, where there is no de Sitter or 
anti-de Sitter solution at the classical level. In the limit, 
we can have a finite solution:
\be
\label{Lamb0}
\e^{-2\phi_0}\rightarrow {(3N+2M)G \over 3}\ ,\ \ 
\e^{2\rho}\rightarrow {(N+M)GC \over 3\cos^2\left(t\sqrt{C}\right)}\ 
\left(R_0\rightarrow {6 \over (N+M)G}\right)\ .
\ee
This tells that the Nariai space can be generated by the quantum 
effects even if $\Lambda=0$.

The obtained solution (\ref{Niv}) (and (\ref{q4})) might appear 
to have a singularity when $\cos^2\left(t\sqrt{C}\right)=0$ 
(for $R_0>0$ case) but the singularity is apparent one. In fact 
the scalar curvature $R_0$ in (\ref{Nv}) is always constant. 
If we change the conformal time coordinate $t$ by the cosmological 
time $\tau$ in (\ref{e10}), we find that the time 
$\cos^2\left(t\sqrt{C}\right)=0$ corresponds to infinite future 
or past. 
In the following, we assume $R_0>0$ for simplicity. 
$R_0<0$ case can be easily obtained by changing the constant 
$C\rightarrow -C$ and analytically continuing solutions.
We now change the time-coordinate by $\tau = \sqrt{2 \over R_0}
\ln\left({1 + \tan\left(t{\sqrt{C} \over 2}\right)
\over 1 - \tan\left(t{\sqrt{C} \over 2}\right)}\right)$. 
Then the time $\cos^2\left(t\sqrt{C}\right)=0$ 
($t\sqrt{C}=\pm {\pi \over 2}$) corresponds to $\tau=\pm\infty$. 
 Using the cosmological time $\tau$, we obtain the following metric
\be
\label{Nvii}
ds^2= - d\tau^2 + {2C \over R_0}\cosh^2\left(\tau\sqrt{R_0 \over 2}
\right)dr^2 + \e^{-2\phi_0}d\Omega^2\ .
\ee
Here $\e^{-2\phi_0}$ is given in (\ref{Niii}). If we assume $r$
has the periodicity of $2\pi$, the metric describes non-singular 
Kantowski-Sachs Universe, whose topology is $S_1\times S_2$.
The radius of the $S_2$ is constant but the radius of $S_1$ has
a minimum when $\tau=0$ and increases exponentially with
the absolute value of $\tau$. 

Hence we found non-singular KS cosmology which exists on classical 
level and which also exists on quantum level (as quantum corrected 
KS cosmology). This metric may be considered as the one obtained 
from Schwarzschild-de Sitter (Nariai) BH 
(for positive cosmological constant) \cite{Nr1}
and from Schwarzschild-anti-de Sitter BH (for negative cosmological 
constant). To make the correspondence one has to interchange time and 
radial coordinates assuming corresponding Wick-rotation. It is very 
interesting that the last case (of negative cosmological constant) 
may be relevant to AdS/CFT  
correspondence (see next chapter). We also found
non-singular KS Universe which does not have the classical limit 
and which is completely induced by quantum effects (even in the 
case of zero cosmological constant). Hence we obtained expanding 
Universe with the radius which is never zero. 
This cosmology may be interesting in frames of inflationary 
Universe as it can describe some sub-stage of inflationary 
Universe where there is effective expansion only along one 
(or two) space coordinates. 

Let us discuss now the situation when we live in the regime 
where quantum (non-local) anomaly induced effective action 
gives major contribution to equations of motion. In other 
words, quantum cosmology is defined completely by quantum effects 
(effective gravity theory which at some point makes transition 
to classical gravity).
As we will see in this case the equations of motion 
admit the analytical solutions which lead to singular KS cosmology.
We consider purely induced gravity, i.e. $N,M\to\infty$ case. 
Then the Einstein action can be dropped away. 
For this case, the field equations admit the following 
integral of motion \cite{NOOO}
\be
\label{integral1}
I_1=\e^\rho\,\left[(N+\frac{2}{3}M)\rho'+2N(\rho+a)\phi'\right]
\ee
Here $'={d \over d\tau}$. 
For the case $\phi=\mbox{const}=\phi_0$, we have the 
following solution of equations 
\be
\label{sol00}
r(\tau)=\mbox{const}=\e^{-\phi_0},
\qquad
f(\tau)=\e^{\rho}=(f'_0\tau+f_0).
\ee
For the solution (\ref{sol00}), the scalar curvature is constant : 
$R=2\,\e^{2\,\phi_0}$. 
For the case $\rho=\mbox{const}=\rho_0$, we have the solution
of equations 
\be
\label{sol0}
\rho_0=-a,
\qquad
r(\tau)=\e^{-\phi}=(c_1\tau+c_2)^{1+\frac{2M}{3N}},
\qquad
c_1,c_2 = \mbox{const}.
\ee
\be
 c_2=\exp\left(\frac{-3\,N\,\phi_0}{2\,M + 3\,N}\right),
\qquad
 c_1=c_2
     {{\frac{-3\,N\,\phi'_0}{\,
          \left( 2\,M + 3\,N \right) }}}
\ee
Here $\phi_0$ and $\phi'_0$ are the values of $\phi$ and $\phi'$ 
at $\tau=0$, respectively. 
For the solution (\ref{sol0}), the scalar curvature 
has a singularity at $\tau=-{c_2 \over c_1}$.
If $\rho\ne\mbox{const}$, then we obtain the following special 
solution
\be
\label{sol1}
f(\tau)=\e^\rho=f'_0\tau+f_0,
\qquad
r(\tau)=\e^{-\phi}
=c_3\left[a+\ln(f'_0\tau+f_0)\right]^{1+\frac{2M}{3N}}.
\ee
Here $f_0$ and $f'_0$ are the values of $f=\e^\rho$ and $f'$ at 
$\tau = 0$ respectively.
For the solution (\ref{sol1}), the scalar curvature 
has a singularity when $\tau=-{f_0 \over f'_0} 
- {\e^{-a} \over f_0}$. Similarly, other cases giving singular solutions
may be discussed.

Generalizing, we should note that the equation of motion 
admit one more integral besides $I_1$ in 
(\ref{integral1}). It may be presented as follows
\be
\label{in2}
0={1 \over 2}(\rho')^2 + V(\rho) \ ,\ \ 
V(\rho)=-{6 I_2 \over N+M}{\rho+a - \alpha \over \rho + a - \beta}
\e^{-2\rho} \ .
\ee 
Here $\alpha\equiv -{I_1^2 \over 8NI_2}$ and 
$\beta\equiv-{3\left(N+{2 \over 3}M\right)^2 \over 2N(N+M)}$. 
Note that $\beta$ is negative and $\alpha$ is positive (negative) 
when $I_2$ is negative (positive). 

Eq.(\ref{in2}) tells that 
there would be a curvature singularity when $\rho + a \rightarrow 
\beta\pm 0$. In fact, when $\rho + a \sim \beta\pm 0$, we obtain 
from (\ref{in2}) $(\rho')^2\sim {A \over \rho + a -\beta}$, 
($A\equiv {12I_2(\beta - \alpha)\e^{2(a-\beta )} \over N+M}$. 
Therefore we find 
$\rho + a = \beta + \left({3 A \over 2}(\tau - \tau_\beta)
\right)^{2 \over 3}$.
Here $\tau_\beta$ is a constant of the integration 
and $\rho + a = \beta$ when $\tau=\tau_\beta$. 
Then the scalar curvature $R$ is always singular when 
$\rho + a \sim \beta\pm 0$ except $\alpha=\beta$ case, when $A$ 
vanishes (we should note that $A$ is finite when $I_2=0$). 

In the case $\alpha=\beta$, (\ref{in2}) can be 
explicitly solved to give 
\be
\label{in7}
\e^\rho = \pm {12 I_2 \over N+M}(\tau - \tau_0)\ .
\ee
Here $\tau_0$ is a constant of the integration.
(\ref{in7}) tells that there is a singularity when $\tau=\tau_0$. 
In case of the expanding universe ($+$ sign in (\ref{in7})), 
$\phi'=0$, i.e., $\phi$ is a constant as in 
the Nariai space \cite{Nr1} 
(note that there is a singularity even in 
this case, which is different from the Nariai space). 
On the other hand, in case of the shrinking universe 
($-$ sign in (\ref{in7})), there is a curvature singularity 
when $\tau-\tau_0=-{N+M \over 12 I_2}\e^{-a}$ besides 
$\tau=\tau_0$. The analysis of all other sub-cases for explicit 
analytical solutions may be presented without problems.

As it follows from above analysis in purely induced gravity 
case when expanding Universe is created by the
 matter quantum effects one always gets the curvature 
singularity like in the case discussed in ref.\cite{GKNO1}.
Nevertheless, it is remarkable that equations of motion in 
this case admit analytical solutions.
Hence, using $s$-wave and large $N$ approximation we 
studied gravitational equations of motion with quantum corrections.
The analytical solutions representing 
(non) singular KS cosmology are found. 

In \cite{GKNO1}, it has been considered $N=1$ 2d dilatonic 
supergravity (SG), 2d dilatonic SG obtained by dimensional 
reduction from $N=1$ 4d SG, $N=2$ 2d dilatonic SG and 
string-inspired 4d dilatonic SG.  We use the conformal anomaly 
induced effective action due to dilaton-coupled conformal 
matter as a quantum correction (for 4d models 
$s$-waves approximation is additionally applied). The solutions 
of the equations of motion have been numerically investigated 
for 2d and 4d FRW or 4d Kantowski-Sachs Universes with 
a time-dependent dilaton. 
The evolution of the corresponding quantum cosmological models 
is given for different choices of initial conditions and 
theory parameters. In most cases one finds quantum singular 
Universes. Nevertheless, there are examples of Universe 
non-singular at early times.  

The general $N=1$ action of 2d dilatonic supergravity is given 
in terms of general functions of the dilaton $C(\phi)$, 
$Z(\phi)$, $f(\phi)$ and $V(\phi)$ in section 3.6. 
As we are going to work in large 
$N$ approximation we may take into account only matter 
quantum effects. Since we presently considering the 
cosmological problem, we assume that all the fields depend 
only on time $t$. Choosing the (super)conformal gauge 
(\ref{IIIiib}), 
equations of motion 
can be solved numerically for several parameters
with the initial condition $\phi=\rho={d\phi \over dt}
={d\rho \over dt}=0$ at $t=0$.
Especially for the supersymmetric extension of 
the CGHS type model where
\be
\label{sCGHS}
C(\phi)=2\e^{-2\phi}\ ,\ \ Z(\phi)=4\e^{-2\phi}\ ,
\ \ V(\phi)=4\lambda\e^{-2\phi}\ .
\ee
Typical graphs are given in \cite{GKNO1} for CGHS type model.
We also calculated the two-dimensional scalar curvature $R$. 
(As again 
solution under discussion may be understood as 4d KS Universe 
or 2d FRW-type cosmology with non-trivial dilaton.) 
In Figures in \cite{GKNO1}, there appeared several types of the behavior of 
the conformal factor $\rho$ and dilaton field $\phi$. In a solution, 
both  the conformal factor $\rho$ and dilaton $\phi$ 
increase monotonically with time and there is a curvature singularity 
in a finite conformal time. We should note that there was not found 
a solution where $\rho$ monotonically increases with time in the 
bosonic model.  If $\rho$ diverges near the singularity, 
the singularity will appear in the infinite future of the 
cosmological time defined by
\be
\label{ct}
d\hat t = \e^\rho dt\ .
\ee
Note $g_{tt}=1$ when we use the cosmological time $\hat t$. It should 
be also noted that there will not appear any singularities in the 
two-dimensional scalar curvature in these solutions.
As a rule, $\phi$ increases monotonically 
in time but $\rho$ increases 
first and decreases later, which means that the universe expands 
first and then later shrinks.  In another solution, 
$\phi$ oscillates and 
$\rho$ decreases with small oscillation which means that an 
oscillating shrinking universe 
is realized. The scalar oscillating curvature goes to zero. 

For the model reduced from 4d Einstein gravity with 
a cosmological term and matter described by $N$ minimal scalars 
coupled to the metric with spherical symmetry, 
we concentrate on the spherically symmetric metrics of 
Kantowski-Sachs form where $g_{\mu\nu}=a^2(t)\eta_{\mu\nu}$. Such 
a metric describes a Universe with a $S^1\times S^2$ spatial 
geometry. Typical graphs for the numerical calculation are also 
given in \cite{GKNO1}.  
The curvature singularity seems always to  
appear, at least in cases under discussion. 
As a rule, both the dilaton field $\phi$ and the conformal 
factor $\rho$ increase monotonically in time as 
in a solution in CGHS type model.  If we regard the solution 
as describing a universe of the Kantowski-Sachs form 
with the topology $S^2\times S^1$, the radius of $S^2$, 
which is given by $\e^{-\phi}$, decreases to zero but 
the radius of $S^1$, which is given by $\e^{2\rho}$, increases. 
Note that the radius of $S^1$ corresponds to the radius of 
the universe when we regard this model as a two dimensional one.  
The four-dimensional scalar curvature increases first and goes 
to minus infinity. In another solution, $\phi$ 
increases monotonically and $\rho$ decreases at first but increases 
infinitely after that.  
This means $S^2$ shrinks but the $S^1$ factor 
shrinks at first and later expands. The four-dimensional 
scalar curvature 
goes to zero at first but increases to infinity later.  The dilaton 
field $\phi$ always increases and runs away to the singularity, 
which means  $S_2$ in Kantowski-Sachs universe shrinks 
to a point, which causes the curvature singularity. When we regard 
the model as a two-dimensional one, the radius of the universe 
goes to infinity at the final stage in the cases under discussion.
We see that at $t=0$ there is no singularity in most of our cases.
However, at late times there is as a rule singularity.
Note that the behavior of our cosmologies at late times 
is not so important 
as other effects should define late time structure of the Universe.

The low-energy effective action of the dilaton gravity sector in the 
 superstring may be given by
\be
\label{dgaction}
S=\int d^4x \sqrt{-g} \e^{-\phi-\phi^*} \left(R-6\nabla_\mu\phi
\nabla^\mu\phi^*\right) + \cdots\ .
\ee
Here $\cdots$ denotes the terms containing moduli, gauge fields, which depend 
on the detail of the compactification, and fermionic fields.  
On the background where the dilaton field $\phi$ is real, 
one takes the non-local effective action $W$. 
We now assume $\phi$ is real and that the conformally flat 
fiducial metric is $g_{\mu\nu}=\e^{2\sigma}\eta_{\mu\nu}$.  
Some examples of the numerical solutions of effective equations are 
given in \cite{GKNO1}. 
An interesting 
thing is that 4d scalar curvature oscillates as a trigonometric 
function. In a solution, both of the dilaton field $\phi$ and the 
conformal factor $\sigma$ slowly increase monotonically 
but the scalar 
curvature vibrates rather quickly like trigonometric 
functions but the amplitude slowly increases.  

Hence, unlike the case of pure Einstein gravity with conformal matter, it is 
more difficult to realize the inflationary Universe in quantum theory 
with dilaton under consideration. Nevertheless, there are cases 
presented above where it is possible.

\subsection{Inflationary Brans-Dicke quantum Universe}

In the present section we apply the anomaly induced effective action 
for dilaton coupled spinor (Section 3.4) in the construction of 
inflationary Brans-Dicke Universe \cite{GOZ1}.

As it has been already mentioned, Brans-Dicke theory 
(for a review, see \cite{will}) represents one of the simplest 
examples of scalar--tensor (or dilatonic) gravities 
where the background is described by the metric and the dilaton. 
Even low--energy string effective action 
(for a recent review, see \cite{polch}) 
may be considered as some kind of BD theory (with higher order terms). 

Let us start from the standard Brans-Dicke 4-dimensional action 
in the Jordan frame:
\be
S_{BD}=\frac{1}{16\pi}\int d^4x \sqrt{-g} \left[ 
\phi R -\frac{\omega}{\phi}
(\nabla_\mu \phi)(\nabla^\mu \phi) \right]+S_M\,,
\label{bdj}
\ee
where $\phi$ is the Brans-Dicke (dilaton) field with
$\omega$ being the coupling constant and $S_M$ is the matter action. 

It has been argued (for a review, see \cite{FGT}) that only 
the action in the Einstein frame is physically relevant. 
For this reason, one may prefer working 
within this frame, performing the following transformations
\be
\tilde{g}_{\mu \nu}=G\phi g_{\mu \nu},\quad 
\tilde{\phi}=\sqrt{{2\omega+3}/(16\pi)} \ln \phi\,\quad
2\omega+3 >0\,. 
\ee
Thus, the classical action in the Einstein frame  reads
\be
S=\int d^4x \sqrt{-\tilde{g}(x)} \left[\frac{\tilde{R}}{16\pi G}
  -\frac{1}{2}
(\tilde{\nabla}_\mu \tilde{\phi})(\tilde{\nabla}^\mu \tilde{\phi})+
\exp{(A\tilde{\phi})}L_M(\tilde{g}) \right]
\,,
\label{bde}
\ee
where $A=-8\sqrt{\frac{\pi G}{2\omega+3}}$.
Below, we consider the theory in the  Einstein frame  as physical 
theory, (no tilde will be written). As matter
Lagrangian we take the one associated with $N$ massless (Dirac) 
spinors, i.e. 
\be
\label{matterOA}
L_M=\sum_{i=1}^N \bar\psi_i\gamma^\mu\nabla_\mu\psi^i\ .
\label{sd}
\ee
The choice (\ref{sd}) is made only for the sake of simplicity.

Our purpose here will be to show the role of quantum effects 
associated with dilaton coupled spinors to cosmological problems 
in BD gravity. Assuming $N$ sufficiently 
large (to allow for the large  $N$ approximation) one can neglect 
in such investigation the  proper quantum gravity corrections. 
The corresponding 4d anomaly induced EA for dilaton coupled 
spinors (section 3.4) will be used. 
Hence, starting from the theory with the action
\be
S=\int d^4x \sqrt{-g} \left[\frac{R}{16\pi G}
  -\frac{1}{2}
(\nabla_\mu \phi)(\nabla^\mu \phi)+
\exp{(A\phi)}\sum_{i=1}^N \bar\psi_i\gamma^\mu\nabla_\mu\psi^i  \right]
\,,
\label{bde1}
\ee
we will discuss FRW type cosmologies
\be
ds^2=-dt^2+a(t)^2 dl^2\,,
\label{st4}
\ee
where $dl^2$ is the line metric element of a 3-dimensional space with 
constant curvature $\Sigma$, namely $k=1$ ($\Sigma=S^3$), $k=0$  
($\Sigma=R^3$) or $k=-1$ ($\Sigma=H^3$).
Introducing the conformal time $\eta$,
one gets a  space--time which is conformally related to an ultrastatic 
space--time with constant curvature spatial section $\overline{M}$, namely
\be
g_{\mu \nu}=e^{2\sigma(\eta)} \bar{g}_{\mu \nu}\,,
\ee

The computation of the anomaly induced EA for the dilaton 
coupled spinor field has been done in Section 3.4: 
\bea
\label{viip}
\lefteqn{\hspace{-.8cm}
W=\int d^4x \sqrt{-\bar g} \bigg\{b \bar F \sigma_1
+ 2b' \sigma_1\Big[ \bar{\Box}^2
+ 2 \bar R^{\mu\nu}\bar\nabla_\mu\bar\nabla_\nu 
  - {2 \over 3}\bar R\bar{\Box}
+ {1 \over 3}(\bar\nabla^\mu\bar R)\bar\nabla_\mu
\Big]\sigma_1}
\nn
&&\!\!\!\!\!\!\! 
+\, b'\sigma_1\Big(\bar G -{2 \over 3}\bar\Box\bar R\Big) 
  -{1 \over 18}(b + b')\left[\bar R - 6 \bar{\Box} \sigma_1
  - 6(\bar\nabla_\mu \sigma_1)(\bar\nabla^\mu \sigma_1)
\right]^2\bigg\}\,,
\eea
where $\sigma_1=\sigma+ A\phi /3$. 
For Dirac spinors
$b={3N \over 60(4\pi)^2}$, $b'=-{11 N \over 360 (4\pi)^2}$.

Generally speaking, it should be noted that the complete 
one-loop EA is given by the anomaly--induced EA presented above, plus some
conformally invariant functional which is the EA computed in the reference
metric $\bar g_{\mu \nu}$. In our case, this second term is rather trivial
(actually, it is a $k$--dependent constant),
being the EA of a free spinor field in an ultrastatic space--time with 
constant spatial section (for a discussion of such effective actions, see 
\cite{Zerbini}). 
For example, for the flat case ($k=0$), which we only consider, 
$\bar g_{\mu \nu}=\eta_{\mu \nu}$, and as a consequence,
$W$ gives the complete non-trivial EA!
Then, the effective action becomes
 \be
\label{1ef}
W=V_3\int d\eta \left\{2b' \sigma_1 \sigma_1''''
  - 2(b + b')\left( \sigma_1'' - {\sigma_1'}^2 \right)^2\right\}\ .
\label{7b}
\ee
Here, $V_3$ is the (infinite) volume of 3-dimensional flat space and
$'\equiv{d / d\eta}$ and $\sigma=\ln a$ where
$a(\eta)$ is the scale factor.

The total one--loop EA may be written adding to $W$ the 
classical action for $k=0$: 
\be
S=V_3\int d\eta \left[ \frac{6}{16 \pi G}(\sigma''+{\sigma'}^2) 
\e^{2\sigma}+\frac{1}{2} {\phi'}^2 \e^{2\sigma}\right]\,.
\label{ca}
\ee
Having in mind the inflationary 
Universe solution, one can get the special solution of 
effective equations as
\be
a(\eta)=\frac{1}{H \eta}\,,\,\,\,\,\,\, 
\phi'(\eta)=\frac{1}{H_1 \eta}\,,
\label{10}
\ee
where $H$ and $H_1$ are some constants. 
Their explicit values are: 
\be
H_1=\sqrt{4\pi G}\left[- \frac{3}{16}\sqrt{2\omega+3}\pm 
\sqrt{\frac{81}{4}(2\omega+3)-\frac{1}{6}} \right]\,, \quad 
H^2\simeq - {1 \over 16\pi G b'} - {1 \over 24 H_1^2 b'}
\label{12b}
\ee
Transforming back to the physical time $t$, one obtains the  
 $a(t)=1/C\exp{(\mp Ht)}$. 
Choosing the plus sign gives
rise to the inflationary universe (in dilaton absence such 
solution has been found in refs.\cite{inf1,inf2}). In this case, $\phi \sim
\frac{H}{H_1}t$, namely the dilaton is expanding with time, but much slower
than the scale factor.

Summarizing, we have presented explicitly one special analytic solution
describing Brans-Dicke non-singular
 Universe with a (much slower) expanding dilaton. 
It should be noted that this is a purely quantum solution which does not 
exist at classical level. In similar way one can study Brans-Dicke 
quantum cosmologies for other types of dilaton coupled matter, 
as well as their properties (like stability) at late times. 
It is interesting that induced dilatonic gravities may 
be studied in a
similar fashion. For example, starting 
 from the Lagrangian of local superconformally 
invariant  ${\cal N} = 4$  super YM  theory in the background 
of ${\cal N} = 4$  conformal supergravity 
(see \cite{KTN1,BRW1,conf} for 
introduction and review) one can find Weyl anomaly \cite{LT1}with 
account of dilaton for such YM theory. Then, the construction of quantum
inflationary Universe 
from super YM theory is also possible.
 
\subsection{Primordial wormholes at the early Universe}

The wormholes are puzzling topological objects (kind of
handles of topological origin) which attract much attention
in General Relativity for years. It is not strange as they may 
be considered as bridges joining two different Universes or two 
separate regions of the same Universe (for an introduction, see 
refs.\cite{MrTh1,TTY1}). There are many speculations related with 
hypothetical effects which may be expected near wormholes.

Moreover, it is known that wormholes usually cannot occur as
classical solutions of gravity (with matter) due to violations
of energy conditions \cite{MrTh1,TTY1}. Nevertheless, one can expect
that primordial wormholes may present at the very early Universe
where quantum effects play an essential role (or yet in
fundamental M-theory, see,for example, \cite{Gbn5}).

Indeed as it was shown in ref.\cite{HPS1} (see also \cite{16})
using quantum stress tensor for conformal scalar on 
spherically symmetric space and in ref.\cite{NOOO3} using
one-loop effective action in large $N$ and $s$-wave approximation 
for minimal scalar there may exist semiclassical quantum solution 
corresponding to a Lorentzian wormhole connecting two asymptotically 
flat regions of the Universe.
The corresponding spherically symmetric wormhole 
solution has been found numerically in both approximations
\cite{HPS1,NOOO3} as well as analytically \cite{NOOO3}. That 
shows the principal possibility of inducing  primordial 
wormholes at the early Universe (in its quantum regime). However 
above discussion has been limited strictly to scalar matter. And 
what happens if other types of matter are included? Hence, the very 
natural question is: Can primordial wormholes be induced from GUTs 
at the early Universe?

In this section, following to ref.\cite{NOOOwh} we  give positive  
 answer to this question for spherically 
symmetric wormholes. As we will see the object under consideration 
may be
understood  as 4d wormhole or as 2d topological object with 
non-trivial dilaton.
We use 4d conformal anomaly induced effective action in one-loop
and large $N$ approximation. Then the effective equations for an arbitrary 
massless GUT
containing conformal scalars, spinors and vectors may be explicitly
obtained for two forms of spherically symmetric background.
Numerical wormhole solution of these equations may 
be obtained 
 for ${\cal N}=4$
super Yang-Mills theory. It is observed that it depends on the
choice of the initial conditions.
 
Let us start from Einstein gravity with $N_0$ 
conformal scalars $\chi_i$, $N_1$ vectors $A_\mu$ and
$N_{1/2}$ Dirac spinors $\psi_i$.
The above matter content is typical for asymptotically free
GUT at high energies as interaction terms
and masses at strong curvature
 are negligible due to asymptotic freedom. For finite GUTs
or asymptotically non -free GUTs
one should not include masses if consider only conformally
invariant theories and interaction plays no role as anyway we consider
purely gravitational background.

The convenient choice for the spherically symmetric
space-time is the following:
\be
\label{CCOIV}
ds^2=f(\phi)\left[
f^{-1}(\phi)g_{\mu\nu}dx^\mu dx^\nu + r_0^2 d\Omega\right]\ .
\ee
where $\mu,\nu=0,1$, $g_{\mu\nu}$ and $f(\phi)$ depend only
from $x^1$ and $r_0^2$ is non-essential constant.

The calculation of matter effective action
on the background (\ref{CCOIV}) may be done in the standard way 
as it was presented in chapter 3. 
In the calculation of effective action, one presents
effective action as :
$\Gamma=\Gamma_{ind}+ \Gamma[1, g^{(4)}_{\mu\nu}]$
where $\Gamma_{ind}=\Gamma[f, g^{(4)}_{\mu\nu}]
  - \Gamma[1, g^{(4)}_{\mu\nu}]$ is conformal anomaly
induced action,
$g^{(4)}_{\mu\nu}$ is metric (\ref{CCOIV}) without
multiplier in front of it, i.e., $g^{(4)}_{\mu\nu}$
corresponds to
\be
\label{CCOVI}
ds^2=\left[
\tilde g_{\mu\nu}dx^\mu dx^\nu + r_0^2 d\Omega\right]\ ,\ \ \
\tilde g_{\mu\nu}\equiv f^{-1}(\phi)g_{\mu\nu} \ .
\ee

The conformal anomaly for above matter is written in section 3.3 (no 
dilaton)
\be
\label{CCOVII}
T=b\left(F+{2 \over 3}\Box R\right) + b' G + b''\Box R
\ee
where $b={(N_0 +6N_{1/2}+12N_1)\over 120(4\pi)^2}$,
$b'=-{(N_0+11N_{1/2}+62N_1) \over 360(4\pi)^2}$, $b''=0$.

Conformal anomaly induced effective action $\Gamma_{ind}$
is:
\bea
\label{CCOVIII}
&& W=b\int d^4x \sqrt{-g} F\sigma
+b'\int d^4x \sqrt{-g} \Bigl\{\sigma\left[
2\Box^2 + 4 R^{\mu\nu}\nabla_\mu\nabla_\nu 
 - {4 \over 3}R\Box + {2 \over 3}(\nabla^\mu R)\nabla_\mu
\right]\sigma \nn 
&& + \left(G-{2 \over 3}\Box R\right)\sigma \Bigr\} 
 -{1 \over 12}\left(b'' + {2 \over 3}(b + b')\right)
\int d^4x \sqrt{-g}\left[R - 6 \Box \sigma
  - 6(\nabla \sigma)(\nabla \sigma) \right]^2
\eea
where $\sigma={1 \over 2}\ln f(\phi)$, and
$\sigma$-independent terms are dropped. All 4-dimensional
quantities (curvatures, covariant derivatives) in
Eq.(\ref{CCOVIII}) should be calculated on the metric
(\ref{CCOVI}).
$\Gamma[1, g_{\mu\nu}^{(4)}]$ is given as follows:
\be
\label{CCSDW}
\Gamma[1, g_{\mu\nu}^{(4)}]=\int d^4x
\sqrt{-g}\left\{\left[{b}F+  {b'}G
+ { 2b\over 3}\Box R\right]\ln { R \over \mu^2}\right\}
+{\cal O}(R^3)
\ee
where $\mu$ is mass parameter, all
the quantities are calculated on the background (\ref{CCOVI}).
The condition of application of above expansion is
$|R|<R^2$ (curvature is nearly constant)..

One can get the equations of motion  from
the above effective Lagrangian $S+\Gamma$ which is effectively 
two-dimensional object. Note that extra scalar (dilaton) appears
after reduction to two-dimensional gravity.
In the following, it is convenient to work in  the conformal gauge 
(\ref{IIIiib}) 
after considering the variation of the effective action
$\Gamma+S$ with respect to two-dimensional metric $g_{\mu\nu}$ 
and $\sigma$.
Note that the tensor $g_{\mu\nu}$ under consideration is
the product of the
original metric tensor and the $\sigma$-function
$\e^{-2\sigma}$, the equations given by the variations over
$g_{\mu\nu}$ are the combinations of the equations given by the
variation over the original metric and $\sigma$-equation.

It often happens that we can drop the terms linear in $\sigma$. 
In particular, one can redefine the corresponding
source term as it is in the case of infra-red sector of 4d 
quantum gravity.
In the following, we only consider this case.
Then the equations of motion given by variations 
of $S+\Gamma_{ind}+\Gamma[1, g^{(4)}_{\mu\nu}]$ are 
written in 
\cite{NOOOwh}( they are too lengthy to write down it here).
Now the real four-dimensional metric is given by
\be
\label{CCmetric}
ds^2=-\e^{2\sigma + 2\rho}dx^+dx^- + r_0^2\e^{2\sigma}d\Omega^2\ .
\ee
It is convenient to make the following change: 
$\partial_r = \e^{\rho+\sigma}\partial_l$. 
Using the new radial coordinate $l$,
the metric (\ref{CCmetric}) is rewritten as follows:
\be
\label{CClmetric}
ds^2= - f(l)dt^2 + dl^2 + r(l)^2 d\Omega^2 \ ,\quad 
f(l)=\e^{2\sigma+2\rho} \ ,\quad 
r(l)=r_0\e^\sigma\ .
\ee
Here $f(l)$ is called a redshift function and $r(l)$ is a shape
function. If $f(l)$ and $r(l)$ are smooth positive-definite
functions, which satisfy
the conditions:
\bea
\label{CCcond}
&f(l)\rightarrow 1,\  r(l)\rightarrow l \ & \mbox{when}\
|l|\rightarrow \infty \nn
&f(l),\ r(l) \rightarrow \ \mbox{finite} & \mbox{when}\
|l|\rightarrow 0\ ,
\eea
the metric expresses the wormhole which connects two asymptotically
flat universes.

In the choice of the metric (\ref{CClmetric}),
the equations of motion contain the 6th order
derivatives of $\rho$, $\partial_l^6\rho$ and the 4th order 
derivatives of $\sigma$.
Therefore we need to impose the
initial conditions including 5th order derivative of $\rho$ and 3rd
order derivative of $\sigma$, say:
\be
\label{CCinicon}
\partial_l\rho |_{l=0}=\partial_l^2\rho |_{l=0}
=\partial_l^3\rho |_{l=0}=\partial_l^4\rho |_{l=0}
=\partial_l^5\rho |_{l=0}
=\partial_l\sigma |_{l=0}=\partial_l^2\sigma |_{l=0}
=\partial_l^3\sigma |_{l=0}=0\ .
\ee
We should also note that all the equations of motion 
are not the dynamical equations of
motion but the combinations with the constraint.
One can prove numerically  that at least for some
initial conditions GUTs at the early Universe may
help in producing of primordial wormholes. It is of course the
open question which particular initial conditions lead to more stable
configuration.
Moreover, due to complicated structure of field equations
and initial conditions itself we cannot classify
from the very beginning the initial conditions as supporting (or not)
wormholes production. Nevertheless for any specific GUT
under discussion above study may be easily repeated
and principal possibility of wormholes inducing may be shown 
at least numerically.

In summary, we discussed the applications of the effective action in
the construction of consistent non-singular Universe induced by 
quantum effects. The examples of such cosmologies even with non-trivial 
dilaton have been indeed presented. Note that the presence of dilaton 
could be consistent with observations as it quickly decays to zero. 
Other cosmological solutions may be constructed in the similar fashion.
In order to understand their relevance to observational data one should study 
their properties in more detail.

\section{Dilatonic gravity and Anti-de Sitter/Conformal Field 
Theory correspondence}

In the present Chapter we discuss the properties of classical 
five dimensional dilatonic gravity (IIB supergravity bosonic 
sector), in particularly, its classical solutions. Using bulk-boundary
correspondence this study may help 
in understanding of strong coupling limit of quantum gauge 
field theory. Hence, volume classical physics turns out to describe 
simultaneously surface quantum physics.
 Note that there exists an excellent review on 
AdS/CFT correspondence \cite{AGMOO} so we concentrate mainly 
on issues where the presence of dilaton is important as it 
changes the properties of theory drastically. 

\subsection{Axion-dilatonic conformal anomaly from AdS/CFT 
correspondence}

In this section, we show how one can find large $N$ conformal 
anomaly with dilaton and axion from 5d AdS dilatonic gravity.
First of all, we remind few simple facts about anti-de Sitter 
space. 
$D$ dimensional anti-de Sitter space can be realized by 
imposing a constraints on $D+1$ coordinates:
$- x_0^2 + x_1^2 + \cdots x_{D-1}^2 - x_D^2= -L^2 $.
 From this realization, it is easy to see that this space 
has $SO(D-1,2)$ symmetry as isometry. 
The algebra of $SO(D-1,2)$ symmetry is  the same as the algebra 
of conformal transformations acting on $D-1$ dimensional 
Minkowski space. 

In an adequate coordinate choice (actually, there are many 
different coordinate choices), the metric on the 
$D$-dimensional anti-de Sitter space is given by
\be
\label{VII3}
ds_{\rm AdS}^2=\rho^{-2}d\rho^2 + \rho^{-1}\sum_{i,j=0}^{D-2}\eta_{ij}
dx^idx^j\ .
\ee
Here $\eta$ is the metric on the flat $D-1$-dimensional Minkowski 
space. We should note that there is a boundary when $\rho$ 
vanishes. The topology of the boundary is almost that of the 
$D-1$ dimensional Minkowski space, or more exactly, 
Minkowski space with a point at infinity, that is, 
topologically compactified Minkowski space.
On the boundary manifold, $SO(D-1,2)$ acts exactly as usual 
conformal transformation. When we consider the surface with 
fixed finite $\rho$, there is a correction proportional $\rho$ 
only in the conformal boost but the correction vanishes just on 
the boundary, that is, in the limit that $\rho$ vanishes.

AdS/CFT correspondence (or bulk/boundary) 
is conjectured in ref.\cite{Mld1}. 
When $N$ p-branes in superstring theory or so-called M-theory 
coincide with each other and the coupling constant is small, 
the classical supergravity on AdS${_{D=d+1=p+2}}$, which is 
the low energy effective theory of superstring or M-theory, 
is, in some sense, dual to large $N$ conformal field theory 
on $M^d$, which is the boundary of the AdS.
For example, $d=2$ case corresponds to (4,4) 
superconformal field theory, 
$d=4$ case corresponds to $U(N)$ or $SU(N)$ ${\cal N}=4$ 
super Yang Mills theory and $d=6$ case to (0,2) 
superconformal field theory. 

The conjecture tells that partition function in 
$d$-dimensional conformal field theory is given in terms 
of the classical action in $d+1$-dimensional gravity theory:
\be
\label{VII4}
Z_d(\phi_0)=\e^{-S_{\rm AdS}\left(\phi^{\rm classical}
(\phi_0)\right)}\ .
\ee
Here $\phi_0$ is the value of the field $\phi$ on the boundary and 
$\phi^{\rm classical}(\phi_0)$ is a field on bulk background, 
which is AdS, given by solving the equations of motion with 
the boundary value $\phi_0$ on $M^d$.
$S_{\rm AdS}\left(\phi^{\rm classical}(\phi_0)\right)$ 
is the classical gravity action on AdS. 
When we substitute the classical solution into the action, 
the action, in general, contains infrared divergences 
coming from the infinite volume of AdS.
Then we need to regularize the infrared divergence. 
It is known that as  
 a result of the regularization and the renormalization
there often appear anomalies.  In ref.\cite{Wtt1}
Witten made the proposal how to calculate the conformal 
anomaly( quantum object) from classical gravity (bulk) side. This proposal
has been 
worked out in detail in ref.\cite{HS}
(for Einstein gravity) where it was shown 
that the conformal (Weyl) anomaly may be recovered after 
regularizing the above infrared divergence (from bulk side-
Einstein theory). The usual result for 
conformal anomaly of boundary QFT thus may be reproduced.
This is a kind of IR-UV duality.

In \cite{NOa}, 
the conformal anomaly from 5d dilatonic gravity 
was investigated. 
Let us briefly describe this calculation. 
We start from the action of $d+1$-dimensional 
dilatonic gravity with boundary terms
\bea
\label{Vib}
S&=&\left.{1 \over 16\pi G}\right[\int_{M_{d+1}} d^{d+1}x \sqrt{-\hat G}
\left\{ \hat R + X(\phi)(\hat\nabla\phi)^2 + Y(\phi)\hat\Delta\phi
+ 4\lambda^2  \right\} \nn 
&& \left. +\int_{M_d} d^dx\sqrt{-\hat g}
\left(2\hat\nabla_\mu n^\mu + \alpha\right)  \right]\ .
\eea
Here $M_{d+1}$ is $d+1$ dimensional manifold, which is 
identified with AdS${_{d+1}}$ and $n_\mu$ is the unit 
vector normal to the boundary manifold $M_d$. 
$\alpha$ is a parameter which is chosen properly. 
The boundary terms play a role for cancellation of  
the leading infrared divergences which guarantee 
the system depends only on the boundary value. 
In Eq.(\ref{Vib}) $X(\phi)$ and $Y(\phi)$ 
are arbitrary functions depending on dilaton $\phi$. Note that 
the arbitrariness of $X(\phi)$ and $Y(\phi)$ can be absorbed into 
the redefinition of the dilaton. In fact, if we define new dilaton 
field $\varphi$ by
\be
\label{Vxviii}
\varphi\equiv\int d\phi\sqrt{2V(\phi)}\ ,\ \ 
V(\phi)\equiv X(\phi)-Y'(\phi)\ ,
\ee
the action (\ref{Vib}) can be rewritten by using partial integration 
as follows:
\be
\label{Vibb}
S={1 \over 16\pi G}\left[\int d^{d+1}x \sqrt{-\hat G}
\left\{ \hat R + 2(\hat\nabla\varphi)^2 + 4\lambda^2  \right\}
 +\int_{M_d} d^dx\sqrt{-\hat g}
\left(2\hat\nabla_\mu n^\mu 
+ \alpha + Y(\phi)n^\mu\partial_\mu\phi \right) \right]\ .
\ee
The $\phi$ dependent term on $M_d$ does not finally contribute to 
Weyl anomaly. We keep, however, $X(\phi)$ and $Y(\phi)$ as 
arbitrary functions for the later convenience. Note also that 
boundary term may be used to present the action as the functional 
of fields and their first derivatives \cite{GH}.

As in \cite{FG}, we choose the metric $\hat G_{\mu\nu}$ on $M_{d+1}$ 
and the metric $\hat g_{\mu\nu}$ on $M_d$ in the following form:
\be
\label{Viib}
ds^2\equiv\hat G_{\mu\nu}dx^\mu dx^\nu 
= {l^2 \over 4}\rho^{-2}d\rho d\rho + \sum_{i=1}^d
\hat g_{ij}dx^i dx^j \ ,\quad 
\hat g_{ij}=\rho^{-1}g_{ij}\ .
\ee
Here $l$ is related with $\lambda^2$ by $4\lambda^2=-d(d-1)/l^2$.
Note that the expression of the metric (\ref{Viib}) has a redundancy.
In fact, the expression (\ref{Viib}) is invariant if we change $\rho$ 
and $g_{ij}$ by 
\be
\label{Vwtr}
\delta\rho= \delta\sigma\rho\ ,\ \ 
\delta g_{ij}= \delta\sigma g_{ij}\ .
\ee
Here $\delta\sigma$ is a constant parameter of the transformation.
The transformation (\ref{Vwtr}) can be regarded as the scale 
transformation on $M_d$.

When $d$ is even, we can expand $\phi$ and $g_{ij}$ 
as power series of $\rho$:
\bea
\label{Vivb}
\phi&=&\phi_{(0)}+\rho\phi_{(1)}+\rho^2\phi_{(2)}
+\cdots \rho^{d \over 2}\phi_{(d/2)}
  - \rho^{d \over 2}\ln\rho \psi + {\cal O}(\rho^{{d \over 2}+1}) \\
g_{ij}&=&g_{(0)ij}+\rho g_{(1)ij}+\rho^2 g_{(2)ij}+\cdots 
+\rho^{d \over 2}g_{(d/2)ij}-\rho^{d \over 2}\ln\rho h_{ij}
+ {\cal O}(\rho^{{d \over 2}+1}) \ . \nonumber
\eea
Here we regard $\phi_{(0)}$ and $g_{(0)ij}$ as independent fields on 
$M_d$ and $\phi_{(l)}$, $g_{(l)ij}$ ($l=1,2,\cdots$), $\psi$ and 
$h_{ij}$ as fields depending on $\phi_{(0)}$ and $g_{(0)ij}$ by 
using equations of motion. Then the action (\ref{Vibb}) diverges 
in general since the action contains the infinite volume integration 
on $M_{d+1}$. The action is regularized by introducing the 
infrared cutoff $\epsilon$\footnote{
From the viewpoint of the supergravity, the cutoff $\epsilon$ 
can be regarded as infrared (IR) cutoff regularizing the infinte 
volume of $M_{d+1}$, which can be AdS$_{d+1}$. From the 
viewpoint of the $d$-dimensional field theory, the cutoff $\epsilon$ 
corresponds to the ultraviolet (UV) cutoff as in the present 
calculation of conformal anomaly. This is nothing but the IR/UV 
duality in AdS/CFT correspondence.
} and replacing 
\be
\label{Vvib}
\int d^{d+1}x\rightarrow \int d^dx\int_\epsilon d\rho \ ,\ \ 
\int_{M_d} d^d x\Bigl(\cdots\Bigr)\rightarrow 
\int d^d x\left.\Bigl(\cdots\Bigr)\right|_{\rho=\epsilon}\ .
\ee
The terms proportional to the (inverse) 
power of $\epsilon$ in the regularized action are invariant under 
the scale transformation 
\be
\label{Vvia}
\delta g_{(0)\mu\nu}=2\delta\sigma g_{(0)\mu\nu}\ ,\ \  
\delta\epsilon=2\delta\sigma\epsilon \ , 
\ee
which corresponds to (\ref{Vwtr}).  Then the subtraction of these 
terms proportional to the inverse power of $\epsilon$ does not break 
the invariance under the scale transformation. When $d$ is even, 
however, the term proportional to $\ln\epsilon$ appears. The term 
is not invariant under the scale transformation (\ref{Vvia}) and the 
subtraction of the $\ln\epsilon$ term breaks the invariance. 
The variation of the $\ln\epsilon$ term under the scale 
transformation (\ref{Vvia}) is finite when $\epsilon\rightarrow 0$ 
and should be canceled by the variation of the finite term (which 
does not depend on $\epsilon$) in the action since the original 
action (\ref{Vibb}) is invariant under the scale transformation. 
Therefore the $\ln\epsilon$ term $S_{\rm ln}$ gives the Weyl 
anomaly $T$ of the action renormalized by the subtraction of 
the terms which diverge when $\epsilon\rightarrow 0$ 
\be
\label{Vvibb}
S_{\rm ln}=-{1 \over 2}
\int d^4x \sqrt{-g_{(0)}}T\ .
\ee

First we consider the case of $d=2$. Choosing $\alpha$ to satisfy 
the equation $\alpha={2 \over l}$ and  using the replacement 
in (\ref{Vvibb}), we find the action (\ref{Vibb}) has the 
following form
\be
\label{Vviib}
S=-{1 \over 16\pi G}{l \over 2}\ln\epsilon\int d^2x
\sqrt{-g_{(0)}}
\left\{ R_{(0)}+ X(\phi_{(0)})(\nabla\phi_{(0)})^2 
 + Y(\phi_{(0)})\Delta\phi_{(0)} \right\} 
+\ \mbox{finite terms}\ .
\ee
Then we find an expression of the Weyl anomaly $T$ by using 
(\ref{Vvibb})
\be
\label{Vixb}
T={l \over 16\pi G} \left\{ R_{(0)}
+ X(\phi_{(0)})(\nabla\phi_{(0)})^2 
+ Y(\phi_{(0)})\Delta\phi_{(0)}
\right\}\ .
\ee
This result could be compared with the UV-calculation
of the conformal anomaly of dilaton coupled $N$ scalars 
and $M$ Majorana dilaton coupled  spinors 
from chapter 3:
\bea
\label{VII10}
&& {l \over 16\pi G} ={2N + M \over 48\pi} \ ,\quad 
{l \over 16\pi G}X(\phi_{(0)})=-{N \over 4\pi}
\left({f'' \over 2f}- {{f'}^2 \over 4f^2}\right) 
  -{M \over 24\pi}\left({g'' \over g} - {{g'}^2 \over g^2}\right)
\ ,\nn
&& {l \over 16\pi G}Y(\phi_{(0)})=-{N \over 4\pi}{f' \over 2f}
  - {M \over 24\pi}{g' \over g}\ .
\eea
The above result should give the conformal anomaly computed from 
the asymptotic symmetry algebra of AdS$_3$ with dilaton, as it 
was the case in the absence of dilaton \cite{JBH}. 

We now consider the case of four dimensions. 
The calculation similar to that in 2 dimensions leads to the 
term $S_{\rm ln}$ proportional to $\ln\epsilon$ in the action, 
which contains $g_{(0)ij}$, $g_{(1)ij}$, $\phi_{(0)}$, and 
$\phi_{(1)}$. Using the equations of motion given by the 
variation of $g_{(1)ij}$ and $\phi_{(1)}$, $g_{(1)ij}$ and 
$\phi_{(1)}$ can be solved with respect to $g_{(0)ij}$ and 
$\phi_{(0)}$. After deleting $g_{(1)ij}$ and $\phi_{(1)}$, 
we obtain the following expression : 
\bea
\label{Vxix}
&& S_{\rm ln}={l^3 \over 16\pi G}\int d^4x \sqrt{-g_{(0)}} 
\left[ {1 \over 8}R_{(0)ij}R_{(0)}^{ij}
  -{1 \over 24}R_{(0)}^2 \right. 
 + {1 \over 2} R_{(0)}^{ij}\partial_i\varphi_{(0)}
\partial_j\varphi_{(0)}  \\
&& \quad \left. - {1 \over 6} R_{(0)}g_{(0)}^{ij}
\partial_i\varphi_{(0)}\partial_j\varphi_{(0)} + {1 \over 4}
\left\{{1 \over \sqrt{-g_{(0)}}} \partial_i\left(\sqrt{-g_{(0)}}
g_{(0)}^{ij}\partial_j\varphi_{(0)} \right)\right\}^2 + {1 \over 3}
\left(g_{(0)}^{ij}\partial_i\varphi_{(0)}\partial_j\varphi_{(0)} 
\right)^2 \right]\ .\nonumber
\eea
Here we choose $\alpha$ to be $\alpha={6 \over l}$ and 
using $V(\phi_{(0)})$, which is defined in (\ref{Vxviii}),  
the field $\varphi_{(0)}$ is defined by 
$\varphi_{(0)}\equiv\int d\phi_{(0)}\sqrt{2V(\phi_{(0)})}$ 
in a similar way to (\ref{Vxviii}). 

The Weyl anomaly coming from the multiplets of ${\cal N}=4$ 
supersymmetric $U(N)$ or $SU(N)$ Yang-Mills conformally 
coupled with ${\cal N}=4$ 
conformal supergravity was calculated in 
\cite{LT1}:
\bea
\label{Vxxi}
T&=&-{N^2 \over 4(4\pi)^2}\left[2\left(R_{ij}R^{ij}
 -{1 \over 3}R^2\right)+F^{ij}F_{ij} \right. \\
&& \left. + 4\left\{ 2\left( R^{ij} - {1 \over 3} Rg^{ij}\right)
\partial_i\varphi^*\partial_j\varphi  
+\left|{1 \over \sqrt{-g}} \partial_i\left(\sqrt{-g}
g^{ij}\partial_j\varphi \right)\right|^2 \right\} + \cdots \right]\ .
\nonumber
\eea
Here $F_{ij}$ is the field strength of SU(4) gauge fields, $\varphi$ 
is a complex scalar field which is a combination of dilaton and RR 
scalar (or axion) and ``$\cdots$'' expresses the terms containing other
fields 
in ${\cal N}=4$ conformal supergravity multiplet and higher powers 
of the fields.

If we choose
\be
\label{Vxx}
{l^3 \over 16\pi G}={2N^2 \over (4\pi)^2}\ ,
\ee
and consider the background where only gravity and the real part 
of the scalar field $\varphi$ in the ${\cal N}=4$ conformal 
supergravity multiplet are non-trivial and other fields vanish 
in (\ref{Vxxi}), Eq.(\ref{Vxix}) exactly reproduces the result 
in (\ref{Vxxi}) by using (\ref{Vvibb}). It is interesting that last 
term in (\ref{Vxxi}) actually gives the correct dilatonic 
dependent contribution to conformal anomaly of dilaton coupled 
electromagnetic field. It corrects the failed result obtained in 
section 3.5 by convenient methods. 

Hence, we got $d=2$ and $d=4$ holographic conformal anomaly for dilaton
coupled theories from supergravity side. The results of this study 
give further check of AdS/CFT correspondence in the presence of 
dilaton. We should note that it has been also found agreement 
\cite{BNG,NO9} between AdS/CFT conformal anomaly and perturbative 
QFT anomaly in ${\cal N}=2$ SCFT even in the next to leading order 
term. 

In \cite{NOOSY1}, it was investigated 5d dilaton-axionic AdS 
gravity, whose action is actually motivated by bosonic sector of IIB 
SG model due to Gibbons, Green and Perry\cite{GGP} which contains 
axion. 
Such investigation is necessary in order to check 
the possibility that the fields besides the graviton and 
dilaton , say axion field, could not spoil 
AdS/CFT correspondence. 
Therefore it would be important that background axion and dilaton 
are kept in the calculation of conformal anomaly.
The starting action in \cite{NOOSY1} is given by
\begin{eqnarray}
\label{Voi}
S = \frac{1}{16\pi G} \left\{ \int _{M_{d+1}} d^{d+1}x \sqrt{-\hat{G}}
({\hat R} + X(\phi ,\chi )(\hat{\nabla } \phi) ^{2} + Y(\phi ,\chi )
\hat{\Delta}\phi \right. &&  \nonumber \\
+Z(\phi ,\chi )(\hat{\nabla } \chi) ^{2} + W(\phi ,\chi )
\hat{\Delta}\chi +4\lambda ^{2} ) 
\left.+\int _{M_{d}}d^{d}x\sqrt{-\hat{g}}
(2\hat{\nabla}_{\mu}n^{\mu}+\alpha) \right \} .&&
\end{eqnarray}
Here $\chi$ is axion (or RR-scalar). 
The bosonic action of IIB SG in \cite{GGP} represents the special 
case of the above action. If we define 
\be
\label{VNi}
\left(\begin{array}{c}X^1 \\ X^2 \\ \end{array}\right)
\equiv \left(\begin{array}{c}\phi \\ \chi \\ \end{array}\right)
\ ,\quad h_{\mu\nu}\equiv \left(\begin{array}{cc}
X -{\partial Y \over \partial\phi} &-{1 \over 2}\left(
{\partial W \over \partial \phi} + 
{\partial Y \over \partial\chi}\right)  \\ 
  -{1 \over 2}\left(
{\partial W \over \partial \phi} + 
{\partial Y \over \partial\chi}\right)
& Z - {\partial W \over \partial\chi} \\ \end{array}\right)\ ,
\ee
the action (\ref{Voi}) can be rewritten, after partial integration, 
in the form of the sigma model, whose target space coordinates are 
$\phi$ and $\chi$. After explicit calculations, similar to pure 
dilaton case above we obtain the following final 
expression for $S_{\ln}$, which is relevant to the anomaly, 
\bea
\label{VNviii}
\lefteqn{S_{\ln}={1 \over 16\pi G}\int d^4 x \sqrt{-g}
\left[l^3\left(-{1 \over 24}R^2 
+ {1 \over 8}R_{ij}R^{ij}\right) + {l^3 \over 4}R^{ij}h_{\mu\nu}
\partial_i X^\mu \partial_j X^\nu\right. } \\
&&  - {l^3 \over 12}R h_{\mu\nu} 
\left(\partial X^\mu\cdot \partial X^\nu\right) 
 -{l^3 \over 24}\left\{ h_{\mu\nu} \left(
\partial X^\mu\cdot \partial X^\nu\right)\right\}^2
+{l^3 \over 8}h_{\mu\nu}h_{\rho\sigma}
\left(\partial X^\mu\cdot \partial X^\rho\right)
\left(\partial X^\nu\cdot \partial X^\sigma\right) \nn
&& \left. + {l^8 \over 8}h_{\mu\nu}
\left\{\triangle X^\mu + \Gamma^\mu_{\rho\sigma}
\left(\partial X^\rho\cdot \partial X^\sigma\right)
\right\}
\left\{\triangle X^\nu + \Gamma^\nu_{\tau\eta}
\left(\partial X^\tau\cdot \partial X^\eta\right)
\right\}\right] \ .\nonumber
\eea
Here $\left(\partial X^\mu\cdot \partial X^\nu\right)
\equiv g^{ij}\partial_i X^\mu \partial_j X^\nu$ 
and $\Gamma^\mu_{\nu\rho}$ is a connection on the 
target manifold:
$\Gamma^\mu_{\nu\rho}={1 \over 2} h^{\mu\tau}\left(h_{\nu\tau,\rho}
+ h_{\rho\tau,\nu} - h_{\nu\rho,\tau}\right)$.
If we put 
\be
\label{VNix}
h_{\mu\nu}=\left(\begin{array}{cc}
2 & 0 \\ 0 & c \\ \end{array} \right)\ , \ 
(c\ \mbox{is an arbitrary constant}),\ 
\quad X^2=0
\ee
the previous result in \cite{NOa} (see Eq.(\ref{Vxix})) 
for dilatonic gravity can be reproduced. 

Furthermore, if we choose the action as motivated by the 
bosonic sector of type IIB supergravity with RR-scalar \cite{GGP}
\begin{eqnarray}
\label{Vov}
S = \frac{1}{16\pi G} \int _{M_{d+1}} 
d^{d+1}x \sqrt{-\hat{G}}
\left( {\hat R} + \frac{1}{2} (\hat{\nabla } \phi)^{2} 
+ \frac{1}{2}\e^{2\phi}(\hat{\nabla } \chi)^{2} \right). 
\end{eqnarray}
the result in (\ref{Vxxi}) can be reproduced if 
we choose
\be
\label{VxxA}
{l^3 \over 16\pi G}={2N^2 \over (4\pi)^2}\ , \quad 
\varphi = \phi + \e^\phi \chi\ , \quad 
\varphi^* = -\phi + \e^\phi \chi
\ee
and consider the background where only gravity and the complex
scalar field $\varphi$ in the ${\cal N}=4$ conformal supergravity 
multiplet are non-trivial and other fields vanish in (\ref{Vxxi}).

Thus, we presented the way to calculate conformal anomaly 
depending not only from gravitational field but also from 
other fields via bulk/boundary correspondence. 
It would be interesting to study the universal structure of 
holographic conformal anomaly in the presence of dilaton in the 
same fashion as it is done in ref.\cite{ISTY1}. 

\subsection{Running gauge coupling and quark-antiquark potential  
 from IIB supergravity} 

AdS/CFT correspondence \cite{Mld1,Wtt1,GKP,AGMOO} 
may also provide new insights to 
the understanding of non-perturbative QCD. For example, 
in frames of Type 0 string theory the attempts
\cite{KT1,KT2,KT3,Mnh1,FM1,AFS1,Ur} have 
been done to reproduce such well-known QCD effects as 
running gauge coupling and possibly confinement. 
It is among the first problems to get the description of 
well-known QCD phenomena from bulk/boundary correspondence. 

In another approach one can consider IIB supergravity vacuum which 
describes the strong coupling regime of a (non)-SUSY gauge theory. 
This can be achieved by the consideration of deformed IIB SG 
vacuums, for example, with non-constant dilaton which presumably breaks the 
conformal invariance and supersymmetry of boundary super YM. 
Such background will be the perturbation of 
${\rm AdS}_5\times{\rm S}_5$ vacuum. The background of such sort 
(with non-trivial dilaton) which interpolates between AdS (UV) 
and flat space with singular dilaton (IR) has been found in 
ref.\cite{NO10}.

This solution of IIB SG \cite{NO10} has been used in ref.\cite{KS1p} 
with the interpretation of it as the one describing the running 
gauge coupling (via exponent of dilaton). 
 QCD-like  properties of such and similar backgrounds have been discussed 
in detail in refs.\cite{GPPZ1,NO11,MPR1,Ghoroku,DZ1,KLM1,BL1}. 
Modifications of IIB SG solution with 
non-constant dilaton \cite{NO10} due to presence of 
constant self-dual vector\cite{LT2} or world volume 
scalar \cite{CM1} 
give further indication on the possible confinement and asymptotic 
freedom of the boundary  gauge theory. 
Unfortunately, situation is very complicated here due to 
double role of IIB SG backgrounds. From one side they may indeed correspond 
to IR gauge theory (deformation of initial SUSY YM theory). 
On the same time such background may simply describe another vacuum of 
the same maximally supersymmetric YM theory with non-zero VEV of some
operator.
Due to fact that operators corresponding to deformation to another gauge
theory
are not known, it is unclear what is the case under discussion
(interpretation 
of SG background). Only some indirect arguments as below may be given. 
As we see these arguments indicate that IIB SG background
 discussed in this section most probably correspond to another vacuum 
of super YM theory under consideration. Then RG flow is induced in the theory 
via giving a nonzero VEV to some operator.

In this section, we review the set of classical 
solutions in IIB supergravity and their properties typical 
for QCD-like boundary gauge theory \cite{NOc}.  

Let us make few remarks on AdS/CFT interpretation of IIB SG 
background.
Choosing the coordinates in the asymptotically ${\rm AdS}_5$ 
spacetime as 
\be
\label{I1b}
ds^2=d\sigma^2 + S(\sigma)\sum_{i,j=0}^{3}\eta_{ij}dx^i dx^j\ ,
\ee
one sees that scalar field $\lambda$, e.g. dilaton, axion or other 
fields, obey the following equation:
${d^2\lambda \over d\sigma^2} + 4 {d\lambda \over d\sigma}=
M^2 \lambda$. Here $M^2$ is the ``mass'' of $\lambda$ and 
$\sigma\rightarrow 0$ corresponds to the boundary of AdS. 
Then $\lambda$ is associated with the operator ${\cal O}_\lambda$ 
with conformal dimension $\Delta = 2+\sqrt{4 + M^2}$. 
The solution of the equation is given by
\be
\label{I3}
\lambda = A \e^{-(4-\Delta)\sigma} + B \e^{-\Delta\sigma}\ .
\ee
The solution corresponding to $A$ is not normalizable but 
the solution to $B$ is normalizable. According to the 
argument in \cite{GPPZ}, the non-normalizable solution is 
associated with the deformation of the ${\cal N}=4$ theory 
by ${\cal O}_\lambda$ but the normalizable solution is associated 
with a different vacuum where  ${\cal O}_\lambda$ has a non-zero 
vacuum expectation value. The behavior of the dilaton  
 is normalizable and seems to be associated with the 
dimension 4 operator, say ${\rm tr} F^2$. Then the arguments in 
 \cite{GPPZ} would indicate that such solution  
 should correspond to another vacuum of 
${\cal N}=4$ theory. Nevertheless,  there might be still 
possibility that 
the solution corresponds to non-SUSY gauge theory. 
Since there occurs the condensation of ${\rm tr} F^2$ 
in the usual non-supersymmetric QCD, in any case, the solution 
given here describes some features typical for  
the non-supersymmetric theory.

The situation is even more complicated due to limits of 
validity of dual SG description.
In order that the classical supergravity description is valid, 
the curvature should be small and the string coupling should be 
also small. If the curvature is large, the $\alpha'$ 
corrections from string theory would appear. In the AdS/CFT 
correspondence, the radius  $R_s$ of the curvature is given by 
$R_s = \left( 4\pi g_s N\right)^{1 \over 4}$. 
Here $g_s$ is the string coupling and $N$ is the number of the 
coincident D-branes. Therefore we should require $g_s N \gg 1$. 
On the other hand, the classical picture works when the string 
coupling is small: $g_s \ll 1$. 
In the solution given here, there appears the curvature 
singularity and $g_s$ depends on the coordinates since the 
dilaton is non-trivial. If we concentrate on the behavior near the 
boundary, which is asymptotically AdS and is far from the 
singularity, the solution is reliable and SG description 
may be trusted.

We start from the following action of dilatonic gravity
in $d+1$ dimensions:
\be
\label{ViR}
S=-{1 \over 16\pi G}\int d^{d+1}x \sqrt{-G}\left(R - \Lambda 
  - \alpha G^{\mu\nu}\partial_\mu \phi \partial_\nu \phi \right)\ .
\ee
This action lies in the class of dilatonic gravity models 
under discussion in this review. 
In the following, we assume $\lambda^2\equiv -\Lambda$ 
and $\alpha$ to be positive. The action (\ref{ViR}) is very general. 
It contains the effective action of type IIB string theory. The 
type IIB supergravity, which is the low energy effective action 
of the type IIB string theory, has a vacuum with only non-zero 
metric and the anti-self-dual five-form. The latter is given by 
the Freund-Rubin-type ansatz:
\be
\label{ViiR}
F_{\mu\nu\rho\kappa\lambda}=-{\sqrt{\Lambda} \over 2}
\epsilon_{\mu\nu\rho\kappa\lambda}\ (\mu,\nu,\cdots=0,1,\cdots,4), 
\quad F_{ijkpq}=-{\sqrt{\Lambda} \over 2}
\epsilon_{ijkpq}\ (i,j,\cdots=5,\cdots,9) \ .
\ee
The vacuum has the topology of ${\rm AdS}_5\times {\rm S}^5$. 
Since ${\rm AdS}_5$ has a four dimensional Minkowski space as a 
subspace, especially on its boundary, ${\rm AdS}_5$ has the 
four dimensional Poincar\'e symmetry $ISO(1,3)$. 
${\rm S}^5$ has, of course, $SO(6)$ symmetry. 

As an extension, we can consider the solution where the dilaton is 
non-trivial but the anti-self-dual five-form is the same as in 
(\ref{ViiR}). Furthermore if we require the solution has the symmetry 
of $ISO(1,3)\times SO(6)$, the metric should have the 
following form as in ref.\cite{NO10}:
\be
\label{ViiiB}
ds^2=G_{\mu\nu}dx^\mu dx^\nu + g_{mn}dx^m dx^n\ ,\quad 
G_{\mu\nu}dx^\mu dx^\nu
=f(y)dy^2 + y \sum_{i,j=0}^{d-1}\eta_{ij}dx^i dx^j \ .
\ee
where $g_{mn}$ is the metric of ${\rm S}^5$.
In order to keep the symmetry of $ISO(1,3)\times SO(6)$, the 
dilaton field $\phi$ can only depend on $y$. Then by integrating 
five coordinates on ${\rm S}^5$, we obtain the effective five 
dimensional theory, which corresponds to $d=4$ and 
$\alpha={1 \over 2}$ case in (\ref{ViR}). We keep working with above 
dilatonic gravity as it will be easy to come to IIB supergravity 
($d=4$, $\alpha={1 \over 2}$) at any step.

The equations of motion given by the variation of the action 
(\ref{ViR}) with respect to the metric $G^{\mu\nu}$ and the dilaton 
$\phi$ can be exactly solved if we assume that $\phi$ depends 
only on one of the coordinate, say $y\equiv x^d$ as in type IIB 
supergravity solution with the symmetry of $ISO(1,3)\times SO(6)$ 
and we can also consider, as a generalization of (\ref{ViiiB}), 
the case that $\eta_{ij}$ in (\ref{ViiiB}) is replaced by $g_{ij}$, 
which is the metric in the Einstein manifold defined by 
 $r_{ij}=kg_{ij}$. 
Here $r_{ij}$ is the Ricci tensor given by $g_{ij}$ and $k$ is 
a constant, especially $k>0$ for sphere and $k=0$ for the flat 
Minkowski space and $k<0$ for hyperboloid. 
The case of $k=1$ is especially interesting as it corresponds 
to gauge theory in de Sitter (inflationary) Universe. 
The solution is given by 
\be
\label{Vxt}
f={d(d-1)  \over 4y^2
\lambda^2 \left(1 + {\alpha c^2 \over \lambda^2 y^d}
+ {kd \over \lambda^2 y}\right)} \ ,\quad 
\phi=c\int dy \sqrt{{d(d-1) \over 
4y^{d+2}\lambda^2 \left(1 + {\alpha c^2 \over \lambda^2 y^d}
+ {kd \over \lambda^2 y}\right)}}\ .
\ee
 From the behavior of $f(y)$, we find that there is a curvature 
singularity at $y=0$. 
The curvature singularity would be generated by the singular 
behavior of the dilaton $\phi$ when $y\sim 0$. 
The curvature singularity tells there should appear 
the $\alpha'$ correction from the string theory 
and the supergravity description would break down 
when $y\sim 0$. Conversely and hopefully, the curvature singularity 
might be apparent and vanish when we can include full string 
corrections. In any case, the dual interpretation of the solution can be 
trusted if we 
investigate the behavior near the boundary ($y\rightarrow +\infty$).

If we change the coordinate $y$ by $\rho$ in (\ref{Vxt}), 
which is defined by
\be
\label{Vxiiitb}
\rho\equiv -\int dy \sqrt{f(y) \over y}
=-\int dy \sqrt{d(d-1)  \over 4y^3
\lambda^2 \left(1 + {\alpha c^2 \over \lambda^2 y^d}
+ {kd \over \lambda^2 y}\right)}\ ,
\ee
the metric has the following form
\be
\label{Vivtbb}
G_{\mu\nu}dx^\mu dx^\nu
=\Omega^2(\rho)\left(d\rho^2 + \sum_{i,j=0}^{d-1}g_{ij}dx^i dx^j
\right) \ .
\ee
Here $\Omega^2(\rho)$ is given by solving $y$ in (\ref{Vxiiitb}) 
with respect to $\rho$: $\Omega^2(\rho)=y(\rho)$.
When $\rho$ is small, $y$ is large and the structure of the 
spacetime becomes AdS asymptotically. 
 From (\ref{Vxiiitb}), we find 
$\rho={\sqrt{d(d-1)} \over \lambda y^{1 \over 2}}\left(1
+{\cal O}\left(y^{-1}\right)\right)$. 
Therefore we find
\be
\label{VoII}
\Omega^2(\rho)=y(\rho)={R_s^2 \over \rho^2}\left(1 
+ {\cal O}\left(\rho^2\right)\right)\ ,\quad 
 R_s\equiv {\sqrt{d(d-1)} \over \lambda }\ .
\ee
 The ${\rm AdS}_5$ part in the solution has the form 
(properly normalized) 
\be
\label{VoIIb}
ds_{{\rm AdS}_5}^2=\left(4\pi 
g_s N\right)^{1 \over 2}\cdot{1 \over \rho^2}
\left(d\rho^2 + \sum_{i,j=0}^{d-1}\eta_{ij}dx^i dx^j
\right) \ .
\ee
Therefore we find 
\be
\label{VoIIc}
R_s=\left(4\pi g_s N\right)^{1 \over 4}\ ,
\ee
where $g_s$ is the string coupling and $N$ is the flux of the 
five-form $F$ in (\ref{ViiR}) through ${\rm S}^5$, which is produced 
by the $N$ coincident D3-branes. 

It is interesting  that there are many  
Einstein manifolds which solve the model. The Einstein equations are given by,
\be
\label{VA1}
R_{\mu\nu}-{1 \over 2}g_{\mu\nu}R+{1 \over 2}\Lambda g_{\mu\nu}
= T^{\rm matter}_{\mu\nu}\ .
\ee
Here $T^{\rm matter}_{\mu\nu}$ is the energy-momentum tensor of 
the matter fields. If we consider the vacuum solution where 
$T^{\rm matter}_{\mu\nu}=0$, Eq.(\ref{VA1}) can be rewritten as
$R_{\mu\nu}={\Lambda \over 2}g_{\mu\nu}$. 
If we put $\Lambda=2k$, this is nothing but Eq. for 
the Einstein manifold. The Einstein manifolds  
are not always homogeneous manifolds like flat Minkowski, 
(anti-)de Sitter space or Nariai space but they can be some 
black hole solutions like Schwarzschild black hole, 
\be
\label{Vschw}
ds_4^2\equiv \sum_{\mu,\nu=0}^3 g_{\mu\nu}dx^\mu dx^\nu
=-\left(1 - {r_0 \over r}\right)dt^2
+{dr^2 \over \left(1 - {r_0 \over r}\right)} + r^2 d\Omega^2\ ,
\ee
or Kerr one for $k=0$\footnote{This type of solutions for $k=0$ case 
has been considered in ref. \cite{Bur}} or Schwarzschild 
(anti-) de Sitter black hole for $k\neq 0$. 
In these solutions, the curvature singularity
at $r=0$ has a form of line penetrating ${\rm AdS}_5$ and 
the horizon makes a tube surrounding the singularity. 
This configuration seems to express D-string whose boundary 
lies on the boundary of  ${\rm AdS}_5$ or possibly D3-brane. 
Especially in case of Kerr or Kerr-(anti-)de Sitter solution, 
the object corresponding to the singularity has an angular 
momentum. 

Consider the running of the gauge coupling, which is given 
by the dilaton field $\phi$ due to AdS/CFT correspondence.   
We  assume the gauge coupling has the following form 
\be
\label{Vrii}
g=g_s\e^{2\beta\sqrt{\alpha \over d(d-1)}
\left(\phi-\phi_0\right)}
=g_s\left\{ 1 - {2\beta c\sqrt{\alpha} \over d\lambda}
y^{-{d \over 2}} + {kd\beta c\sqrt{\alpha} \over (d+2)\lambda^3}
y^{-{d \over 2}-1} + \cdots \right\}\ .
\ee
In case of type IIB supergravity ($\alpha={1 \over 2}$), 
$\beta=\sqrt{d(d-1) \over 2}$. 
If we define a new coordinate $U$ by $y=U^2$, 
$U$ expresses the scale on the (boundary) $d$ dimensional space
(due to holography \cite{SW,PP1}). Following the correspondence 
between 
long distances/high energy in the AdS/CFT scheme, $U$ can be 
regarded as the energy scale of the boundary field theory. 
Then we obtain the following 
renormalization group equation
\bea
\label{Vriv}
& \circ\ k\neq 0\quad 
\beta(U)&\equiv U{dg \over dU} 
= -d (g-g_s) - {2kd\beta c\sqrt{\alpha} \over (d+2)\lambda^3}
\left({d\lambda \over 2\beta c\sqrt{\alpha} }
\right)^{{2 \over d}+1} {(g - g_s)^{{2 \over d}+1} 
\over {g_s}^{2 \over d}} + \cdots\ \nn
&\circ\ k=0 \quad \beta(U) &= -d (g-g^*) 
  - \left(d - {d^2 \over 2\beta} \right) {(g - g_s)^2 \over g_s}
+ \cdots
\eea
The leading behavior is the same as in refs. 
\cite{KS1p,GPPZ1,NO11,MPR1,LT2,NOc} and tells that 
QFT on the boundary has the non-trivial 
UV-fixed point.

Hence, we found that beta-function explicitly depends on the 
curvature of four-dimensional manifold. Of course, curvature 
dependence is not yet logarithmic as it happens with usual 
QFTs (perturbative consideration) in curved spacetime \cite{BOS1}. 
The power-like running of gauge coupling is much stronger than 
in $k=0$ case. Note that previous discussion of power-like running 
includes GUTs with large internal dimensions 
\cite{TV1,Ant1,Wtt2,Lkk1,DDG1,Bch1}. 

One may also comment about 
the static potential between ``quark'' and 
``anti-quark''\cite{Mld2}. We evaluate the following Nambu-Goto action 
\be
\label{Vrg5}
S={1 \over 2\pi}\int d\tau d\sigma \sqrt{\det\left(g^s_{\mu\nu}
\partial_\alpha x^\mu \partial_\beta x^\nu\right)}\ .
\ee
with the ``string'' metric $g^s_{\mu\nu}$, which 
could be given by multiplying a dilaton function $k(\phi)$ to 
the metric tensor in (\ref{ViiiB}). Especially we choose $k(\phi)$ 
by
\be
\label{Vrg6}
k(\phi)=\e^{2\gamma
\sqrt{\alpha \over d(d-1)}\left(\phi-\phi_0\right)}
= 1 -  {2\gamma c\sqrt{\alpha} \over d\lambda y^{d \over 2}}
+ \cdots \ .
\ee
In case of type IIB supergravity, $\gamma=\beta
=\sqrt{d(d-1) \over 2}$. 
We consider the static configuration $x^0=\tau$, 
$x^1\equiv x=\sigma$, $x^2=x^3=\cdots=x^{d-1}=0$ and $y=y(x)$.  
The orbit of $y$ can be obtained by minimizing the action $S$ 
or solving the Euler-Lagrange equation 
${\delta S \over \delta y}- \partial_x\left({\delta S 
\over \delta\left(\partial_x y\right)}\right)=0$. 
 Substituting the configuration into the action (\ref{Vrg5}), 
we obtain the total energy of the system $E(L)$ by 
$S={T \over 2\pi}E(L)$. Here $T$ is the length of the region 
of the definition of $\tau$. The energy $E(L)$, 
however, contains the divergence due to the self energies of the 
infinitely heavy ``quark'' and ``anti-quark''. The sum of their 
self energies can be estimated by considering the configuration 
$x^0=\tau$, $x^1=x^2=x^3=\cdots=x^{d-1}=0$ and $y=y(\sigma)$ (note 
that $x_1$ vanishes here) 
and the minimum of $y$ at $y_D$ where brane would lie.
Then the finite potential $E_{q\bar q}(L)\equiv
E(L) - E_{\rm self}$ between ``quark'' and ``anti-quark'' is 
given by
\bea
\label{Vrg15}
&&\circ\ k\neq 0\quad E_{q\bar q}(L)
={1 \over A}\left({C_{3 \over 2} \over AL}\right)\left\{D_0 
+ B\left({C_{5 \over 2} D_0  \over C_{3 \over 2}}
+ D_2 \right)
\left({AL \over C_{3 \over 2}}\right)^2 + \cdots \right\} \nn
&& A \equiv{2\lambda \over \sqrt{d(d-1)}} \ ,\quad 
B\equiv -{kd\gamma \over 4\lambda^2} \ , 
\quad C_a\equiv \int_{-\infty}^\infty dt \cosh^{- a}t
= {2^{(a-1)} \Gamma\left({a \over 2}\right)^2 
\over \Gamma(a)} \nn
&& D_d \equiv 2\int_0^\infty dt \cosh^{-{d+1 \over 2}}t\,
\e^{-t} + {4 \over d-1}
={2^{d-3 \over 2} \Gamma\left({d-1 \over 4}\right)^2 
\over \Gamma\left({d-1 \over 2}\right)} \\
&& \circ\ k=0\quad E_{q\bar q}(L)
=\left({C_{3 \over 2} \over AL}\right)\left\{D_0 \pm B_0 
\left(D_d+F_d+{S_d  D_0 \over C_{3 \over 2}} 
\right)\left({AL \over C_{3 \over 2}}\right)^d
+ {\cal O}(L^{2d})\right\} \nn
&& S_d\equiv\int_{-\infty}^\infty dt \cosh^{{1 \over 2}}t 
\sinh^{-2}t \left(1 - \cosh^{- {d \over 2}} t \right),\ 
F_d\equiv \int_{-\infty}^\infty dt \sinh^{-2}t\cosh^{1 \over 2}t
\left(1 - \cosh^{-{d \over 2}}t\right) .\nonumber
\eea
Here $L$ is the distance between ``quark'' and ``anti-quark'' 
and we neglected the $L$ independent terms. 
Note that leading and next-to-leading term 	for $k\neq 0$ 
does not depend on the parameter $\gamma$ in (\ref{Vrg6}). The 
leading behavior is 
 attractive since $D_0 = - 2.39628...$ but we 
should note that next-to-leading term is linear in $L$ for 
$k\neq0$, which does not depend on the dimension $d$. 
Since ${C_{5 \over 2} D_0 
\over C_{3 \over 2}} + D_2 =3.49608>0$ and $B$ is negative if 
$k$ is positive and vice versa. 
Therefore the linear potential term in 
(\ref{Vrg15}) is repulsive if $k>0$ (sphere, i.e. gauge theory 
in de Sitter Universe) and attractive if 
$k<0$ (hyperboloid). 

Of course, the confinement depends on the large $L$ behavior of 
the potential. When $L$ is large, however, the orbit of the string 
would approach to the curvature singularity at $y=0$, where the 
supergravity description would break down. 
Despite of this, there might be interesting to investigate the 
large $L$ behavior. From the behavior of $f(y)$ 
and the dilaton $\phi$ when $y$ is small, we find 
the integrand in the Nambu-Goto action behaves as
\bea
\label{lL1}
&& k\left(\phi(y)\right) y \sqrt{
{f(y) \over y}\left(\partial_x y\right)^2 + 1}
\sim y^{{\rm sgn}(c)\sqrt{d(d-1) \over 2} +1}
\sqrt{ {d(d-1) \over 4\alpha c^2}y^{d-3}
\left(\partial_x y\right)^2 + 1}  \nn 
&& \quad = \sqrt{ {d(d-1) \over 4\alpha c^2\left( 
{\rm sgn}(c)\sqrt{d(d-1) \over 2} + {d+1 \over 2} \right)^2}
\left(\partial_x \tilde U \right)^2
+ \left(\tilde U\right)^{\gamma_0} }\ .
\eea
Here $\tilde U\equiv y^{{\rm sgn}(c)\sqrt{d(d-1) \over 2} 
+ {d+1 \over 2}}$ and 
$\gamma_0\equiv {2{\rm sgn}(c)\sqrt{d(d-1) \over 2} 
+ 2 \over {\rm sgn}(c)\sqrt{d(d-1) \over 2} 
+ {d+1 \over 2}}$. 
When $d=4$, $0<\gamma_0<2$ when $c>0$ and $\gamma<0$ when $c<0$. 
According to the analysis in \cite{GPPZ}, the orbit of the 
string goes straight to the region $y\sim 0$ when $c>0$ 
($0<\gamma_0<2$) and the 
potential becomes independent of $L$. In this case, however, 
the potential would receive the $\alpha'$ correction from 
the string theory. On the other hand, when $c<0$ ($\gamma<0$), 
there is an effective barrier which prevents the orbit of string 
from approaching into the curvature singularity and the potential 
would not receive the $\alpha'$ correction so much and the 
supergravity description would be reliable. Furthermore 
$c<0$ ($\gamma<0$) case predicts the confinement. 
Similarly, one  can  evaluate the potential between monopole and 
anti-monopole by using the Nambu-Goto action for $D$-string 
instead of (\ref{Vrg5}) (cf.ref.\cite{GO1}) and study its behavior.

One can include the axion field $\chi$ into the action of type IIB 
supergravity ($\alpha={1 \over 2}$) in (\ref{ViR}),
following ref.\cite{GGP,Tsy1}
\be
\label{Vaxi}
S=-{1 \over 16\pi G}
\int d^{d+1}x \sqrt{-G}\left(R + \lambda^2 
  - {1 \over 2} G^{\mu\nu}\partial_\mu \phi \partial_\nu \phi 
+ {1 \over 2}\e^{2\phi}
G^{\mu\nu}\partial_\mu \chi \partial_\nu \chi \right)\ .
\ee
Again the solution may be found and its boundary gauge theory 
interpretation presented.
By properly choosing the constants of the integration, 
the boundary field theory can be regarded as 
asymptotically free.
We also find the dilaton field behaves, when $y$ is small as 
$\phi \sim - \sqrt{d(d-1) \over 2}\ln y$, 
which corresponds to $c<0$ case in the pure dilaton case. 
Since the behavior of the metric is 
essentially identical with the pure dilaton case, 
the supergravity description would be valid even for large $L$ 
and the confinement would be predicted.
Especially, one can investigate the supersymmetric background. 
Since that is background where the fermion 
fields, that is, dilatino $\xi$ and gravitino $\psi_\mu$ vanish, if 
the variation under some of the supersymmetry transformations of 
these fermionic fields vanishes the corresponding supersymmetries 
are preserved. We find that although all supersymmetries 
break down in general, half of the supersymmetries survives.
This situation does not depend on $k$. Such solution corresponds 
to super YM theory with asymptotically free gauge coupling 
and two supersymmetries broken. 
 In the same way, one can study 
dilatonic AdS-like backgrounds in the presence of other fields, 
like fermionic ones. Our next problem will be to take into account 
temperature in the above picture.

\subsection{Dilatonic AdS black holes: running gauge coupling and 
$q\bar q$-potential at non-zero temperature} 

It has been realized that planar 
${\rm AdS}_5$ BH is dual to a thermal state of ${\cal N}=4$ 
super YM theory. The corresponding coupling constant 
dependence has been studied in ref.\cite{GKT,TY1} based on 
earlier study of SG side free energy in ref.\cite{GKPt}. 
Spherical AdS BH shows the finite temperature phase transition 
\cite{HP} which may be used to realize the confinement 
in large $N$ theory at low temperatures \cite{Wtt1}.
In this section, based on \cite{NOt}, we give the example of 
the deformation of IIB SG ${\rm AdS}_5\times{\rm S}_5$ 
background with non-trivial dilaton where AdS sector is described 
by BH (hence, temperature appears). Then, running gauge coupling 
of gauge theory at non-zero temperature may be defined by exponent of 
dilaton. We give some approximate solutions of 
IIB SG\footnote{These solutions presumably describe thermal states 
of (non)-SUSY gauge theory which descends from ${\cal N}=4$ super YM 
after breaking of SUSY and conformal invariance or another vacuum 
of the same super YM theory.} with such 
properties where running coupling shows power-like behavior in the 
temperature (in the expansion on radius).

We start from the action of dilatonic gravity
in $d+1$ dimensions in (\ref{ViR}). 
We assume the $(d+1)$-dimensional metric is given by
\be
\label{ViiT}
ds^2=-\e^{2\rho}dt^2 + \e^{2\sigma}dr^2 
+ r^2 \sum_{i,j=1}^{d-1}g_{ij}dx^i dx^j\ .
\ee
Here $g_{ij}$ does not depend on $r$ and it is the metric in the 
Einstein manifold. We also introduce new variables $U$ and $V$ by
\be
\label{Vxb}
U\equiv \e^{\rho + \sigma}\ ,\quad 
V\equiv r^{d-2}\e^{\rho - \sigma}\ . 
\ee
When the dilaton vanishes (or constant), the solution is given by
\be
\label{VxviiiB}
U=1\ , \quad V=V_0
\equiv {kr^{d-2} \over d-2} + {\lambda^2 \over d(d-1)}r^d 
  - \mu\ .
\ee
Here $\mu$ corresponds to the mass of the black hole. $k=0$,
positive or negative corresponds to planar, spherical or
hyperbolic AdS BH, respectively. 
When $\mu=0$, the solution is isomorphic to AdS. If we choose 
$k<0$, the metric has the following form:
\be
\label{VxixB}
ds^2=-{(r^2 - r_0^2) \over l^2}dt^2 + {l^2 \over (r^2 - r_0^2) }dr^2 
+ r^2 \sum_{i,j=1}^{d-1}g_{ij}dx^i dx^j\ .
\ee
Here
\be
\label{VxxT}
l^2\equiv {d(d-1) \over \lambda^2}\ ,\quad 
r_0\equiv l \sqrt{-{k \over d-2}}\ .
\ee
The obtained AdS metric has a horizon at $r=r_0$. 
 From the behavior of the metric at $r\sim r_0$ 
after Wick-rotated, we find a temperature $T$: 
\be
\label{Vxxiv}
T={r_0 \over 2\pi l^2}={1 \over 2\pi l}\sqrt{-k \over d-2}\ .
\ee

One can now  consider the perturbation by assuming the 
dilaton is small \cite{NOt}. 
We will concentrate on the case of type IIB SG in $d=4$, 
by putting $\alpha={1 \over 2}$. Note that in this approximation 
the radius is away from horizon. Near-horizon regime may be 
discussed independently. 

For $\mu=0$ and $k<0$ case, the leading term for the dilaton $\phi$ 
is is found as 
\bea
\label{Vxxv}
\phi&=&\phi_0 +cl^2\left\{{1 \over 2r_0^4}\ln\left(1 
  - {r_0^2 \over r^2}\right) + {1 \over 2r_0^2 r^2}\right\} \nn
&=& \phi_0 + c\left\{{1 \over 2l^6(2\pi T)^4}\ln\left(1 
  - {l^4 (2\pi T)^2 \over r^2}\right) 
+ {1 \over 2l^2(2\pi T)^2 r^2}\right\}\ .
\eea
which gives the temperature dependent running dilaton. We should 
note that there is a singularity in the dilaton field at the horizon 
$r=r_0=2\pi l^2 T$. The fact that dilaton may become singular at 
IR has been mentioned already in two-boundaries AdS solution of 
IIB SG in ref.\cite{NO10}. It is also interesting that when $r$ is 
formally less than $r_0$ 
then dilaton (and also running coupling) 
becomes imaginary.

Since the string coupling is given by $g=g_s\e^\phi$ ($g_s$ : 
constant), 
we find the behavior of $g$ when $r$ is large and $c$ is small. 
Since $r$ is the length scale corresponding to the radius of the 
boundary manifold, $r$ can be regarded as the energy scale of the 
field theory on the boundary \cite{SW,PP1}. Therefore the 
beta-function is given by \cite{NOt}
\be
\label{Vciii}
\beta(g)=r{dg \over dr}=-4\left(g-g_s\right)
+ {2^{5 \over 2} \over3}\left(2\pi T\right)^2l^3 g_s 
\left( {g_s - g \over c g_s} \right)^{3 \over 2}\ .
\ee
The first term is usual and universal. 
The second term defines the temperature dependence.

Let us comment on the case of high $T$. As we consider the 
behavior near the boundary, first we take $r$ to be large. After 
that we consider the case of high $T$. In this case $r\gg Tl^2$ 
and we can consider the large $T$ case in the expression 
(\ref{Vciii}). The problem might happen when $r\sim Tl^2$. In this 
case, we need to solve Eq.(\ref{Vxxv}) with respect to $r$ as a 
function of $T$ and $\phi$ or coupling: $r=r(g,T)$. 
Then from (\ref{Vxxv}), we find the following 
expression of the beta-function:
\be
\label{VgTii}
\beta(g) \sim \left.r{dg \over dr}\right|_{r=r(g,T)}
={g_s c l^2 \over r(g,T)^4 
\left(1 - {l^4 (2\pi T)^2 \over r(g,T)^2}\right) }\ .
\ee
In case $r$ is large, the above equation reproduces (\ref{Vciii}).
We can also consider the case that the last term 
in (\ref{Vxxv}) is larger than the second term which contains 
$\ln (\cdots)$. In this case, the leading behavior of 
the beta-function has the following form:
$\beta(g)\sim - 2 \left(g-g_s\right) + \cdots$.
This beta-function presumably defines strong coupling regime
of non-SUSY gauge theory at high temperature.
It is interesting to note that in perturbative gauge theory 
at non-zero 
temperature the running gauge coupling contains not only 
standard logarithms of $T$ but also terms linear on $T$  
(see ref.\cite{volodya} and references therein).
Of course, in our case we have not asymptotically free theory 
but the one with stable fixed point.

Now we consider the correction for $V$ and $U$ , writing them 
in the following form: $V=V_0+c^2 v$, $U=1+c^2 u$, 
where $v=v(s)$, $u=u(s)$ are given explicitly in \cite{NOt}. 
Here $s=-{2 r^2 \over kl^2}$. Then we find that $U$ and $V$ or 
$\e^{2\rho}$ and $\e^{2\sigma}$ have the singularity at the 
non-perturbed horizon corresponding to $s=1$. Eq.(\ref{Vxxv}) tells 
also that the dilaton field is also singular there. In other words, the 
expansion with respect to $c^2$ breaks down when $s\sim 0$. Therefore 
the singularity in $U$, $V$ would not be real one.
 
In order to investigate the behavior in near-horizon regime 
we assume that the radius of the horizon is large 
and use ${1 \over r}$ expansion: 
$V={r^4 \over l^2} + {kr^2 \over 2} + {a \over r^4} + 
{\cal O}\left(r^{-6}\right)$.
We put the constant term to be zero  assuming that the black hole 
mass vanishes. Then $a={c^2l^2 \over 48}$ 
and $\e^\phi$, $V$ and $U$ have the following forms:
\bea
\label{Vr3}
&& \e^\phi=\e^{\phi_0}\left(1 - {cl^2 \over 4 r^4} 
+ {\cal O}\left(r^{-6}\right)\right) \nn
&& V={r^4 \over l^2} + {kr^2 \over 2} + {c^2l^2 \over 48 r^4} + 
{\cal O}\left(r^{-6}\right) \ ,\quad 
U=1-{c^2l^4 \over 192 r^8} + {\cal O}\left(r^{-10}\right)\ .
\eea
 From the equation $V=0$ we find the position of the horizon 
and the correction to the temperature:
\be
\label{Vr4}
r=r_h\equiv l\sqrt{-{k \over 2}}\left(1 
  - {c^2 \over 6 k^4 l^4}\right)\ ,\quad 
T={1 \over 2\pi l}\sqrt{-{k \over 2}} 
  - {c^2\left(-{k \over 2}\right)^{-{7 \over 2}} 
\over 192 l^5}\ .
\ee

We can also analyze the potential between quark 
and anti-quark \cite{Mld2} by evaluating the Nambu-Goto action in 
(\ref{Vrg5}) as in the previous section. 
We consider the static configuration $x^0=\tau$, 
$x^1\equiv x=\sigma$, $x^2=x^3=\cdots=x^{d-1}=0$ and $r=r(x)$. 
Choose the coordinates on the boundary manifold so that the 
line given by $x^0=$constant, 
$x^1\equiv x$ and $x^2=x^3=\cdots=x^{d-1}=0$ is geodesic and 
$g_{11}=1$ on the line. 
The orbit of $r$ can be obtained by minimizing the action $S$ 
or solving the Euler-Lagrange equation. By the similar procedure 
in the previous section, we find the following expression for  
quark-antiquark potential \cite{NOt}
\bea
\label{Vrg15B}
E_{q\bar q}(L)&=&{lAC \over L} - {l^3 \left(2\pi T\right)^2 \over 2}
\left({BC \over A^2} + {D \over A}
\right)L + {\cal O}\left(L^3\right) \\
C&=&2\int_0^\infty dt\,\left\{ 
{\cosh^2 t \over \left(\cosh^2 t + 1\right)^{1 \over 2}} 
  -\sinh t\right\} -2 
=-1.19814... \nn
D&=& 2\int_0^\infty {dt 
\over \left(\cosh^2 t + 1\right)^{3 \over 2}}
=0.711959 \ .\nonumber
\eea
Here we neglected the $L$ independent term.
We should note that next-to-leading term is linear in $L$, which 
might be relevant to the confinement. 
For the confinement, it is 
necessary that the quark-antiquark potential behaves as 
$E_{q\bar q}\sim a L$ 
with some positive constant $a$ for large $L$. For high temperature, 
it is usually expected that there occurs the phase transition to 
the deconfinement phase, where the potential behaves 
as Coulomb force, $E_{q\bar q} \sim {a' \over L}$. 
Since ${BC \over A^2} + {D \over A}>0$ and 
$k<0$, the potential is repulsive. 
The leading term expresses the repulsive but shows the 
Coulomb like behavior. The next-leading-term tells that 
the repulsive force is long-range than Coulomb force. 

The expression (\ref{Vrg15B}) is correct even at high 
temperature if $L$ is small. If $L$ is large and the orbit 
of string approaches to the horizon and/or enters inside 
the horizon, the expression would not be valid. Since the 
horizon is given by (\ref{VxxT}), the expression (\ref{Vrg15B}) 
would be valid if $L\ll A\sqrt{-{2 \over k}}=2\pi A l T$. 
The above condition makes difficult to evaluate 
the potential quantitively by the analytic calculation when $L$ 
is large and numerical calculation would be necessary. 
In order to investigate the qualitive behavior of the potential 
when $L$ is large, we consider the background where the dilaton 
is constant $\phi=\phi_0$, which would tell the effect of the 
horizon or finite temperature. As $c=0$ when the dilaton is 
constant, we can use the solution in (\ref{VxviiiB}). Then by the 
calculation similar to (\ref{Vrg15B}) but without assuming $L$ 
is small, we obtain the following 
expression of the quark-antiquark potential:
\be
\label{VPotlL}
E_{q\bar q}=r_{\rm min}
\int_{-\infty}^\infty dt \sinh t \left\{ \left( 1 - {1 \over 
\cosh^2 t}\cdot {1-{r_0^2 \over r_{\rm min}^2} \over 
\cosh^2 t - {r_0^2 \over r_{\rm min}^2}} \right)^{-{1 \over 2}}
  -1 \right\} + 2\left(r_D - r_{\rm min}\right)\ .
\ee
Here $r_{\rm min}$ is the minimum of $r$ and 
constant $-1$ in $\{\ \}$ and the last term correspond to 
the subtraction of the 
self-energy. The integration in (\ref{VPotlL}) converges and the 
integrand is monotonically decreasing function of 
${1 \over r_{\rm min}}$ if $r_{\rm min}$ is larger than the radius 
of the horizon $r_0$ : $r_{\rm min}>r_0$ and vanishes in the limit of 
$r_{\rm min}\rightarrow r_0$. Therefore if $r_{\rm min}$ 
decreases and approaches to $r_0$ when $L$ is large, which seems to 
be very natural, the potential $E_{q\bar q}$ approaches to a 
constant $E_{q\bar q}\rightarrow 2\left(r_D - r_0\right)$ 
and does not behave as a linear function of $L$ \cite{NOt}. 
This tells that the 
quark is not confined. This effect would correspond to deconfining 
phase of QCD in the finite temperature. 

Similarly, one can consider another interesting case like 
that $k=0$ and $\mu\neq 0$ \cite{NOt}, 
which corresponds to the throat limit of D3-brane 
\cite{GKT,GKPt}. In a way similar to $k<0$ and $\mu=0$ case, 
the corrections coming from the dilaton were calculated 
for the beta-function and quark-antiquark potential 
with the indication to possible confinement \cite{NOt}. 
One can consider even more general case where either  $k$ or $\mu$ 
do not vanish (for details, see \cite{NOt}). 
The analysis of this case may be done in the same way as above.

It is also extremely interesting to study thermodynamics of 
obtained dilatonic AdS BHs. 
After Wick-rotating the time variable by $t\rightarrow i\tau$, 
the free energy $F$ can be obtained from the action $S$ in (\ref{ViR}) 
where the classical solution is substituted after the 
Wick-rotation. 
The expression of $S$, however, contains the divergence 
coming from large $r$. In order to subtract the divergence, we 
regularize $S$ by cutting off the integral at 
a large radius $r_{\rm max}$. After that we subtract the 
divergent part.
In case of $k<0$ and $\mu=0$, we subtract it by using 
the extremal solution with $c=0$. Then we obtain 
the following expression 
\cite{NOt}
\be
\label{VF6}
F=-{V_3 \over 2\pi G l^2}\left(
{ 5l^8\left(2\pi T\right)^4 \over 32} 
+ {c^2 \over 768 l^4 \left(2\pi T\right)^4}\right)\ .
\ee
In order to get the entropy ${\cal S}$, 
we need to know $T$ dependence of $V_3$ although $V_3$ is infinite. 
Since $k$ is proportional 
the curvature, $V_3$ would be proportional to $k^{-{3 \over 2}}$. 
Therefore the entropy ${\cal S}$ and mass (energy) $E$ are given by
\bea
\label{VF8}
{\cal S}&=&-{dF \over dT}={V_3 \over 2\pi G l^2 T }\left(
{ 25l^8\left(2\pi T\right)^4 \over 64} 
+ {49c^2 \over 1536 l^4 \left(2\pi T\right)^4}\right) \nn
E&=&F+T{\cal S}={V_3 \over 2\pi G l^2}\left(
{ 15 l^8\left(2\pi T\right)^4 \over 64} 
+ {47 c^2 \over 1536 l^4 \left(2\pi T\right)^4}\right)\ .
\eea
In terms of string theory correspondence\cite{GKT}, 
the parameters $G$ and $l$ 
are given by
$l^4=2g_{YM}^2 N{\alpha'}^2$ and 
$Gl={\pi g_{YM}^2{\alpha'}^2 \over N}$.
Here the Yang-Mills coupling $g_{YM}$ is given by the string 
coupling $g_s$: $g_{YM}^2=2\pi g_s$ and $N$ is the number of the 
coincident D3-branes.
As $V_3$ is now dimensionless, we multiply $l^3$ with $V_3$:
$\tilde V_3\equiv l^3 V_3$. Then Eqs.(\ref{VF6}) and (\ref{VF8}) 
can be rewritten as follows:
\bea
\label{VF10b}
F&=&-{\tilde V_3 \over 2\pi^2 }\left(
{ 5N^2 \left(2\pi T\right)^4 \over 16} 
+ {c^2 \over 3072 l^4 g_{YM}^6 N{\alpha'}^6
\left(2\pi T\right)^4}\right)\nn
{\cal S}&=&{V_3 \over 2\pi^2 T }\left(
{ 25N^2\left(2\pi T\right)^4 \over 32} 
+ {49 c^2 \over 6144 g_{YM}^6 N{\alpha'}^6
\left(2\pi T\right)^4}\right) \nn
E&=&{V_3 \over 2\pi G l^2}\left(
{ 15 N^2\left(2\pi T\right)^4 \over 32} 
+ {47 c^2 \over 6144 g_{YM}^6 N{\alpha'}^6
\left(2\pi T\right)^4}\right)\ .
\eea

For $k=0$ and $\mu>0$ case, we can obtain the thermodynamical 
quantities in a similar way, where we  
regularize $S$ by subtracting the solution 
with $\mu=0$ and $c=0$. The results are given in \cite{NOt}. 
The leading behavior of $F$ and $S$ are consistent with \cite{GKT}.
As we used ${1 \over r}$ expansion, the next to leading terms become 
dominant when the radius of horizon $r_h$ is large 
and the parameter $c$ is not very small. 
One has to remark that leading term in above free energy describes 
the strong coupling regime free energy for ${\cal N}=4$ 
super YM theory with 
the usual mismatch factor 3/4 if compare with perturbative free 
energy. 

Thus, we discussed applications of dilatonic AdS gravity for 
construction of dilatonic AdS BHs which may be used for 
description of gauge theory at non-zero temperature.

\subsection{Strong coupling limit of ${\cal N}=2$ superconformal field theory 
free energy from AdS black hole thermodynamics}

In the previous section we gave one example where AdS BH gives the 
strong coupling limit of ${\cal N}=4$ super YM theory free energy 
\cite{GKPt,GKT} which differs by factor $3/4$ with perturbative 
result (boundary QFT) in the leading order of $1/N$ 
expansion. Note that thermodynamics of  super YM theory 
in relation with AdS/CFT correspondence has been discussed in 
large number of  works\cite{Wtt3,TY1,IMSY1,RTL1,FM1,BISY1,CEJM,HHP1,BKR1,
HO1,
KrS1,PR1,KSS1,GOl1,Lnd1,EGM1,CG1,KrTy1,NOt,BCM,Snd1,LL1,HrOb}
(see also references therein). 
It is quite interesting to understand what happens in the next 
order of $1/N$ expansion. Clearly that such analysis should be 
related with higher derivatives (HD) terms on SG side. In more 
general framework HD terms may help in better understanding of 
AdS/CFT correspondence or even in formulation of new versions 
of bulk/boundary conjecture.

That is the purpose of this section to show the role of HD terms 
in bulk action in AdS/CFT correspondence as it was done in \cite{NOs}. 
The conformal anomaly of 
$d=4$, ${\cal N}=2$ and ${\cal N}=4$ SCFTs (bulk side calculation) 
has been found up to next to leading order in the $1/N$ expansion 
in refs. \cite{BNG,NO9}. Even in next to leading order term it 
coincides with QFT result (for gravity side derivation of trace 
anomaly, mainly in ${\cal N}=4$ super YM case, see also 
refs.\cite{NOa,NT1,BK1,MV1,MN1,Ho}). The ${\cal N}=2$ theory 
with the gauge group 
$Sp(N)$ arises as the low-energy theory on the world volume on $N$ 
D3-branes sitting inside 8 D7-branes at an O7-plane 
\cite{Sen1,BDS1,ASYT1,DLS1,AFM1}. 
The string theory dual to this theory has been conjectured to be 
type IIB string theory on $AdS_5\times X^5$ where $X_5=S^5/Z_2$ 
\cite{FS}, whose low energy effective action is given by
\be
\label{Vbng3} 
S=\int d^5x \sqrt{g}\left\{{N^2 \over 4\pi^2} \left(R + 12 \right) 
+ {6N \over 24\cdot 16\pi^2}R_{\mu\nu\rho\sigma}
R^{\mu\nu\rho\sigma}\right\}\ .
\ee
The overall factor of the action is different by ${1 \over 2}$ 
from that of the action which corresponds to ${\cal N}=4$ $SU(N)$ 
gauge theory. The latter action is given by type IIB (compactified) 
string theory on $AdS_5\times S^5$. The factor ${1 \over 2}$ comes 
from the fact that the volume of $X^5=S^5/Z_2$ is half of $S^5$ due 
to $Z_2$. Note that Riemann curvature squared term in the above bulk 
action is deduced from heterotic string via heterotic-type I duality 
\cite{Arkadii} (dilaton is assumed to be constant). 
The interesting fact about theory (\ref{Vbng3}) is that it may be 
rewritten also as dilatonic gravity.

Then the equations of motion have the following form:
\be
\label{Vaiii}
0=-{c \over 2}g_{\zeta\xi}\left(
R_{\mu\nu\rho\sigma} R^{\mu\nu\rho\sigma}
+ {1 \over \kappa^2} R - \Lambda \right)
+ c R_{\zeta\mu\nu\rho} R_\xi^{\mu\nu\rho}
+ {1 \over \kappa^2} R_{\zeta\xi} 
  -4 c D_\nu D_\mu R^{\mu\ \nu}_{\ \zeta\ \xi}\ .
\ee
Here ${1 \over \kappa^2}={N^2 \over 4\pi^2}$, 
$c={6N \over 24\cdot 16\pi^2}$, 
$\Lambda=-{12 \over \kappa^2}=-{12N^2 \over 4\pi^2}$. 
One can treat the next-to-leading term of order $N$ as perturbation 
of order $N^2$ terms. 
Solving the equation perturbatively, one finds that BH-like metric 
is modified by 
\bea
\label{Vsp1bb}
&& ds^2=g_{\mu\nu}dx^\mu dx^\nu
=-\e^{2\rho}dt^2+\e^{-2\rho}dr^2 + r^2\sum_{i=1}^3\left(
dx^i \right)^2 \nn
&& \e^{2\rho}={1 \over r^2}\left\{-\mu + \left(1+{2 \over 3}\epsilon 
\right)r^4 +2\epsilon {\mu^2 \over r^4} \right\}\ , \quad 
\epsilon \equiv c\kappa^2 = {1 \over 16N}\ .
\eea
The radius $r_h$ of the horizon and the temperature $T$ 
are given by
\be
\label{Vsp3}
r_h\equiv \mu^{1 \over 4}\left(1 - {2 \over 3}\epsilon \right)\ ,
\quad T={\mu^{1 \over 4} \over \pi }\left(1 - 2\epsilon\right)
={\mu^{1 \over 4} \over \pi }\left(1 - {1 \over 8N}\right)\ .
\ee
It is interesting to calculate 
the thermodynamical quantities like free energy.
After Wick-rotating the time variables by $t\rightarrow i\tau$, 
the free energy $F$ can be obtained from the action $S$ in 
(\ref{Vbng3}) where the classical solution is substituted:
$F={1 \over T}S$. 
Using (\ref{Vaiii}) and (\ref{Vsp1bb}), we find
\bea
\label{Vsp5}
S&=& {N^2V_3 \over 4\pi^2 T}\int_{r_h}^\infty dr r^3 
\left\{ 8 - {2\epsilon \over 3}\left(40 + {72 \mu^2 \over r^8}
\right)\right\}\ .
\eea
Here $V_3$ is the volume of 3d flat space and we assume $\tau$ has 
a period of ${1 \over T}$. The expression of $S$ contains the 
divergence coming from large $r$. After subtraction of this divergence 
one gets \cite{NOs} 
\be
\label{Vsp8}
F=-{N^2V_3\mu \over 4\pi^2 }\left(1 + {3 \over 4N}\right)\ .
\ee
The entropy ${\cal S}$ and the mass (energy) $E$ are given by
\be
\label{Vsp9}
{\cal S}={N^2V_3L^6\left(\pi T\right)^4  \over \pi^2 T }
\left(1 + {3 \over 4N}\right) \ ,\quad 
E={3N^2V_3L^6\left(\pi T\right)^4  \over 4\pi^2 }
\left(1 + {3 \over 4N}\right)\ .
\ee

We now compare the above results with those of field theory of 
${\cal N}=2$ $Sp(N)$ gauge theory. ${\cal N}=2$ theory contains 
$n_V=2N^2+N$ vector multiplets and $n_H=2N^2+7N-1$ hypermultiplets. 
Vector multiplet consists of two Weyl fermions, one 
complex scalar and one real vector which gives 4 bosonic (fermionic) 
degrees of freedom on shell and hypermultiplet contains two complex 
scalars and two Weyl fermions, which also gives 4 bosonic (fermionic) 
degrees of freedom on shell \cite{SbW}. Therefore there appear 
$4\times \left(n_V + n_H\right)=16 \left(N^2 + 2N 
  -{1 \over 4}\right)$ boson-fermion pairs. In the limit which we 
consider, the interaction between the particles can be neglected. 
The contribution to the free energy from one boson-fermion pair in 
the space with the volume $V_3$ can be easily estimated 
\cite{GKPt,GKT}. Each pair gives a contribution to the free energy of 
${\pi^2 V_3T^4 \over 48}$. Therefore the total free energy 
$F$ should be \cite{NOs}
\be
\label{Vff1}
F=-{\pi^2 V_3 N^2 T^4 \over 3}\left(1+{2 \over N}-{1 \over 4N^2}
\right)\ .
\ee
Comparing (\ref{Vff1}) with (\ref{Vsp8}), there is the difference 
of factor ${4 \over 3}$ in the leading order of $1/N$ as observed 
in \cite{GKPt,GKT}. 

Hence, we presented the strong coupling limit of free energy in 
${\cal N}=2$ SCFT from SG side up to next to leading order term 
(it was generated by Riemann curvature squared term). Its weak 
coupling limit (\ref{Vff1}) obtained from QFT side cannot be 
presented as strong coupling limit free energy multiplied to some 
constant. This only holds for leading order terms where mismatch 
multiplier is 3/4.  The next to leading term in (\ref{Vff1}) should 
be multiplied to $9/32$ in order to produce the corresponding term 
in (\ref{Vsp8}). 

To conclude, in this Chapter, we discussed the role of classical 
solutions of dilatonic 
gravity  (when dilaton is not constant) in bulk/boundary 
correspondence for getting strong coupling limit in (super) gauge 
theory. Of course, we only discussed few topics related 
with our main purpose: to describe some properties and 
consequences of (quantum) dilatonic gravity. To understand other 
topics in this quickly growing field of research one is 
advised to consult recent review \cite{AGMOO}.

\section{Discussion}

Dilatonic gravity (or scalar-tensor gravity) is the generalization 
of Einstein theory where scalar partner is added to usual tensor 
graviton. In this review, we discussed various aspects, methods, 
results and examples related with dilatonic gravity in quantum 
regime (or, in classical framework where dual, quantum 
interpretation may work). The quantum effective action 
calculated in the models of 2d and 4d dilatonic gravity (with dilaton 
coupled matter) helped to give the answers to number of questions. 
In particular, the renormalizability and quantum equivalence of 
dilatonic gravities, the existence of fixed points in 
renormalization group study near two dimensions, quantum 
corrections to thermodynamics of (dilatonic) 
black holes and to early Universe evolution have been investigated. 
In all cases, the corresponding results and examples are given with 
sufficient details and may be easily understood and generalized for 
another gravitational backgrounds. The number of interesting 
phenomena caused by quantum effects like Hawking radiation in 
dilatonic black holes, anti-evaporation of multiply horizon 
black holes, inducing of primordial wormholes in Grand Unified 
Theories and realization of non-singular inflationary Universe 
are presented.

The study of classical Anti-de Sitter-like solutions of 5d 
dilatonic gravity (bosonic sector of IIB supergravity) may give 
the answers to questions in the boundary supersymmetric quantum 
gauge theory via bulk/boundary correspondence. In this way, 
dilatonic gravity explains how to find the quark-antiquark 
potential (with indication to possible confinement), running 
gauge coupling and free energy (or entropy) of boundary SUSY 
gauge theory.

We were addressing mainly the aspects of quantum dilatonic 
gravity. As a result, there are many problems left outside of 
this review. First of all, we did not discuss the comparison 
with observational data (macroscopic manifestations of dilaton) 
indicating sometimes to difficulties related with dilatonic gravity. 
The reason is very natural: one can always suppose 
that dilatonic gravity is acceptable only near Planck scale 
and after that dilaton should quickly decay by some mechanism. From 
another side, as it was already explained earlier the dilaton may 
be often regarded as purely technical tool to simplify the 
consideration (say, 4d Einstein gravity after reduction looks like 
effective 2d dilatonic gravity, etc.). 

Second, while we mentioned that dilaton is unavoidable element 
of string theory there was no discussion of dilaton from stringy 
point of view (forgetting the fact that some of the theories under 
consideration may be string-inspired ones). However, one should 
note that there may be mechanisms (see, for example, \cite{damour}) 
how to make the presence of even massless dilaton in stringy 
gravity to be consistent with observational cosmology. 

Third, it would be really interesting to develop further the 
elements of inflationary dilatonic cosmology (fifth chapter) 
in order to find the deviations from the standard cosmological 
predictions. One example of this sort comes from the calculation 
of cosmic background radiation spectrum from the dilatonic theory 
where dilaton appears as conformal sector of infrared  quantum 
gravity \cite{mottola}.

Fourth, as it follows from fourth chapter the qualitatively 
new phenomena (like the anti-evaporation of black holes) may 
be expected in the quantum considerations around (dilatonic) 
black holes. As so far, the corresponding treatment is 
semi-classical , the new strong coupling methods to study 
the gravitational physics at extremal conditions should be developed. 
As it was shown in last chapter one can often realize the classical
gravitational systems (like Anti-de Sitter black holes) as 
quantum non-gravitational systems 
(supersymmetric gauge theory) using AdS/CFT correspondence. 
Hence, it is quite possible that right strong coupling methods 
in quantum gravity may appear from deeper study of above 
bulk/boundary correspondence. In particular, the hopes are 
related with the possibility to construct the complete,  non-local 
but finite gravitational effective action (calculated using kind 
of strong coupling expansion or applicable at extreme conditions). 
This one and many more similar questions in dilatonic quantum 
gravity await their final resolution.

\subsection*{Acknoweledgements.} 

We would like to thank I. Brevik, 
I.L. Buchbinder, A.A. Bytsenko, R. Bousso, S.J.Gates, B. Geyer, 
S.W. Hawking, S. Ichinose, P.M. Lavrov, E. Mottola, 
I. Oda, O. Obregon, V.V. Obukhov, K.E. Osetrin, A. Sugamoto, 
A.A. Tseytlin, P. van Nieuwenhuizen and S. Zerbini for useful 
discussions of related questions. The research by SDO has been 
supported in part by CONACyT(CP, ref.990356 and grant 28454E) 
and in part by RFBR.

\end{document}